\documentclass[a4paper,10pt]{article}

\usepackage{amssymb}
\usepackage{amsmath}
\usepackage[usenames,dvipsnames]{color}
\usepackage{hyperref}
\usepackage[left=.8in,right=.8in,top=.8in,bottom=.8in]{geometry}              

\usepackage{xcolor}
\usepackage{mathpazo}
\usepackage[normalem]{ulem}
\usepackage{palatino}
\usepackage{mathrsfs}
\usepackage[mathscr]{euscript}
\usepackage{lscape}
\usepackage{cancel}
\usepackage{bm}
\usepackage{bbm}
\usepackage{multirow} 
\usepackage{amsfonts}
\usepackage{amssymb}
\usepackage{graphicx}
\usepackage{amsmath}
\usepackage[usenames,dvipsnames]{color}
\usepackage{hyperref}
\usepackage{epstopdf,epsfig}
\usepackage{color}

\usepackage{cite}

\hypersetup{
    colorlinks=true,         % false: boxed links; true: colored links, false is default
    linkcolor=MidnightBlue,          % color of internal links, red is default
    citecolor=BrickRed,        % color of links to bibliography, 'green' is default
    urlcolor=MidnightBlue             % color of external links, cyan is default
}

\def\be{\begin{equation}}
\def\ee{\end{equation}}
\def\bea{\begin{eqnarray}}
\def\eea{\end{eqnarray}}

\def\fL{{\mathfrak{L}}}
\def\fI{{\mathfrak{I}}}
\def\d{{\mathrm{d}}}

\renewcommand{\hat}{\widehat}
\renewcommand{\bar}{\overline}

\newtheorem{defi}{Definition}
\newtheorem{theo}{Theorem}

\newcommand{\diby}[2]{\ensuremath{\frac{\delta #1}{\delta #2}}}

\newcommand{\order}[1]{\ensuremath{\mathcal{O}(#1)}}
\newcommand{\st}[1]{\textrm{\tiny{#1}}}
\renewcommand{\tilde}{\widetilde}

\def\be{\begin{equation}}
\def\ee{\end{equation}}
\def\bea{\begin{eqnarray}}
\def\eea{\end{eqnarray}}

\def\sfrac{\textstyle \frac }
\title{Quantum gravity  in timeless configuration space}
\author{\bf Henrique Gomes\footnote{\href{mailto:gomes.ha@gmail.com}{gomes.ha@gmail.com}}\\\it Perimeter Institute for Theoretical Physics\\ \it 31 Caroline Street, ON, N2L 2Y5, Canada}
\begin{document}
\maketitle

\begin{abstract}
On the path towards quantum gravity we find friction between temporal relations in quantum mechanics (QM) (where they are fixed and field-independent), and in general relativity (where they are field-dependent and dynamic). This paper aims to attenuate that friction, by encoding gravity in the timeless configuration space of spatial fields with dynamics given by a path integral.  The framework demands that boundary conditions for this path integral be uniquely given, but unlike other approaches where they are prescribed --- such as the no-boundary and the tunneling proposals ---   here I postulate basic principles  to identify boundary conditions in a large class of theories. Uniqueness arises only if a reduced configuration space can be defined and if it has a profoundly asymmetric fundamental structure. These requirements place strong restrictions on the field and symmetry content of theories encompassed here; shape dynamics  is one such theory.  When these constraints are met, any emerging theory will have a Born rule given merely by a particular volume element built from the path integral in (reduced) configuration space. Also as in other boundary proposals, Time, including space-time, emerges as an effective concept; valid for certain curves in configuration space but not assumed from the start. When some such notion of time becomes available, conservation of (positive) probability currents ensues. I show that, in the appropriate limits, a Schroedinger equation dictates the evolution of weakly coupled source fields on a classical gravitational background. Due to the asymmetry of reduced configuration space, these probabilities and currents avoid a known difficulty of standard WKB approximations for Wheeler DeWitt in minisuperspace: the selection of a unique Hamilton-Jacobi solution to serve as background. I illustrate these constructions with a simple example of a full quantum gravitational theory (i.e. not in minisuperspace) for which the formalism is applicable, and give a formula for calculating gravitational semi-classical relative probabilities in it.% Although this simple model gives the same likelihood for the evolution of all TT gravitational modes, there is evidence that a slightly more complicated model would favor modes with the smallest eigenvalues of the Laplacian and thus could drive towards homogeneity.
\end{abstract}

\tableofcontents
\section{Introduction.}

\subsection{Motivation}
Conventional approaches to quantum gravity suffer from  serious conceptual and technical problems. On the technical side, most focus has been concentrated on the perturbative regime and  on  extending the theory to higher energy domains  while maintaining some semblance of unitarity. Conceptual problems on the other hand, are starker when trying to make sense of the {non-perturbative theory}. There we witness a violent clash between, on the one hand, the standard \textit{quantum mechanical} properties of: 
\begin{itemize}
\item[{\bf 1a)}]  the quantum superposition principle
\item[{\bf 2a)}] the non-locality of instantaneous measurements; 
\end{itemize}
and,  on the other, the \textit{general relativistic} properties of
\begin{itemize}
\item[{\bf 1b)}] a fixed causal structure %which we do not know how to describe in a state of superposition 
 \item[{\bf 2b)}]   space-time covariance
 \end{itemize}
 These two pairs still leave out difficulties quantum cosmology faces arising directly from the measurement problem in the foundations of quantum mechanics, which the present work also aims to address. Of course, the non-perturbative regime poses infinitely more challenging obstacles to quantitative treatment. Nonetheless, resolving a conceptual incongruence between full quantum mechanics and general relativity might point us to different approaches, friendlier to quantum gravity.  The aim of this paper is indeed to point out a framework in which such incongruences are resolved. The framework suggests the adoption of  alternative formulations of gravity at a fundamental level, such as shape dynamics \cite{SD_first} (see also \cite{Flavio_tutorial} and references therein). 

The first clash, between {\bf 1a}  and {\bf 1b},  arises because, intuitively, a quantum superposition  makes sense for space-like separated components of a system -- i.e. belonging to a single constant-time hypersurface. But, to be meaningful, the label \lq{}space-like\rq{} already requires a {fixed, unique},\footnote{Unique up to conformal transformations.}  gravitational field.

Accounting for back-reaction, it becomes obvious a problem lurks here. We understand how non-back-reacting matter degrees of freedom can be in states of superposition in  a background spacetime, e.g. what we usually mean when we write something of the sort $|\psi\rangle=|\psi_1\rangle+|\psi_2\rangle$ in the position space representation. But due to the universality of the gravitational interactions, back-reaction would necessarily put the gravitational field in a state of superposition as well. How does a superposed state gravitate -- and therefore affect the causal structure? The problem is also present in covariant axiomatic QFT language, where one demands that operators with space-like separated support, $\mathcal{O}_1, \mathcal{O}_2$ commute, $[\mathcal{O}_1,\mathcal{O}_2]=0$. Again, \lq{}space-like\rq{} presumes a definite property for the very field we would like to be able to superpose.\footnote{For  QFT in curved spacetimes, different \lq\lq{}time\rq\rq{}  problems are present. There, the issue rears its head in the absence of Killing vector fields. The issue is that one should find solutions of the dynamical equations $f_\alpha$, which are also eigenfunctions of the time derivative (i.e. along the Killing vector), $\square f_\alpha=0 ~,\,~\mbox{and}\, ~\partial_t f_\alpha=-i\lambda_\alpha f_\alpha ~,\,~\, ~\partial_t f^*_\alpha=i\lambda_\alpha f^*_\alpha$. 
One then uses this eigenfunction split into positive and negative frequencies to construct associated raising and lowering operators, a Fock space, etc.  In the absence of  a preferred time-like direction in space-time,  raising and lowering operators become less meaningful (see \cite{DeWitt_book} for a more complete treatment of this point). } What is the meaning of this demand when the operators are acting on states representing superpositions of gravitational fields?  It is hard to tell, and it seems to me there are very few ways forward in thinking about these problems.\footnote{See \cite{Bianca_perturbative} for a notable exception, using the formalism of partial and complete observables \cite{Rovelli_partial} (see \cite{Rovelli_book, Tambornino} for reviews).}

Further tension lies between {\bf 2a}  and {\bf 2b}. It stems from distinct roles of (global) time in
quantum theory and in general relativity. 
% As masterfully put by Wald and Unruh in \cite{Wald_Unruh}
 In quantum mechanics, all measurements are
made at \lq\lq{}instants of time\rq\rq{}, and in this sense only quantities referring to the
instantaneous state of a system have physical meaning. Dynamics is most naturally understood as an evolution from one instantaneous state to another.  On the other hand, in general relativity, a global \lq\lq{}time\rq\rq{} is
an arbitrary label assigned to spacelike hypersurfaces, and physically meaningful quantities are required to be independent
of such labels.  In other words, in GR, only the spacetime geometry
is measurable, and thus only the  histories of the Universe have physical meaning. The dynamics of the gravitational field must be carved out of the theory by subjecting it to an initial value formulation, an unnatural procedure from the covariant standpoint.

When  (naively)
merging quantum theory and general relativity, it should cause no surprise that  the only physically meaningful surviving quantities  are those which are
both: 1) instantaneously measurable (i.e. refer to quantities defined on a spacelike hypersurface) and 2)  depend \textit{only} on the spacetime geometry (i.e. are independent of a choice of spacelike hypersurface)  (see e.g. \cite{Isham_POT, Wald_Unruh}). In a full geometrodynamical theory, this creates great constraints on what constitute reasonable quantum observables \cite{Tim_chaos, Tim_chaos2}.

   Lastly lies a further distinction (but not necessarily a clash) between the notions of local time, or duration,  in the two theories. In classical GR, there exists \textit{one}  notion of local time inherent in the theory, independently of the space-time metric:  proper time. Proper time  gives an approximate notion of duration along a world-line, it is a dimensionful quantity that is not given in relation to anything else. It has this role whether a space-time satisfies the Einstein equations or not, and is therefore a kinematical, general proxy for elapsed time.  On the other hand,  in modern relational approaches to quantum mechanics \cite{Page_Wootters, Dolby}, one can parse two notions of time:  time as measured by relations between subsystems and just an external evolution time. One of these subsystems is usually called the \lq\lq{}clock\rq\rq{}; its emerging notion of local time is  akin to duration, but only becomes available through relational dynamics. The other, global evolution time, is indeed not dynamic, but  it also has no meaningful parallel in GR. 

These issues surface on the technical front as well; efforts to obtain a viable notion of quantum evolution in quantum gravity -- even at the semi-classical level -- founder for a variety of reasons \cite{Kuchar_Time}.\footnote{At least outside of drastic truncations of field space, such as minisuperspace. In many cases, such truncations do not commute with quantization \cite{Kuchar_valid}.} My claim is that these reasons have a common birthplace. They arise from the distinction between  unitary evolution and reduction processes in quantum mechanics, and from the confluence of dynamics and kinematics in general relativity.

\subsection{A diplomatic resolution}

 This paper outlines a proposal to address such questions. Although it is still of largely qualitative character, the aim of this first paper is to discuss the mechanisms by which one could implement said proposal.

My present attempt to untangle both the problem of the superposition of causal structures and the problems of time consists in basing gravitational  theories, {at the kinematical level}, on fields for which there is no notion of causality. Causality emerges, but only dynamically.  In this way, physics should be encoded in a path integral operating in timeless configuration space, embodying only compatible \textit{spatial} symmetries and with prescribed boundary conditions. Such symmetries allow us to define and exploit a reduced space of physical configurations. No such reduced configuration space exists for theories with local refoliation invariance --- such as GR --- even abstractly.  

 The boundary conditions required for the constructions here are extremely special. They are based on a fundamental asymmetry of the reduced configuration space (configuration space modulo gauge-symmetries). Namely a reduced configuration space must not only exist, but also have a unique, most homogeneous element. This requirement places extreme constraints on the field content, on the symmetries, and on the topology of the manifold, but its satisfaction is what will allow most of the constructions here to be defined unambiguously. Anchoring the transition amplitude on this unique boundary,\footnote{Or more precisely, this unique \lq{}corner\rq{}, as we will see below.}  the present framework provides a \lq{}first principles\rq{}, non-covariant  alternative to the Hartle-Hawking prescription of the global wavefunction of the Universe \cite{Hartle_Hawking}.
 
 It also addresses questions that need to be answered in any fully relational, timeless theory referring to instantaneous states, such as Hartle-Hawking and Vilenkin\rq{}s tunneling proposals. These proposals yield a wave-function whose argument is an instantaneous field content, e.g. $\psi(g, \varphi)$, where $g$ is the 3-metric and $\varphi$ some spatial matter field. In this respect, such theories thus share many of the features of the framework constructed here. The difference, again,  is that in such approaches, the wave-functions still need to correspond to covariant (i.e. space-time) quantities. Outside of minisuperpsace approximations, the requirement of covariance (at a fundamental level) does not allow such theories to be describable in a physical configuration space, even abstractly, and thus they  do not assuage the conflicts between QM and GR I have outlined in the previous section.  
 
   As I will show, the present setup yields a single static wavefunction and a corresponding \textit{volume-form in physical configuration space}. The passage of time is then abstracted from a particular notion of \lq{}records\rq{} -- subsets of configuration space where the volume form concentrates. Causality emerges thus as an approximate pattern encoded in the volume-form when it is well-described semi-classically. In this sense, no separate causal structures \lq{}superpose\rq{} --- the frozen patterns we associate with them can at most be said to mesh and interfere in configuration space. 
    Unlike previous work in which one attempts to emulate the effects of wave-function collapse within standard unitary evolution, here both evolution and reduction should be emergent from a fundamentally timeless description of the entire Universe.

      Here,  space-times don\rq{}t exist a priori, but require a  relational construction of duration -- or definitions of clocks -- on classical solutions. I.e. in general one must define clocks and an experienced  duration through the classical evolution of relational observables; I call such an emergent notion of time \textit{duration-time}.  Indeed, the price to pay in the present, purely relational framework, is apparent in trying to recover, in the appropriate limits, the  local,  experienced time of GR. In general relativity, experienced time is directly related to proper time, but the  gravitational theories compatible with the framework  have no universal replacement for proper time.   This feature might not be so damaging; it is also the case in relational approaches to quantum mechanics that one constructs clock-time from relations between subsystems \cite{Page_Wootters}, as discussed in the previous section.  
      
      The challenge to the sort of theories encompassed by the present work thus  shifts, from the standard fundamental problems mentioned above --- superposition of causal structures, untangling evolution and symmetries, and reduction vs unitary evolution --- to one of recovering a smooth space-time description.
While it may indeed be difficult to recognize classical relativity of simultaneity from the global evolution-time --- i.e. the global time parametrization that will follow from the construction of records ---  it can be gleaned when using duration-time, as we will see.

      In sum, the claim here is that, although unorthodox, this approach  resolves issues both with the superposition of causal structures and the \lq{}evolution vs reduction\rq{}  dichotomy, while bringing the notion of duration in gravity closer to the non-unique versions of \lq{}clock systems\rq{} in quantum mechanics.  
The constructions advocated in this paper  can similarly work with other types of timeless configuration space; they do not limit themselves to the geometrodynamic setting in which they are mostly pursued here. Even the  consequences of this approach to the construction of gravitational models will be only touched on here, being further pursued in an upcoming publication, as well as in \cite{Conformal_geodesic}.

     %Moreover, with a positive definite action in physical configuration space, there are no problems with requiring complex integration contours, an ambiguity that plagues the Hartle-Hawking construction.  ii) Armed solely with a global wavefunction in the space of physical degrees of freedom, I introduce a definition of \lq{}records\rq{}, which are features of this wavefunction. Such records encode the notion of conditional probabilities: the record-holding configurations are conditional on the recorded ones. And thus a primitive notion of time emerges from a fundamentally timeless setting.   iii) I show that with records one can still recover some notion of conservation of probabilities. This is done by the use of a \lq{}screen\rq{}, which is a well-behaved substitute for the problematic \lq{}equal-time  surface in superspace\rq{} for the Wheeler-DeWitt equation. iv) Finally, I introduce a simple geometrodynamical toy model to exemplify these structures. Within it, I\rq{}m able to explicitly calculate the relative semi-classical quantum gravity probabilities for a screen close to the round 3-sphere, showcasing the viability of this approach. By adding a matter field to serve as a \lq{}local clock\rq{}, I reconstruct an experienced space-time along the classical solutions of the theory.    
   
  \paragraph*{Roadmap}         
     I will now describe the main constructions of this paper and its relevance for quantum mechanics and quantum gravity. I will start by making clear what are the main assumptions, or axioms, of the work. This is done in section \ref{sec:axioms}. The non-relativistic setting will  in many ways resemble standard particle quantum mechanics. Leaving at most a single reparametrization constraint, the assumptions are then perfectly compatible with past work on relational dynamics and quantum mechanics, which I very briefly review in section \ref{sec:config_space}. At the end of this section, I include a proof that the principles single out the Born rule as a measure in configuration space.   Finally, I discuss another problem that is not fully addressed in relational quantum mechanical approaches within the context of path integrals: a replacement for the role of the epistemological updating of probabilities in the timeless context, through the notion of \lq\lq{}record-holding submanifolds\rq\rq{},  in section \ref{sec:records}. Lastly, I sketch a gravitational dynamical toy model based on strong gravity, to illustrate the structures introduced here.

\section{Axioms}\label{sec:axioms}
   I would like to examine  the conceptual picture that emerges from  theories for which time plays \textit{no} role at the kinematical level.    
 I will first clarify the axioms I am led to adopt in order to have a geometric theory of gravity without such kinematical causal relations. 

%The original formulation of quantum mechanics cannot be found on most textbooks. At first, Schroedinger wrote a time-independent equation \cite{Schroedinger}, only later becoming preoccupied with time-varying potentials. These were treated as an external \emph{classical} influence on the system.  It was also clear to other founders of quantum mechanics, such as Born, Heisenberg and Jordan, that time came in through a classical treatment of the external environment \cite{three-man}.  This difficulty with dealing with time in a quantum mechanical manner has been present ever since. It has mutated into difficulties of associating an operator to time evolution, $\hat{t}$, conjugate to the energy (which should be a constant for closed systems). 

% The simplest way of understanding the mismatch is to consider the transition amplitude $\langle q_1, t_1|e^{-i\hat{H}t}|q_2,t_2\rangle=G(q_1,t_1\,;\,t_2,q_2)$  evolves a future outcome backwards in time to find its overlap with the current wavefunction using the Heisenberg picture of nonrelativisitic quantum mechanics.  However, the extension of this concept to general relativity is nontrivial because the key requisite feature of quantum mechancis -- unitarity -- becomes muddied in the presence of alternate causal structures. 

The five given structures  that the present work is based on are the following: 
\begin{enumerate}
\item {\bf{A closed topological manifold, $M$.}} Prior to defining our fields, we require some weak notion of locality, which I will take to mean \lq{}open neighborhoods\rq{}. In the gravitational case, I will take this to be provided by the closed \emph{topological manifold} $M$ (i.e. compact without boundary), of dimension $n=3$.\footnote{ The demand of being closed is a consequence of relationalism. It does not imply that through evolution we couldn\rq{}t get causal horizons, or that there are no meaningful definitions of effectively open subsystems within $M$.} If one wanted to apply the following constructions to a fundamentally discrete theory, $M$ could be replaced by a lattice, or piecewise linear manifold, on top of which the physical degrees of freedom live.

\item {\bf The kinematic field space $\mathcal{Q}$.} In the field theory case, this is the infinite-dimensional space of  field configurations over $M$. Each configuration -- the  \lq{}instantaneous\rq{} field content of an entire Universe-- is represented by a point of $\mathcal{Q}$, which we denote by $\phi$. These should correspond to relational data, on which the dynamical laws act.  For defiteness,  I will take field configurations to be  sections of  tensor bundles over $M$.  The example for pure gravity would be $\mathcal{Q}=\{g\in C^\infty_+(T^*M\otimes_S T^*M)\}$, the space of positive smooth sections on the symmetrized (0,2)-covariant tensor bundle.   In principle the constructions here should apply for non-causally related observables of any kind, such as e.g.  field values on a lattice. 

 \item {\bf \lq\lq{}The Past Hypothesis\rq\rq{} -- boundary (or \lq{}initial\rq{}) conditions for the Universe}:  The requirement of such boundary conditions for the wave-function goes in line with the fact that, if time plays no fundamental role, the wave-function of the Universe must  be given uniquely, and only once. Thus I require unique boundary conditions in order to anchor the construction of a unique transition amplitude. Unlike what is the case with the \lq{}no-boundary\rq{} proposal \cite{Hartle_Hawking}, this will be a choice of \lq{}the most homogeneous\rq{} configuration, $\phi^*$, with respect to which the wave-function is defined, $\psi(\phi)=W(\phi^*, \phi)$, see equation \eqref{anchor_path}, below.  As I will show,  $\phi^*$ furthermore plays a fundamental part in the definition of \lq{}records\rq{}, and for \lq{}records\rq{} to function in the capacity  their name suggests, the boundary conditions should correspond to field configurations which are as \lq{}structureless as possible\rq{}.  Below, I give a criterion that can, in some circumstances, select a unique such boundary state. 
 
\item {\bf{An action functional on curves on $\mathcal{Q}$.}} I.e. $S(\gamma)$, for $\gamma: [0,1]\rightarrow \mathcal{Q}$, invariant wrt the gauge group $\mathcal{G}$ acting on $\mathcal{Q}$, $\mathcal{G}\times\mathcal{Q}\rightarrow \mathcal{Q}$, and (up to boundary terms) depending only on first derivatives in the time parametrization. The action should be invariant with respect to the given gauge symmetry group $\mathcal{G}$, defining $\tilde{S}([\gamma])$, for $[\gamma]:[0,1]\rightarrow \mathcal{Q}/\mathcal{G}$. For this definition, the gauge-symmetry group needs to have a pointwise action on $\mathcal{Q}$.  Such an action functional will be used for the transition amplitude between two physical configurations, for $[\phi]\in \mathcal{Q}/\mathcal{G}$, and   $[\gamma](0)=[\phi^*], [\gamma](1)=[\phi_2]$:
\be\label{anchor_path}\Psi([\phi])\equiv W([\phi^*],[\phi]):=\int^{[\phi]}_{[\phi^*]} \mathcal{D}\gamma \exp{(i \tilde S[\gamma])/\hbar}
\ee
To avoid cluttered notation, I will drop the square brackets in most of the paper, calling attention when the distinction between full configuration space and the reduced one becomes material. 
%This requires a gauge-fixing procedure, which we take to be the arc-length one for an action that can be put into a metric form,\footnote{In general I will call \emph{a Jacobi metric}, any  metric in configuration space whose geodesics are extremal paths of the action (see appendix \ref{app:Jacobi_metric}).} (see appendix \ref{app:reparametrization}). 

\item {\bf{A positive scalar function  $F:\mathbb{C}\rightarrow \mathbb{R}^+$}}.  I will call this  function the \textit{pre-probability density},  $F$, as it will give a measure in configuration space. To serve our purposes, $F$ must preserve the multiplicative group structure, 
 \be\label{equ:factorization_density} F(z_1z_2)=F(z_1)F(z_2)\ee  For future reasons I will term \eqref{equ:factorization_density} the \emph{factorization property} of the density. It is a required  condition for the definition of record that I introduce here to have physical significance.\footnote{And for it to have cluster decomposition properties \cite{Locality_riem}).} It is the specific form  $F(z)=|z|^2$ of this $F$ obeying \eqref{equ:factorization_density} which encodes the Born rule.

% \item  {\bf{Unique quantum measure (optional).}} Here, I use Gleason's result combined with a preferred `in' state for $\mathcal{Q}$. The choice of preferred state utilized here is the completely degenerate metric (alongside zero sections for all the fields on $M$). Taking partial traces  
\end{enumerate}
These axioms are quite standard implicit choices in most work related to field theory, here I am only making these choices explicit. Axiom 5 is implicit in the Born rule, and 3 is usually explicitly stated in the literature, either as boundary conditions, or as some initial condition.

The only structure required for doing physics, arising from the five premises above, is the density over $\mathcal{Q}$, given by 
\be\label{measure_prob}P(\phi):=F(W(\phi^*,\phi))\mathcal{D}\phi\ee
where $F$ is restricted by axiom 4, $\phi^*$ is given by axiom 5, and $W(\phi^*,\phi)$ is defined from the action functional given in axiom 3, and \eqref{equ:path_integral} and \eqref{equ:measure_timeful}.  This measure gives  a volume form on configuration space. It gives a way to ``count" configurations, $P(\phi)$, and thereby the likelihood of finding certain relational observables within $\mathcal{Q}$.  It is assumed to be a  positive functional of the only non-trivial function we have defined pointwise on $\cal{Q}$, namely, $W(\phi^*,\phi)$. %\footnote{  This measure assumes a very similar role to Page\rq{}s, on his work  on Sensible Quantum Mechanics \cite{Page}.  There, the measure refers to expectation values ofcertain preferred operators (`awareness operators,' one for each possible conscious perception).} 
 I should stress that roughly the same or similar assumptions are implicit in both Hartle-Hawking and Vilenkin\rq{}s tunneling proposals.

 Here, we have a pre-conceived notion of space, or rather, \lq\lq{}of things which are not causally related\rq\rq{}, but not necessarily of time, in either its absolute or relativistic forms. We have the space of relational objects on which dynamical laws should act. 
 
 At the end of the day, we have at our disposal the density over field space, $P(\phi)$. But without space-time, and without any notion of absolute time, can we still extract some physics from the formalism? I would like to show that there can still be enough structure in the timeless path integral in timeless configuration space for doing just that.\footnote{Moreover, this formalism could apply to any relational system with a specified configuration space, action over curves on it, and unique maximally homogeneous point. For example, certain types of tensor network models.}  This is what I will focus on for most of the paper.

  \subsection{Relationalism and the symmetry group ${\mathcal{G}}$.}\label{sec:symmetry}
 Intimately related to our clash {\bf 2a}  and {\bf 2b} is what is known as the \lq{}problem of time\rq{}. The GR Hamiltonian mingles local gauge symmetries and evolution \cite{Kuchar_Time, Isham_POT}.\footnote{I note that while a relativistic particle also has a Hamiltonian which generates a single reparametrization  invariance, it does not mix this symmetry with local gauge transformations \cite{Kuchar_Time}. Moreover, as a subsystem, it is easily describable by relational observables.   I give a broader analysis of the time problems in appendix \ref{app:scalar_ADM}, and related ones in their formulation of a transition amplitude in superspace, in appendix \ref{app:issues}.} At least without the use of matter fields there is no preferred manner to split the Hamiltonian constraints such that all but one are fixed by a \lq\lq{}definition of simultaneity\rq\rq{}-- a partial gauge-fixing of the Hamiltonian well defined everywhere in phase space. 
 
 Although the main constructions of the paper should be applicable in a more general setting, let\rq{}s investigate the standard case for gravity in closer detail. 
\paragraph{The standard ADM action and symmetries.}
 
 Upon a Legendre transformation, the vacuum Einstein-Hilbert action yields primary and then secondary first-class constraints, $H^\alpha=(H^\perp, H^a)$. 
The action can then be put in the form 
$$S[h_{ab}, \pi^{ab}, N^\alpha]=\int_{t_i}^{t_f} dt \int d^3x\, \left(\pi^{ab}\dot h_{ab}-H^\alpha N_\alpha\right)$$
where $N_\alpha$ are Lagrange multipliers and
 \begin{eqnarray} H^\perp&=&\frac{\pi^{ab}\pi_{ab}-\frac{1}{2}\pi^2}{\sqrt{g}}-R\sqrt{g}\\
 H^a&=&-\nabla_b \pi^{ab}
 \end{eqnarray}
   with some algebra given by 
 $$\{H_\alpha(x), H_\beta(y)\}=\int d^3 x\rq{} U^{\gamma(x \lq{})}_{\alpha(x) \beta(y)}H_{\gamma}(x\rq{})
 $$
% The only non-zero  elements of the ADM constraint algebra matrix are:
%\begin{eqnarray*} U_{H^\perp(x) H^\perp(y)}^{H_ a(z)}&=&g^{ac}\left(\delta(z,x)\nabla_c\delta(z,y)-\delta(z,y)\nabla_c\delta(z,x)\right)\\
  %U_{H^\perp(x) H_a(y)}^{H^\perp(z)}&=&-\delta(z,y)\nabla_a\delta(z,x)\\
 %U_ {H_a(x) H_b(y)}^{H_c(z)}&=&\delta^c_b\delta(z,x)\nabla_a\delta(z,y)-\delta^c_a\delta(y,z)\nabla_b\delta(z,x)
%\end{eqnarray*} 
It has been shown that the local constraints above are essentially unique if the algebra they generate is required to mimic the commutation algebra of vector fields orthogonally decomposed along a hypersurface of space-time (the hypersurface deformation algebra). Thus the Hamiltonian constraint, and therefore the local Wheeler-DeWitt equation, are intimately tied to a covariant picture of space-time.
 
 Under the transformation generated by the flow of the constraints, 
 \begin{eqnarray}
 \delta_\epsilon h_{ab}(x)&:=\{ h_{ab}(x), \int d^3 x H^\alpha\epsilon_\alpha\}\\
  \delta_\epsilon \pi^{ab}(x)&:=\{ \pi^{ab}(x), \int d^3 x H^\alpha\epsilon_\alpha\}
 \end{eqnarray}
 and under the more ad hoc
\be\delta_\epsilon N^\alpha(x)=\dot \epsilon^\alpha(x) -\int d^3x\rq{} d^3 y\rq{} U^{\alpha(x)}_{\gamma(x\rq{}) \beta(y\rq{})}N^{\gamma}(x\rq{})\epsilon^{\beta}(y\rq{})\label{Lagrange_transf}\ee
 the action transforms by  a boundary term: 
\be\label{boundary_transf}
\delta_\epsilon S= \left(\int d^3x \epsilon^\alpha \left(\int d^3y\, \pi_{ij}(y)\frac{\delta H_\alpha(x)}{\delta \pi^{ij}(y)}-H_\alpha(x)\right)\right)\Big|^{t_i}_{t_f}
\ee
which clearly vanishes for constraints linear in the momenta, or if the generator of refoliations, $\epsilon^\perp$, vanishes at the initial and final surfaces. 

Whereas, 
$$\delta_{\epsilon^a}g_{ab}=\mathcal{L}_{\vec{\epsilon}}g_{ab}$$
has pointwise  dependence  in $\mathcal{Q}$, the transformation 
$$\delta_{\epsilon^\perp}g_{ab}=\frac{2\epsilon^\perp (\pi_{ab}-\frac12 \pi g_{ab} )  }{\sqrt{g}}$$ depends not only on the metric, but also on the momenta. We could also have obtained a similar result directly from the ADM decomposition (with no Legendre transformation): decomposing a vector field along the time direction and tangential components to the hypersurface, $X^\mu(t, x)=(s(t,x),\xi^a(t,x))$, one obtains (assuming zero shift):
\be\label{refol}\delta_{X^\mu(t, x)}g_{ab}=s \dot g_{ab}+\mathcal{L}_\xi g_{ab}\ee
where $\mathcal{L}_\xi$ is the Lie derivative along the vector field tangential to the hypersurface. This part of the action is easy to make sense of: it is the infinitesimal action of a spatial diffeomorphism, acting on the spatial metric through pull-back: Diff$(M)\times \mathcal{Q}\ni (f, g_{ab})\mapsto f^*g_{ab}$. 

 On the other hand, equation \eqref{refol} shows us that refoliations act on the metric in a manner also depending on its velocity.\footnote{In fact, the strict relationship between the Hamiltonian constraint and refoliations holds only on-shell.} 
It is the action of refoliations that makes it difficult to meaningfully define surfaces and points in superspace, the quotient space $\mathcal{Q}/$Diff$(M)$. For suppose one has two different curves intersecting at $g^0_{ab}$, $\gamma^{(i)}_{ab}(t)=g^0_{ab}+t^{(i)} v^{(i)}_{ab}$, $i=1,2$, for $v_{ab}$ a positive-definite symmetric $(0,2)$ tensor (they intersect at $t^{(i)}=0$). Since the action of the spatial diffeomorphisms depends solely on the metric, the two curves will still intersect after the action of $\xi^a$ (or $\delta_{\epsilon^a}$), as they will be jointly moved. But with the action of $\epsilon^\perp$, the curves will be shifted: $\tilde\gamma^{(i)}_{ab}(0)=g^0_{ab}+s(0,x) v^{(i)}_{ab}$, and will not intersect anymore. Taking the quotient by (spatial) diffeomorphisms will not help since it is easy to find $v^{(i)}$ such that $f^*(g^0_{ab}+ s(0,x)v^{(1)}_{ab})\neq (g^0_{ab}+s(0,x) v^{(2)}_{ab})$ for all $f\in$ Diff$(M)$.\footnote{That this must be so is easy to ascertain from a simple degree of freedom count: the space of symmetric $(0,2)$ tensor has 6 degrees of freedom per space point, whereas the generators of spatial diffeomorphisms can only account for three.}

Even though $\mathcal{Q}$ is an infinite-dimensional space, there is at least one situation in which  such curves can still intersect at different $t^{(i)}$. If the transformation corresponds merely to a reparametrization along each curve,  $s(x,t^{(i)})=s(t^{(i)})=s_0+s_1 t^{(i)}$, the transformed curves merely shift their time-parameters:\footnote{This would happen order by order for higher contributions in powers of $t$.} $\tilde\gamma^{(i)}_{ab}(t)=\gamma^{(i)}_{ab}(s_0+s_1 t^{(i)})$, and thus the curves will still intersect, but now  at $t^{(i)}=-s_0/s_1$. However, if the refoliation  is not spatially constant, generically  the curves will miss each other, also in superspace. 

This fact makes it doubtful that one can implement boundary conditions on superspace for the path integral that are physically meaningful from a covariant space-time perspective.

\paragraph{The present requirement on  symmetries.}
 
The  alternative explored here requires its fundamental symmetries to be generated by local constraints linear in the momenta. The treatment of such linear symmetries, including their quantization, is then much more straightforward than for local constraints quadratic in the momenta.  I will show how this non-fundamentally-covariant setting still allows the emergence of an on-shell \emph{refoliation invariance}.\footnote{I should note that in the ADM Hamiltonian formalism for GR \cite{ADM}, refoliations only generate a symmetry on-shell in any case \cite{Wald_Lee}.}
\begin{figure}[!ht]\label{figure:multiple}
\begin{center}
\includegraphics[width=10cm, height=7cm]{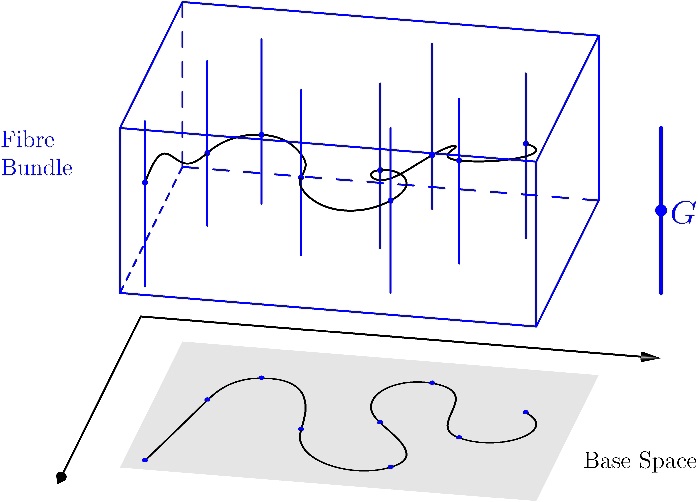}
\caption{ A horizontal lift of a curve in a principal fiber bundle setting. The lift corresponds to a connection. }
\end{center}
\end{figure}

Although the principle of having a pointwise action in configuration space is generically applicable, in the case of gravity in metric variables, i.e. $\mathcal{Q}=$Riem(M),  the symmetries which satisfy this criterion are found by requiring a phase space representation of first-class constraints linear in the momenta, 
$$ \chi(g_{ab},\pi^{ab})\approx 0,\qquad   \chi(g_{ab},\alpha\pi^{ab})= \alpha\chi(g_{ab},\pi^{ab}),\qquad  \{\chi(g_{ab},\pi^{ab})(N_1), \chi(g_{ab},\pi^{ab})(N_2)\}\approx0
$$
where $\alpha\in \mathbb{R}$ is a constant in  $M$,  $N_1$ and $N_2$ are appropriate (tensor) smearings of the local constraint densities, $\chi$, e.g. 
$$\chi(g_{ab},\pi^{ab})(N)=\int d^3x\,  \chi(g_{ab},\pi^{ab},\, x)_{cdef}N^{cdef}(x)$$
     the conjugate momentum to the metric is $\pi^{ab}$, and curly brackets denote the Poisson bracket wrt this canonical relation. Linearity in the momenta implies that the symmetries have an intrinsic action on configuration space, as required, i.e. 
     $$\{\chi(g_{ab},\pi^{ab})(N), g_{cd}(y)\}= \diby{\chi(g_{ab},\pi^{ab})(N)}{\pi_{cd}(y)}=F(g_{ab}, N, y)
     $$
     i.e. its action on $g_{ab}$ depends only on $g_{ab}$ and the smearing. The first class property implies that these constraints correspond to symmetries. It turns out that, under some assumptions, that the allowed non-trivial  symmetries that  act as a group in the metric configuration space --- and thus fulfill our axioms --- are diffeomorphisms and scale transformations  (see \cite{Conformal_geodesic} for more details).

 In the field theoretic framework proposed here, relational observables are those that live in the quotient $\mathcal{Q}/\mathcal{G}$ -- e.g.: those whose spatial position and scale are only defined relationally -- and thus I will bypass a more complicated relational phrasing of the amplitudes by just assuming that all statements are suitably translated when necessary, from $\mathcal{Q}$ to $\mathcal{Q}/\mathcal{G}$. In the presence of this structure, boundary conditions for the path integral will have physical meaning. %The relational probability would then become merely $P([A]|[B])$, or, to unclutter  it even further, just $P(A|B)$, assuming $A$ and $B$ are given by relational observables. 

The demands of our axioms thus severely restrict the form of the infinite-dimensional groups $\mathcal{G}$, and imply  that configuration space form a principal fiber bundle $\mathcal{G}\hookrightarrow  \mathcal{Q}\rightarrow\mathcal{Q}/\mathcal{G}$,\footnote{This is not quite true, for the base space $\mathcal{Q}/\mathcal{G}$ may not form a manifold, as it doesn\rq{}t for the spatial diffeomorphisms. However, insofar as I will use this structure for a field-space connection 1-form -- which I associate with an abstract observer --  the group can be restricted by a further condition: only allow the diffeomorphisms that maintain a given (observer location) $x_o\in M$ and a observer frame at $x$, $e^a_{|x_o}$ fixed. Calling this restricted set of diffeomorphisms Diff$_{x_o}$, then the space $\mathcal{Q}/$Diff$_{x_o}$ is indeed a manifold \cite{Ebin} and one can use its principal fiber bundle structure in the usual ways.\label{PFB_footnote} } making the quantum treatment of gauge-symmetries straightforward. For instance, unlike what is the case for ADM \cite{ADM}, in which one requires the more complicated use of the FBV formalism \cite{Fradkin_Vilkovisky, Hartle_Halliwell}, standard Fadeev-Popov is sufficient to give rise to a well-defined BRST charge; the Fadeev Popov determinant is a true determinant, which implements gauge-covariance of the gauge-fixed path integral.   It is the principal fiber bundle structure that  allows us to  unambiguously consider individual points and dynamics in the quotient space $\mathcal{Q}/\mathcal{G}$, an impossibility when gauge symmetries become tangled with dynamics (as is the case with ADM). 
 
%Normally, in  finite-dimensional relational dynamics, one posits two or more subsystems, whose relations are the only observables of the theory. Thus one would ask for the probability $P(A_2~~ \mbox{when}~B_2~| ~A_1 ~~\mbox{when}~B_1)$, which can be roughly translated, in a semi-classical approximation, as \lq\lq{}the proportion of dynamical paths in configuration space for which, having $A_1$ occurred at the same instant as $B_1$, then $B_2$ occurs at the same instant as $A_2$\rq\rq{} \cite{Page_summary, Dolby}.   For theories in which the action and measure are invariant wrt these symmetries,  the  \lq{}transfer matrix\rq{}, or amplitude kernel,  does  act on gauge-invariant \lq{}in\rq{} and \lq{}out\rq{} states, allowing us to write ${W([g_1],\,[g_2])}$.

 In the 3+1 path integral setting here, one can make use of the principal fiber bundle structure for a straightforward  treatment of gauge-symmetries, with the use of a a field-space connection 1-form, $\varpi:T\mathcal{Q}\rightarrow T{\mbox{\tiny Id}}\mathcal{G}$, as described in appendix \ref{app:PFB}. This field-space connection 1-form selects a way to horizontally lift to $\mathcal{Q}$  a given curve in $\mathcal{Q}/\mathcal{G}$. For instance, take $\mathcal{G}=$Diff$(M)$ and $\mathcal{Q}=$Riem(M), $T{\mbox{\tiny Id}}\mathcal{G}=C^\infty(TM)$, the vector fields with the algebra being just the commutator algebra, acting on the metric with the Lie derivative. The horizontal projection of a field-space vector, $\dot g_{ab}$ is given by $\dot g_{ab}^H:=\dot g_{ab}-\mathcal{L}_{\varpi(\dot g)} g_{ab}$. I.e. it acts as a field-dependent shift in the ADM formalism.\footnote{The connection 1-form itself can be related to a choice of equilocality and  associated to abstract, non-back-reacting, observers \cite{Aldo_HG} (see also footnote \ref{PFB_footnote}). In a completely gauge-covarint way, it selects a manner in which an observer dynamically distinguishes physical transformations from pure gauge transformations, along time.  This is a generalization of Barbour\rq{}s concept of \lq\lq{}best-matching coordinates\rq{} \cite{Flavio_tutorial}. } According to \eqref{eq_fundamental2}, under a time-dependent diffeomorphism, $f(t)\in$Diff$(M)$, generated by the vector field $\xi^a(t)$,\footnote{Strictly speaking, we are only considering diffeomorphisms connected to the identity. Outside of this domain, many problems appear. For instance, there are always diffeomorphisms arbitrarily close to the identity which are not connected to it through the exponential flow of some vector field. This poses problems for the path integral.} we have the following transformation of $\varpi$:
     $$\delta_{\vec\xi}(\varpi(\dot g))^a=\dot \xi^a -[\varpi, \xi]^a$$
     which is precisely the transformation required of the shift from \eqref{Lagrange_transf} under time-dependent spatial diffeomorphisms.

% This procedure eliminates the degenerate directions of the action functional, and does not suffer from Gribov ambiguities (see \cite{Conformal_geodesic} for more details).  In other words, one can make sense of \lq\lq{}instantaneous\rq\rq{}  observables, even in pure gravity. In the WdW equation (see appendix \ref{app:scalar_ADM}), one would at the very least require extra matter fields that effectively define a foliation, but, again, there is no choice of such a good \lq{}clock\rq{} throughout field space. This discussion concludes considerations about gauge-symmetry on the path integral itself. 

 %  Fixing the initial point, $\phi_1$, in the absence of local reparametrization invariance the theory can have a well defined reduced configuration space, and thus a well-posed transition amplitude between gauge-invariant states. This is true even in the presence of global reparametrization invariance, as we discuss below (see also appendix \ref{app:arc_length}). 

\paragraph*{Summary of this section.}
In this subsection I have sketched how the local symmetry groups can be more or less uniquely selected in such a manner as to have a pointwise representation in configurational field space. In this manner, a well-defined reduced configuration space exists, and I guarantee that local symmetries will be distinguished from dynamics. Moreover,  a gauge-invariant path integral may be constructed making use of the connection-form, possible from the emerging principal fiber bundle structure (see also \cite{Aldo_HG, Conformal_geodesic} for a more complete account). 
     
  \subsection{Uniqueness of preferred boundary conditions}
  The principle I will use to select boundary conditions is based, roughly, on information content. I would like to set the anchor of the transition amplitude \eqref{anchor_path} to be the configuration (or reduced configuration) that carries the least amount of information. What I mean by that is that it carries the most amount of symmetries, under my group of transformations. I.e. it is the most \lq\lq{}homogeneous\rq\rq{} configuration. 
  
As mentioned in footnote \ref{PFB_footnote}, it is not true, for groups which don\rq{}t act freely on configuration space, that the quotient $\mathcal{Q}/\mathcal{G}$ forms a manifold. Indeed, if there are elements of $\mathcal{Q}$ that remain fixed under the action of subgroups of $\mathcal{G}$, such as is the case with $\mathcal{Q}$ and Diff$(M)$ for instance, then the quotient can be at most a \textit{stratified manifold} \cite{Fischer}. A stratified manifold is basically a union of manifolds of different dimensions, with concatenated boundaries of boundaries. The standard example is a cube, with its bulk being of dimension 3, and its boundary being composed of a further union of manifolds; a face of dimension two, with its boundaries composed of a further union of manifolds; edges of dimension one, with its boundaries composed of a further union of (zero-dimensional) manifolds.  The points with the highest isotropy group will correspond to the lowest dimensional boundaries (of all the other boundaries). In other words, 
$$\mathcal{U}_m={q\in \mathcal{Q}}\, |\, ~ \mbox{dim(Iso}_q(\mathcal{G}))=m\}$$
we have: 
$$\mathcal{Q}/\mathcal{G}=\bigcup_{m=m_{\mbox{\tiny max}}}^{m=0} [\mathcal{U}_m]$$
    where each $ [\mathcal{U}_m]:=\mathcal{U}_m/\mathcal{G}$ is a manifold with boundaries, such that $[\mathcal{U}_m]\subset \partial [\mathcal{U}_{m\rq{}}]$ for $m>m\rq{}$. 
    
            Given $\mathcal{G}$, let $\Phi_o\subset \mathcal{Q}$ be the set composed of all the most homogeneous field configurations, i.e. the subset of elements with the largest dimensional isotropy subgroup of $\mathcal{G}$. This will stand in for the configurations being \lq{}as structureless as possible\rq{}.     
      For example, let us take   $\mathcal{Q}=\mbox{Riem}(M)$ and $\mathcal{G}=\mbox{Diff}(M)$, acting through pull-back $A_f (g_{ab})\mapsto f^*g_{ab}$. Then let
$$\Phi_o=\{ g\in \mbox{Riem} ~|~\mbox{Iso}_g\subset \mbox{Diff}(M)~~ \mbox{has maximum dimension} \}$$
 Allowing for degenerate metrics, we have a unique such point,   $\Phi_o=\{\phi^*\}=\{0\}$, the completely degenerate metric, since it has the full Diff(M) as an isotropy group.
 
 In accordance with the symmetry principles of the theory, as described above, one could extend the group Diff$(M)$ to the one given by the \lq{}maximum geometric group\rq{}, Diff$(M) \ltimes \mathcal{C}$,  with $\mathcal{C}$ the Weyl group of conformal transformations (through pointwise multiplication of the metric by positive scalar functions $\alpha\in C_+^\infty(M)$, and a semi-direct product between the two groups) \cite{FiMA77, Conformal_geodesic}.  

 In fact, the case of Diff$(M) \ltimes \mathcal{C}$ does not require further specification of the field space boundary at all (as is required, for example in Hartle-Hawking  \cite{Hartle_Hawking}). This property can be seen either by parametrizing physical space $\mathcal{Q}/\mathcal{G}$ with unimodular metrics, $\tilde g_{ab}:=g^{-1/3} g_{ab}$ or using the horizontal lifts of the previous section. In the extended case, the orbit in $\mathcal{Q}/\mathcal{G}$ corresponding to the completely degenerate metric is not continuously path-connected to the rest of $\mathcal{Q}/\mathcal{G}$.  For $M=S^3$, this leaves only $\Phi_o=\{\phi^*\}=\{d\Omega^3\}$, the round metric, as generating the unique allowed past orbit.

In the first case, using unimodular metrics as the conformal section of the principal fiber bundle $\mathcal{C}\hookrightarrow \mathcal{Q}\rightarrow \mathcal{Q}/\mathcal{C}$,\footnote{Note that since this group is Abelian, there is no Gribov problem \cite{Singer_Gribov}.} for a curve of metrics in $\mathcal{Q}$ to change signature, the determinant must become degenerate, which disconnects the physical spaces of positive definite signatures from those with other signatures. In other words, the  \lq{}cone\rq{}\footnote{Riem is a cone inside the affine space of sections of symmetric (0,2)-covariant tensors, $C^\infty(TM\otimes_S TM)$, in that the sum of two metrics is still inside Riem, but not their difference \cite{Fischer}. }  which makes up the boundary of Riem inside the affine space $C^\infty(TM\otimes_S TM)$ becomes unreachable coming from $\mathcal{Q}/\mathcal{G}$.

In the horizontal lift picture, the orbits are defined by vectors $u_{ab}\in T_g\mathcal{Q}$ of the form $\alpha g_{ab}$. For any ultralocal supermetric in configuration space of the form 
\be\label{equ:inner_prod_superspace}\langle u, v\rangle_g= \int d^3 x \sqrt{g} F(g) G^{abcd}_\lambda u_{ab}v_{cd}\ee where $G^{abcd}=g^{ac}g^{bd}-\lambda g^{ab}g^{cd}$ where $\lambda\neq 1/3$, and $F(g)$ is a function of the metric and its (spatial) derivatives,  the 
orthogonal vectors to the orbits are going to be of traceless form. Thus the standard horizontal lifts orthogonal to the Weyl orbits (through any  standard canonical supermetric in Riem) imply a traceless velocity, $g_{ab}\dot g ^{ab}=0$, which does not change the volume-form $\sqrt{g}\, d^3x$. Thus such curves  cannot reach metrics with different signatures than the initial ones, as this would require going through a zero in the determinant of the metric. We can thus formulate the path integral (see section below) in terms of horizontal lifts and without stipulating further boundary conditions.

\paragraph*{Summary of this section.} In either the cases of  Diff$(M)$ or Diff$(M) \ltimes \mathcal{C}$, the quotient space $\mathcal{Q}/\mathcal{G}$ is only a stratified manifold, and the orbits corresponding to $\Phi_o$ are the lowest strata,  the \lq{}ultimate boundaries\rq{}, or the lowest dimensional corners, of $\mathcal{Q}/\mathcal{G}$, and the natural place to set up boundary conditions. When there is a unique least structured configuration, which  is also the corner of corners in reduced configuration space, the boundary conditions can be uniquely specified. This is the case for Diff$(M) \ltimes \mathcal{C}$ on $M=S^3$. Moreover, no further boundary conditions for the path integral on field space need to be specified. 

  In  other words, the specification of such a unique initial point in the amplitude kernel is sufficient to fully determine the entire wave-function of the Universe. 
Moreover, as mentioned above, the existence of such a preferred point will be fundamental in defining a \lq{}record\rq{}, which, in its turn, will replace standard notions of time.

%At this point it is worthwhile to call attention to two structural differences from the standard covariant transition amplitude \eqref{transition_amplitude}: i)  in the standard examples, we will choose a simple arc-length parametrization for the paths. In the case of a geodesic action for some Jacobi metric -- which is reparametrization invariant --  this gives a constant Fadeev-Popov determinant (see \eqref{FP_arc}). Now, the path integral in \eqref{equ:path_integral} is oscillatory, and thus has convergence issues, as does its covariant counter-part. In the covariant case, a Wick rotation is problematic, as there is no notion of a \lq\lq{}preferred time direction\rq\rq{} which would work for any space-time (see e.g. \cite{Wiltshire_intro}). Here however, the path curves don\rq{}t change under the symmetry, and thus we can equally formally perform the Wick rotation $\lambda\rightarrow i\lambda$ for any parametrization, to make the path integral (formally) convergent.  

\section{Path integrals in configuration space}\label{sec:config_space}
There are different ways of relating the standard relational setting for quantum mechanics to the path integral formalism. In the end, the constructed wave-function should obey some form of the reparametrization constraint. In appendix \ref{app:timeless_transition} I report on work of Chiou showing, for a relational particle model, how a rigorously defined timeless path integral regains this constraint, and how it reduces to the standard transition amplitude in the presence of a subsystems that behaves like a clock.    The mechanisms used in this proof (e.g. the Riemann-Stieltjes integral in terms of a mesh) can be recycled for building the path integral with  only global reparametrization constraints, as we have here.  

Equation \eqref{equ:deparametrizable} guarantees that \textit{in the presence of a reliable clock in a given portion of configuration space}, one can recover the standard quantum mechanics transition amplitude purely relationally.  But it is silent in what regards the existence of such subsystems. As we will see, the asymmetry of reduced configuration space will give us enough structure to build such time functions in the appropriate approximations, even for the (infinite-dimensional) gravitational case.

   \subsection{The basic definitions}
     Given an action functional $S(\gamma)$ as above, a connection-form $\varpi$, and a preferred `in' configuration $\phi^*$ (see The Past Hypothesis, below), the (timeless) transition amplitude (or propagator) to the orbit of the configuration $\phi$ is given by a timeless Feynman path integral in configuration space:\footnote{Rigorously, I should start with a phase space action, and only if the momenta can be integrated out of the path integral -- which up to the measure amounts to a Legendre transformation -- move onto a configuration space action. Here I overlook these issues. }
     \be\label{equ:path_integral}W(\phi^*,[\phi])=A \int_{\mathcal{P}} \mathcal{D}\gamma \mathcal{D}g \exp{[iS[\gamma_H(\phi^*, g\cdot \phi)]/\hbar]}\ee 
 where here $g\in\mathcal{G}$ acts on an arbitrary representative of the final point of the transition amplitude, $\phi$, and is integrated over with some measure.  A Haar measure is not required here;  unlike the standard case, we are not doing a group averaging procedure, each path on the base space corresponds to at most one $g$. If the curvature of the field-space connection form is zero, there is no relative holonomy on the fiber for two paths  $\gamma_1, \gamma_2$, interpolating between $\phi^*$ and the orbit of $[\phi]$. Thus all the lifts for the paths will end up in a single height of the orbit, let\rq{}s say $\tilde g \cdot \phi$. Then the path integral will acquire a functional delta: $\delta  (g,\tilde g)$, cancelling the  integral over $ \mathcal{D}g$. 
 
 In the case of gravity, we would have: 
\be\label{equ:path_integral_gravity}W(g_0,[g])=A \int_{\mathcal{P}} \mathcal{D}\gamma \mathcal{D}f \exp{[iS[\gamma_H(g_0, f^*g)]/\hbar]}\ee 
%Here $\Gamma(\phi^*,\phi)$ is taken to be the space of (piece-wise) unparametrized smooth paths without self-intersection between the two points.
The class of paths $\mathcal{P}$ under which this is integrated over are the horizontal lifts of paths in $\mathcal{Q}/\mathcal{G}$, as explained in the previous section. I.e.  $\gamma_H(g_0, f^*g)$ is a path that has a horizontal velocity (i.e. it is a horizontal lift through $g_0$), ending at $f^*g$. The implementation of horizontality will in general incur a Jacobian, which substitutes the standard Fadeev-Popov determinant. For the purposes of this paper, the specification is not necessary.\footnote{In the interest of completeness, horizontality by being orthogonal to the fibers generated by the group of conformal diffeomorphisms,  Diff$(M)\ltimes \mathcal{C}$, as discussed below,  for the standard supermetric in $\mathcal{Q}$, will implement transverse and traceless conditions on the metric velocities, $\dot g_{ab}=\dot g_{ab}^{TT}$. The appropriate Jacobian is calculated in appendix \ref{app:Jacobian}. We note that in the case of odd-dimensional Weyl symmetries, there is  no conformal anomaly, and thus the measure can be suitably made Weyl-invariant in conjunction with the action. This is also true in the Hamiltonian setting if the anomalies have a local representation \cite{SD_Weyl}.} This procedure eliminates the degenerate directions of the action functional --- and makes the projection of the Liouville measure non-degenerate ---  and does not suffer from Gribov ambiguities (see \cite{Conformal_geodesic} for more details).

Regarding the measure $\mathcal{D}\gamma$,  Barvinsky  (see \cite{Barvinsky2}, Sec II) has shown how to split such functional measures into  a lower dimensional functional integral over spatial fields, and then another integral over parametrizations. The first order part is just the projected Liouville measure:
 \be\label{equ:measure_timeful}\mathcal{D}\gamma=\prod_{\tau,i}d\phi^i(t)[\det (\mathbf{a})]^{1/2}(\tau)+\order \hbar\ee
where we are using DeWitt notation, so that $i$ here runs over both the tensor indexes and the spatial continuous ones, and 
\be\label{measure}\mathbf{a}=a_{ik}=\frac{\partial^2 \mathcal{L}}{\partial \dot\phi^i\partial \dot\phi^k}\ee
and $\partial$ is the functional partial derivative. { This is nothing but the invertibility matrix between velocity and momenta of course (and it is only non-degenerate in the absence of gauge symmetries).} Of course, by using the horizontality conditions, here one would replace $\dot\phi^i\rightarrow \dot\phi^i-\varpi(\dot\phi)\cdot \phi$. Suppose some gravitational action is given (again) by \eqref{equ:inner_prod_superspace} 
$$ L[g]= \int d^3 x \sqrt{g} F(g) G^{abcd}_\lambda \dot g^H_{ab}\dot g^H_{cd}+V(g)
$$
where $V(g)$ does not depend on the velocities, and $ \dot g^H_{ab}= \dot g_{ab}-\mathcal{L}_{\varpi(\dot g)} g_{ab}$,  according to \cite{Aldo_HG, Gauge_riem}. Then 
$$\mathbf{a}\rightarrow a^{abcd}(x,y)= \sqrt{g} F(g) G^{abcd}_\lambda \delta(x,y)$$
which has non-zero determinant. As mentioned above, a choice of a connection form is largely equivalent to a choice of gauge, but more suited to the context we are applying here. For a very simple connection form for the diffeomorphisms, one would get that horizontal vectors are those for which $\nabla^a\dot g_{ab}=0$, i.e. the transverse ones \cite{Gauge_riem}.

% % This definition is roughly equivalent to defining a Gaussian measure for a perturbation field (in our case, the velocities) \cite{Mottola}, a procedure we will explicitly use in calculating the Jacobians for our toy model path integral (see appendix \ref{app:Jacobian})
 Going back to the more general case, when combining the measure with the one-loop determinant integrated over time parametrizations, one finds the amplitude: 
$\det\left(-\frac{\delta^2 S({\gamma^\alpha_{\mbox{\tiny{cl}}}})}{\delta {\phi^a_i}\delta\phi_f^b}\right)$, which is only a determinant over the spatial fields, with no time integration, but for the on-shell action (equation 3.24 in \cite{Barvinsky2}). The semi-classical approximation will be recapitulated in section \ref{sec:config_space}.

The measure for the spatial diffeomorphisms (connected to the identity), are given by first replacing the diffeomorphisms by the path-ordered exponential of vector fields. I.e. by vector fields $X^a$ such that $X^i(t, x)=\frac{d}{dt} f^i(t, x)$ (here on the rhs $f^a$ actually indicates the coordinates of the image of $x$ under $f$), and such that $f(0,x)=x$, i.e. $f^i(0,x)=x^i$. Calling again $U^a_{bc}$ the structure constants of the commutator algebra of diffeomorphisms, given implicitly in e.g. \eqref{Lagrange_transf}, and introducing the continuous matrix $\Omega^a_b=X^c U^a_{bc}$, Teitelboim has shown that \cite{Teitelboim_path}:
$\mathcal{D}f=det\left(\frac{1-e^\Omega}{\Omega}\right)\mathcal{D}X^a$.
Finally, if the Weyl group $\mathcal{C}$ is part of $\mathcal{G}$, as it is in the example given, $\mathcal{G}=$Diff(M)$\ltimes \mathcal{C}$, the integration over the conformal factor cancels out with a functional delta,  because the conformal field-space connection-form is flat; all lifted paths end up at the same height in the conformal orbit, as mentioned above (between eqs \eqref{equ:path_integral}-\eqref{equ:path_integral_gravity}). 

For more on the folding properties of the path integral, and other technical details, we can follow Teitelboim in the procedure given in \cite{Teitelboim_path}, where only the conditions on the shift are local. These details, for a specific conformally  invariant model, are made explicit in \cite{Conformal_geodesic}.

  The horizontal lift and integration over the final point guarantees that the action functional is invariant wrt to gauge transformations everywhere but the initial point (which can be fixed by boundary conditions).
  %The thus constructed amplitude $W(\phi^*, \phi)$ obeys the respective functional versions of the constraints:
%\be\label{wave_constraint}\int d^3x\, \diby{\left(\chi(\phi,\pi_\phi)(N)\right)}{\pi_\phi(x)}\diby{W(\phi^*, \phi)}{\phi(x)}=0\ee
% This follows from the linearity of the constraints, equation \eqref{boundary_transf}, showing that there are no boundary terms for such constraints, and \eqref{final_func_conserv} in the appendix \ref{app:issues} (see also \cite{Hartle_Halliwell}). 
  Note also that there is no redundancy in this integral. Each path in the reduced configuration space is lifted uniquely. If the field-space connection had zero associated curvature, no integration over $\mathcal{D}f$ would be necessary, for all horizontal lifts would end up at the same height in the fiber over the final point $[g]$. 

\paragraph*{Summary of this subsection:} In this subsection, we have briefly and formally sketched how the field space path integral can be defined in the timeless context. We have explored the fact that the local gauge groups allow the construction of a reduced configuration space, to formulate a natural \lq\lq{}particle-like\rq\rq{} path integral formally defining a gauge-invariant wave-function. We have largely used the example of gravity with spatial diffeomorphisms and Weyl transformations. From now, unless otherwise specified, I will keep a more abstract approach that ignores the presence of gauge-symmetry. 
% I will assume that $\mathcal{D}\phi$ is the canonical measure in configuration space, arising from projection of the Liouville measure on phase space (for actions in which the momenta can be integrated out) and the path integral measure $\mathcal{D}\gamma$ arises from it through the  cylindrical measure construction  given in \cite{Cecille} and expanded to Riem in \cite{Clarke}.

 \subsection{The semi-classical transition amplitude}\label{sec:semi_classical}
Now we move back to the more general field configuration space specified above. 
First, I should note the ubiquity of interference experiments relying solely on multiple extremal paths; double slit and all sorts of interferometers rely on no dynamical information besides a semi-classical approximation with multiple extremal paths interpolating between initial and final configurations.

Indeed, here I will mostly study transition amplitudes between configurations that have at least one extremal (classical) path interpolating between them, and only briefly touch on more general cases in the accompanying \cite{path_deco}.\footnote{In the usual quantum mechanics setting, if the energy of the particle is lower than a potential barrier,  then the fixed energy transition amplitude from one side to the other is exponentially decaying in the phenomenon known as tunneling. One can model the transition using imaginary time, or an Euclidean version of the path integral.  Having said this, much progress has been made in explaining tunneling in the usual (or real time) path integral \cite{Tanizaki, Turok_real_time}. Moreover, we can separate fields, as I will discuss in section \ref{sec:conservation}. One field can be in a semi-classical approximation and serve as background, while the other undergoes evolution through the Schroedinger equation in this background.} In the context of path integrals in configuration space, I will be in the \emph{semi-classical} (or WKB, or saddle point), approximation (in the oscillatory domain). 
 
 Explicitly, the setting is given by a path integral in configuration space, \eqref{equ:path_integral}, for (locally) extremal paths parametrized by the set $I\subset \mathbb N$,  $\{\gamma_{\mbox{\tiny{cl}}}^\alpha\}_{\alpha\in I}$, between an initial and a final field configuration $\phi^a_i,\, \phi^a_f$, where here $a$ stands for both the tensorial and continuous indices. I will denote the on-shell action for these paths as $S_{\gamma_{\mbox{\tiny{cl}}}}$. The expansion, which is accurate for $1<< S_{\gamma_{\mbox{\tiny{cl}}}}/\hbar$ in arc-length parametrization,  is then:  
\be\label{equ:semi_classical_exp} W_{{\mbox{\tiny{cl}}}}({\phi_i}, \phi_f)= A \sum_{\alpha\in I}(\Delta_{\gamma^\alpha_{\mbox{\tiny{cl}}}})^{1/2}\exp{\left(i S({\gamma^\alpha_{\mbox{\tiny{cl}}}})/\hbar\right)}
 \ee 
 where $A$ is field-independent, and  the Van Vleck determinant has been defined as
 \be\label{equ:path_Van_vleck}
 \Delta_{\gamma_{\mbox{\tiny{cl}}}}^\alpha(\phi_i,\phi_f):=\det\left(-\frac{\delta^2 S({\gamma^\alpha_{\mbox{\tiny{cl}}}})}{\delta {\phi^a_i}\delta\phi_f^b}\right)= \det\left(\frac{\delta\pi^{(\gamma)}_b({\phi_i})}{\delta {\phi^a_i}}\right)
 \ee
 We here write the action as a functional of its initial and final points along ${\gamma_{\mbox{\tiny{cl}}}}$. To clarify the notation, if we wanted to use continuous indices explicitly, we would have e.g.: the \textit{on-shell momenta}  defined as 
\be\label{mom_VV}\pi^{(\gamma)}_a[{\phi_i}; x):= \frac{\delta S_{\gamma^\alpha_{\mbox{\tiny{cl}}}}[{\phi_i},\phi_f]}{\delta \phi^a_f(x)}\ee
where we used DeWitt's mixed functional/local  dependence notation $[{\phi_i}; x)$. For a proof of \eqref{equ:semi_classical_exp} in finite dimensions in this context, see \cite{Chiou}, and in the Euclidean field theory setting, see \cite{Barvinsky2}, Sec III. 

  I do not concern myself with the normalization factor $A$ for the moment, nor with the effect of the Maslov index (which emerges after focusing points only).\footnote{Equation \eqref{equ:semi_classical_exp} is valid up to the point where the first eigenvalue of $\frac{\delta^2 S}{\delta \phi\delta \phi'}[\phi_{{\mbox{\tiny{cl}}}}]$ reaches an isolated zero, which is a focal point of the classical paths. At such points the approximation momentarily breaks down. However, it becomes again valid after the focal point, acquiring the phase factor known as the Maslov index $\nu$, which is basically the Morse index (given by the signature of the Hessian) of the action.} The field determinant contained in the Van Vleck factor is the only element that would require regularization.
    As previously mentioned, this Van-Vleck determinant arises from the combination of the projected Liouville measure and the one-loop determinant. In more generality, these semi-classical prefactors can be translated into each other also in the field theory setting using what are known as ``reduction methods for functional determinants" \cite{Barvinsky2}, for which standard regularization methods can be applied.

It is fruitful to present this object in the context of Riemannian geometry, for later purposes. There, its given by a time integral of the expansion scalar along a geodesic congruence. I.e. let the  Lagrangian be just the infinitesimal line element of a curve, in a $d$-dimensional (semi) Riemannian manifold, and the action is the length of the curve, $S(\gamma)=\int d\gamma$. Let $t^\mu$ then be a geodesic congruence, defining the expansion as $\nabla_\mu t^\mu=\theta$, the Van-Vleck can be written as: 
$$\Delta_\gamma(x,y)=s_\gamma(x,y)^d\exp{\left(-\int_\gamma \theta ds\right)}
$$ 
   where $s_\gamma(x,y)$ is the proper distance along $\gamma$ between $x$ and $y$, and $s$ is the arc-length parameter along $\gamma$. The geometric version of  the Van-Vleck can be seen as quantifying  the focussing or defocussing of classical trajectories interpolating between $x$ and $y$ and around $\gamma$.  It provides a useful analogy to dynamical systems in configuration space, and a bridge between classical and quantum behavior, which we will explore later in section \ref{sec:gravity}.

Indeed, the standard  interpretation of  \eqref{equ:path_Van_vleck} is as an indicator of the spread of classical trajectories in configuration space. That is, the Van Vleck determinant measures how the dynamics expand or contract extremal paths along configuration space, i.e. how densely the initial configurations are transported by the equations of motion to the final configurations.\footnote{The interpretation of this fact is  quite ubiquitous, for instance, the Raychaudhuri equation -- which describes the spreading of geodesic congruences -- has a compact formulation in terms of the Van Vleck determinant \cite{Visser_Van_Vleck}. The same is true of the Jacobi fields used to perform the semi-classical expansion of the path integral \cite{Cecille}.} As in the particle case, I assume that the classical field history maps an infinitesimal configuration space density $\rho( {\phi_i})$ to an infinitesimal configuration space density $\rho_\alpha(\phi_f)$, and the Van Vleck determinant gives a ratio of these densities as propagated by paths around $\gamma_{\mbox{\tiny{cl}}}^\alpha$:
\be\label{equ:Van_vleck_volume}
\Delta_{\gamma_{\mbox{\tiny{cl}}}}^\alpha(\phi_i,\phi_f)=\frac{\rho_\alpha(\phi_f)}{\rho (\phi_i)}
\ee

 Under the assumption  that there exists at least one classical  (locally extremal) path between the two configurations, we find for the absolute value squared of the amplitude: 
\be\label{equ:semi_field_interference}
|W_{{\mbox{\tiny{cl}}}}(\phi_i,\phi_f)|^2=A\left(\sum_{\alpha\in I}\Delta_{\gamma^\alpha_{\mbox{\tiny{cl}}}}+2\sum_{\alpha\neq\alpha\rq{} }|\Delta_{\gamma^\alpha_{\mbox{\tiny{cl}}}}\Delta_{\gamma^{\alpha\rq{}}_{\mbox{\tiny{cl}}}}|^{1/2}\cos{\left(\frac{ S_{\gamma^\alpha_{\mbox{\tiny{cl}}}}-S_{\gamma^{\alpha\rq{}}_{\mbox{\tiny{cl}}}}}{\hbar}\right)}\right)
\ee 
Here, to unclutter notation, I have omitted the dependence of the Van-Vleck  on the configurations.  Interference terms can be clearly identified from \eqref{equ:semi_field_interference}. 

 The most appropriate way of extending equation \eqref{equ:semi_classical_exp}-\eqref{equ:semi_field_interference} to higher order of approximations was described in  \cite{Cecille}.  This is still based on extremal paths, and it can be incorporated with a piecewise approximation as well. Although I will not technically require or use these higher approximations, they provide -- at least in principle -- a way of extending the analysis done here to a more general context. 
 %In general, I will assume that the cases under study here have $\Delta_\gamma=\Delta_{\gamma'}$, which will cancel out in taking relative transition amplitudes in section \ref{sec:records}. 
\paragraph*{Summary of this subsection.} 
I have here summarized necessary concepts regarding semi-classical approximations for oscillatory path integrals in configuration space. Most important is the appearance in second order of the Van-Vleck determinant; it has the interpretation of a \lq{}focussing\rq{} of extremal trajectories. 

\subsection{The Born rule}\label{sec:Born}

In the absence of a separation into system and apparatus, the meaning of measurements and of the Born rule become more nebulous. Since my ultimate aim is to apply the constructions here to the whole Universe and to cosmology, I need to address the emergence and meaning of the Born amplitude in the present context. 

I have left the form of the function $P(\phi):=F[W(\phi^*, \phi)]\mathcal{D}\phi$, which ``counts" the number of configurations in small regions undetermined. We can associate the Born rule with a relative `density of observers' if we interpret  the likelihood of finding oneself e.g. in a region around configuration $\phi_1$ relative to configuration $\phi_{2}$ as the relative volume of these two regions.  I now turn to this.

Decoherence is a form of dynamical diagonalization of the (reduced) density matrix.  I will use a form of diagonalization appropriate to timeless configuration space, assuming there  is no significant interference between different extremal trajectories.\footnote{This is the description of decoherence used extensively in the consistent histories formulation \cite{CH_review}.} In this regime, using \eqref{equ:semi_field_interference}, we can compare the density function $P(\phi)$ to the classically propagated  volume elements in configuration space. 

From \eqref{equ:Van_vleck_volume}, 
\be\rho(\phi_f):=\rho(\phi_i)\Delta_{\gamma_\alpha}(\phi_i,\phi_f)
\ee
 Where there is only one  extremal path  connecting $\phi_i$ and $\phi_f$, namely $\gamma_\alpha$. More generally, one would have multiple  extremal paths between $\phi_i$ and $\phi_f$. This fact forbid us to take $\Delta$ to define the densities, since it is path dependent, giving only $\rho_{\alpha}(\phi_f)$, not  $\rho(\phi_f)$ -- which needs to take into account interference from other paths.  
 Nonetheless, for the semi-classical kernel, from the no-interference terms in \eqref{equ:semi_field_interference}, we also have
 \be \label{equ:Born_semi} \Delta_{\gamma_\alpha}(\phi_i,\phi_f) \approx |W_{\mbox{\tiny cl}}(\phi_i,\phi_f)|^2
\ee
Thus if we demand that our density functional gives the classically propagated densities in the no-interference limit, i.e.  
$$F[W_{\mbox{\tiny cl}}(\phi_i,\phi_f)]\propto \rho(\phi_f)=\rho(\phi_i)\Delta_{\gamma_\alpha}(\phi_i,\phi_f)$$ we find that from \eqref{equ:Born_semi}
\be\label{Born_semi2} F[W_{\mbox{\tiny cl}}(\phi_i,\phi)]:=\,\rho(\phi_i)|W_{\mbox{\tiny cl}}(\phi_i,\phi)|^2\ee
Since we will have a unique choice of $\phi_i$, given by the preferred \lq{}vacuum\rq{}, or in-state $\phi^*$, we can absorb $\rho(\phi^*)$ into the normalization $A$. This proves that, semi-classically, the Born rule emerges from the classical propagation of volumes in configuration space.

But even outside of the semi-classical regime, demanding positivity of $F$, it is easy to see that if one expands the function into polynomials of $z\bar z$, i.e. $F(z)=\sum_i a_i(z\bar z)^ i$,  one would obtain:
\be\label{Born_pol}F(z_1z_2)=\sum_i a_i(z_1z_2\bar z_1\bar z_2)^ i\ee and, on the other hand
\be\label{Born_pol2}F(z_1)F(z_2)= \sum_i a_i(z_1\bar z_1)^i \sum_j a_j(z_2\bar z_2)^j \ee
Only the diagonal terms of \eqref{Born_pol2} can match the polynomials of \eqref{Born_pol}. Therefore, only one $a_i$, say for $i=k$ can survive, and only if $a_k=1$. By the semi-classical limit above, \eqref{equ:Born_semi}, $k=1$. 

Alternatively, taking $F(z_1 z_2)=F(z_1)F(z_2)$, by differentiating both sides wrt $z_1$ and setting $z_1=1, z_2=z$ we obtain:
$$zF\rq{}(z)=F\rq{}(1)F(z)
$$ and a similar equation for the complex conjugates, $\bar z$, which means that the function is homogeneous: $F(z)= z^\alpha\bar z^\beta$, and demanding positivity, $\alpha=\beta$.\footnote{I thank W. Wieland for these remarks.} By the above classical limit, we must finally have that  $\alpha=1$. 
 
   Therefore, extending \eqref{Born_semi2} to the full  volume form, from the \emph{static wave-function over configuration space} $\psi(\phi):= W(\phi^*,\phi)$,\footnote{Alternatively, we could have defined a \lq\lq{}vacuum state\rq\rq{} $|\phi^*\rangle$ existing over a trivial (one complex dimension) Hilbert space $H_{\phi^*}$ with the usual complex space inner product,  over $\phi^*$,  and  then taking $\hat W(\phi^*,\phi):H_{\phi^*}\rightarrow H_{\phi}$ as an operator such that $|\psi(\phi)\rangle:= \hat W(\phi^*,\phi)|\phi^*\rangle$.} gives
  \be\label{Born} P(\phi)=|W(\phi^*,\phi)|^2\ee
   Equation \eqref{Born} -- i.e. the Born rule --  is  the extension of $F[W(\phi^*,\phi)]$ to arbitrary points $\phi$. 
   
   Note that no mention of \lq\lq{}measurements\rq\rq{}, or \lq\lq{}observations\rq\rq{} need to be made. Volumes in configuration space suffice. 
   % % I will now set for convenience, $A=1$,  reinstating it when attacking conservation of probability (as we will see, this amounts to setting the volume of configuration space to unity).
 
 \paragraph*{Summary of this subsection.}
 The decomposition property of the function $F$ is necessary if we want to translate certain statements about the amplitude to statements about probability. Here I have shown that given just this property and implementing a semi-classical limit, one can uniquely derive the Born rule as giving the volume-form in configuration space.

\section{Records}\label{sec:records}

%I will now attempt to describe the usual experimental setup from a timeless configuration space point of view.  Here, a laboratory  experiment can be generally described as a two-step process: i) first, define the experimental setting, call it $S$, and then ii) investigate outcomes after a certain time $t$, as measured by a clock in the laboratory. 

Given the Past Hypothesis, even allowing for an objective meaning to a relational transition amplitude, $W(\phi_i, \phi_f)$,  we should be able to do science from assumptions about relative number of configurations in $\mathcal{Q}$. It is true that while  examining the results of an experiment, all we have access to for comparison are the memories, or records, of the setup of the experiment.

   Since we are in a context that cannot rely on  absolute time, the meaning  of measurements and the updating of probabilities needs to be significantly modified.  Here, I will give a semi-classical definition of records. This definition should be seen as a mathematical pre-requisite for any functioning role for records; they do not, however, pinpoint the physical encoding of information in physical structures. 

%Probabilities will be interpreted as  the relative frequency of  configurations holding the same records, a completely Bayesian point of view. One does science by evaluating the consistency  of multiple redundant records of the same events \cite{Riedel}. 

   %, and only have access to  `the quantum state of the Universe' -- whatever we might mean by that -- over those configurations. 

\subsection{Semi-classical records} 
  I will denote a recorded configuration as $\phi_r$, and the manifold whose elements have $\phi_r$ as a record, as $\mathcal{Q}_{(r)}$.  If these represent experiments,  coexisting with the given record, each copy of \lq\lq{}the experimenter\rq\rq{} will find itself in one specific configuration, $\phi\in \mathcal{Q}_{(r)}$. The manifold $\mathcal{Q}_{(r)}$, consisting of all those configurations with the same records, then represents the setup of the experiment. In the cosmological setting this could be, for instance, all of the configurations that contain a \lq{}record\rq{} (as defined below) of the surface of last scattering. The post-selection is locating your configuration within $\mathcal{Q}_{(r)}$.

Loosely,   a configuration $\phi$ will be defined to  hold a  \emph{record} of (the recorded) configuration $\phi_r$ if,  for a given  `in' configuration ${\phi^*}$, all extremal paths from $\phi^*$ to $\phi$ go through  $\phi_r$.  This definition is  meant to embody the idea that $\phi_r$'s ``happening" is encoded in the state $\phi$. At least semi-classically, `branches' of the wavefunction contributing to the amplitude of $\phi$ have \lq\lq{}gone through\rq\rq{} $\phi_r$. 
 
 One note of caution: in the oscillatory semi-classical regime, when one speaks of the major contribution  to the amplitude as coming from extremal paths, what is really meant is that there is a coarse-graining of paths, seeded by the extremal ones, in which the paths close to the extremal ones enjoy constructive interference, whereas ones that deviate much have their interference wash out. To be rigorous, in the companion paper \cite{Deco_foundations} I have formulated the constructions here directly from specific types of coarse-grainings of paths. The results below go through without problems.\footnote{ In that more rigorous setting, for the semi-classical record to be of order $\hbar$,   the length of the extremal paths that seed the coarse-graining, $\gamma_\alpha^{\mbox{\tiny{cl}}}$, need to obey  $S_\alpha:=S(\gamma_\alpha^{\mbox{\tiny{cl}}})>>\hbar$, and, up to this order of approximation, the preferred extremal coarse-graining defined in \cite{Deco_foundations} cannot resolve  if $\phi_r$ lies along the actual extremal paths, or just around them. Moreover, to extend this notion to that of piece-wise extremal paths, I need to use that notion of extremal coarse-grainings as well.}

 \begin{defi}\label{def:full_record} Given $\phi^*$, the action $S(\gamma)$ on curves in configuration space, and the collection of parametrized extremal paths  $\{\gamma^\alpha_{\mbox{\tiny{cl}}}\}_{\alpha\in I}:[0,1]\rightarrow \mathcal{Q}$ such that $\gamma_{\mbox{\tiny{cl}}}^\alpha(0)=\phi^*$, $\gamma_{\mbox{\tiny{cl}}}^\alpha(1)=\phi\,,\, \forall \alpha\in I $, then $\phi$ is said to have a semi-classical record of $\phi_r$ if for each , $\alpha\in I$ there exists a  $t_\alpha\in [0,1]$, such that $\gamma_{\mbox{\tiny{cl}}}^\alpha(t_\alpha)=\phi_r$ and $S[\gamma_{\mbox{\tiny{cl}}}^\alpha(1)]>>\hbar$. 
 \end{defi}
 And with this definition we can prove the following
\begin{theo}\label{theo:record}
Given a configuration $\phi$ with a semi-classical record of $\phi_r$, then
\be\label{equ:semi_classical_record} W_{{\mbox{\tiny{cl}}}}({\phi^*},\phi)=  W_{{\mbox{\tiny{cl}}}}({\phi^*},  \phi_r)W_{{\mbox{\tiny{cl}}}}( \phi_r,  \phi)+\order{\hbar^2}
  \ee
Combined with property \eqref{equ:factorization_density} for the density functional, this means that the equation for the probability of $\phi$ automatically becomes an equation for conditional probability on $\phi_r$, 
\be
\label{equ:conditional} P(\phi)=P(\phi_r)P(\phi|\phi_r)
\ee  where $P(\phi|\phi_r)=|W(\phi_r,\phi)|^2$.  
\end{theo}
   Using the semi-classical approximation  \eqref{equ:semi_classical_exp}, 
$$ W_{{\mbox{\tiny{cl}}}}({\phi^*}, \phi)=  \sum_{\gamma_{\mbox{\tiny{cl}}}}(\Delta_{\gamma_{\mbox{\tiny{cl}}}})^{1/2}\exp{\left(i S_{\gamma_{\mbox{\tiny{cl}}}}({\phi^*},\phi)/\hbar\right)}+\order{\hbar^2}$$
If the system is deparametrizable, this means that one of the configuration variables can be used as \lq\lq{}time\rq\rq{}, and extremal trajectories can be monotonically parametrized by this variable. Then we could use \eqref{equ:deparametrizable},  and the semi-classical composition law for the semi-classical transition amplitude  (see \cite{Kleinert})\footnote{Here we are crucially assuming the \lq{}folding property\rq{} for the path integral. See \cite{Teitelboim_path}.}  to write: 
$$ W_{{\mbox{\tiny{cl}}}}({\phi^*},\phi)= \int \mathcal{D}\bar\phi_m W_{{\mbox{\tiny{cl}}}}((t_*,{\bar\phi^*}),  (t_m,\bar\phi_m))W_{{\mbox{\tiny{cl}}}}( (t_m,\bar\phi_m),  (t,\bar\phi))
$$
for a given  intermediary time $t_m$. Choosing $t_m=t_r$, there is a unique configuration through which all of the paths go through, $\phi_r=(\bar\phi_r, t_r)$.  The integral gains a  $\delta(\bar\phi_r)$, since extremal paths pass only through that point at $t_r$, and thus:
$$ W_{{\mbox{\tiny{cl}}}}({\phi^*},\phi)=  W_{{\mbox{\tiny{cl}}}}((t_*,{\bar\phi^*}),  (t_r,\bar\phi_r))W_{{\mbox{\tiny{cl}}}}( (t_r,\bar\phi_r),  (t,\bar\phi)) ~~~~ \square
$$
 In the more general case however, there is more work to be done.

Using \eqref{equ:semi_classical_exp}, we first write the rhs of \eqref{equ:semi_classical_record},
\be\label{rhs}  \sum_{\alpha_1}\Delta_{\alpha_1}^{1/2}\exp{\left(i S_{\alpha_1}({\phi^*},\phi_r)/\hbar\right)}\sum_{\alpha_2}\Delta_{\alpha_2}^{1/2}\exp{\left(iS_{\alpha_2}(\phi_r,\phi)/\hbar\right)}  +\order{\hbar^2}
\ee
where $\alpha_1$ are the sets of extremal paths interpolating between $\phi^*$ and $\phi_r$ and $\alpha_2$ those between $\phi_r$ and $\phi$. 
 %\footnote{In the standard particle path integral scenario, the configuration $\phi$ would be replaced by a position and a time. In that context, equation \eqref{equ:semi_classical_comp} means that the $\phi^\gamma_m$ are chosen at an intermediary time. One could integrate over the entire time interval, but that would only multiply the  normalization constant by the time interval, $A\rightarrow \Delta T A$.  In our context, it would be more convenient to use the arc-length gauge-fixing, \eqref{equ:arc_length}.} For the coarse-graining to be exclusive, they cannot share elements, thus we also demand that $S_\alpha(\phi,\phi_r)\,,\, S_\alpha(\phi_r,\phi_r)<<\hbar$
What we want to show is that to order $\order{\hbar^2}$, the contributing paths will be those that are continuous at  $\phi_r$. After this is done, we need to show that the Van Vleck determinants have the right composition law. This is done in appendix \ref{app:semi_classical_proof}.

%It is these two composition properties which ensure that the leading order of the multiple sum (implicit in each kernel) in the rhs of \eqref{equ:semi_classical_comp} reduce to a single sum on the lhs. In the particle case the last equality of equation \eqref{equ:semi_classical_comp} is a consequence of the fact that  the intermediary integration is at fixed time -- which in our case is included in the configuration. This means that only one representative of each path can be taken. 

\paragraph*{Strings of records} 
 
 For  more than one recorded configurations, say $\phi_r^1, \phi_r^2$,   definition \ref{def:full_record} demands that each  extremal path $\gamma_\alpha\in \Gamma({\phi^*},\phi)$ go through  $\phi_r^1, \phi_r^2$ in one order or another. Let $\phi$ contain multiple semi-classical records $\phi_r^i$, i.e.   $\phi\in \bigcap_i\mathcal{Q}_{(r^i)}$. Then an ordering of the records must exist for each extremal curve.  For each  $\alpha$, the set $\{\phi_r^i\}_{i=1\cdots n}$ is  ordered: $(\phi_r^1,\, \cdots\,,\,\phi_r^n)$. 
 
  In the presence of a single element $\alpha$,\footnote{And remembering that the on-shell action of $\gamma_\alpha$ between $\phi_r^i$ and $\phi^{i+1}_r$ must be much larger than $\hbar$ for records to be defined.} we can concatenate $P_{{\mbox{\tiny{cl}}}}({\phi^*},\phi^{i+1}_r)=P_{{\mbox{\tiny{cl}}}}({\phi^*},\phi^{i}_r) P_{{\mbox{\tiny{cl}}}}({\phi^i_r},\phi^{1+1}_r)$. The decomposition of the  density  follows:
 $$P_{{\mbox{\tiny{cl}}}}({\phi^*},\phi)\approx  P_{{\mbox{\tiny{cl}}}}({\phi^*},  \phi_{r}^{i})P_{{\mbox{\tiny{cl}}}}(\phi_{r}^i,  \phi_{r}^{i+1})P_{{\mbox{\tiny{cl}}}}( \phi_{r}^{i+1},  \phi)
 $$ 
 In other words, if there is one single $\alpha$ (no interference), and the semi-classical limit holds (i.e. the action spacing between the records is larger than $\hbar$), then the records in $\phi\in \mathcal{Q}_{(r)}$ yield a coarse-grained classical history of the field. In this case \emph{we recover a (granular) notion of classical Time}.\footnote{In a context of consistent histories, the separation of order greater than $\order\hbar$ between records will avoid the argument of Halliwell concerning the quantum Zeno effect \cite{Halliwell_Zeno}.} 

 We can still obtain   a consistent string of records for many interfering branches if the ordering $(\phi_r^{\alpha_1},\, \cdots\,,\,\phi_r^{\alpha_n})$ given for each $\alpha$, coincide, i.e. if for all $i$ and any $\alpha, \alpha'$, we have $\phi_r^{\alpha_i}=\phi_r^{\alpha'_i}=\phi_r^ i$. 
 Then kernel still can be decomposed:
  \be\label{equ:records_string}
  P_{{\mbox{\tiny{cl}}}}({\phi^*},\phi)\approx  P_{{\mbox{\tiny{cl}}}}({\phi^*},  \phi^1_r)P_{{\mbox{\tiny{cl}}}}( \phi^1_r,  \phi_r^2)\cdots P_{{\mbox{\tiny{cl}}}}( \phi^n_r,  \phi)
  \ee  This matches the analysis performed by Halliwell, in which he recovers \lq\lq{}time\rq\rq{}  from a simple Hamiltonian Mott bubble chamber ansatz,\footnote{Where spontaneous emission of $\alpha$-particles from a source ionize bubbles in a chamber of water vapor \cite{Mott}.} finding that the amplitude for $n$-bubbles to be excited, with the $n$-th bubble configuration being $q_f$, is given by:
 $$\langle q_f|\psi_n\rangle\propto \int d^Nq_n\cdots d^Nq_1 G(q_f,q_n)f_n(q_n)\cdots G(q_2,q_1)f_n(q_1)$$
 where $G$ are Green\rq{}s functions (for the free Hamiltonian), $f_i(q^i)$ are projections onto small regions of configuration space surrounding the $i$-th configuration, and $N$ is the dimension of configuration space. For this derivation, Halliwell notes that it is essential that there is some asymmetry in configuration space, marked by the source of the $\alpha$-particles. It is the same here: we require axiom 5 of an\rq{}origin\rq{} of configuration space (see also \cite{path_deco} for an extended version of this relationship).

 %However, we have reason to believe that in most cases a total ordering will exist, even if there are still   interfering extremal coarse-grainings (ECs). 

The definition implies that a  configuration can hold many records, and an ordering among these, with earlier recorded configurations being themselves recorded in later recorded configurations. Through this ordering,  a semblance of global time emerges from a fundamentally timeless theory. Again, this is in line with Halliwell\rq{}s reconstruction of time in the Mott chamber context \cite{Halliwell_Mott}. 
Configurations that are far from  the given initial configuration $\phi^*$ -- but still connected to it by extremal paths -- will in general have concentrated amplitude, and hold more records.

 \paragraph*{Relative probabilities of regions in $\mathcal{Q}$ with the same records.}   
The setup of an experiment implies the fixing of a submanifold in configuration space, $\mathcal{Q}_{(r)}$, characterized by its points all possessing the same records. In this abstract simple example, the submanifold is characterized by a single configuration $\phi_r$, which could represent for instance, \lq\lq{}the whole laboratory setup of a double slit experiment  and the firing of the electron gun\rq\rq{}, or, \lq\lq{}the cosmological surface of last scattering\rq\rq{}.

We can compare transition amplitudes for configurations with the same records in the following way, given $\phi_1, \phi_2 \in \mathcal{Q}_{(r)}$, for any initial ${\phi^*}$:
\be\label{equ:relative_prob}
\frac{P({\phi^*}, \phi_1)}{P({\phi^*}, \phi_2)}\approx \frac{P({\phi^*},\phi_r)P(\phi_r, \phi_1)}{P({\phi^*},\phi_r)P(\phi_r, \phi_2)}= \frac{P(\phi_r, \phi_1)}{P(\phi_r, \phi_2)}
\ee
This ratio gets rid of a common quantity to both density functions. 

In most experimental settings then,  equation \eqref{equ:relative_prob} justifies  the practical use of the record configuration $\phi_r$ as the effective initial point in the kernel. It is as if \emph{a measurement occurred, determining $\phi_r$}.  While it is true that the theory is timeless, we can still give a meaning to active verbs such as \lq\lq{}updating\rq\rq{} (e.g. of our confidence level): to the extent that humans are  classical systems, extremal paths in configuration will reflect anything that the equations of motion predict, including rational (and irrational) \lq\lq{}updating\rq\rq{} of our theories.

\paragraph*{Summary of this subsection.}
 I have here defined records. These are structures that can arise given  our volume-form on $\mathcal{Q}$, and which have many interesting properties. They yield  conditional probabilities (in much the same way that the Mott bubbles do); as correlated volumes in configuration space that emulate a notion of \lq\lq{}causation\rq\rq{}. I have only discussed these structures in the semi-classical regime, whence one obtains at most a \lq{}granular\rq{} history \footnote{With records separated by action of  order greater than $\hbar$. In the arc-length parametrization of the following sections, this is a separation in superspace. } 
 
 It is important to note here that our arguments from section \ref{sec:symmetry}, regarding the gauge-dependent character of intersecting curves in superspace if refoliations are allowed as a local symmetry. I.e. two curves might intersect before, but not after the action of a refoliation. This is not the case if one allows only the sort of local symmetries we have considered here, and/or global reparametrizations. This is the reason why the relativistic particle and minisuperspace models would not present such a problem. But it shows a disconnect, in so far as the concept of records explored here is concerned, between minisuperspace and the full theory with local refoliations. 

\subsection{Records and conservation of probability}\label{sec:conservation}
The generic case should be one of  redundancy of records in $\mathcal{Q}_{(r)}$.\footnote{ I mean this in two ways: first, that many different subsystems will have redundant records of the same \lq\lq{}event\rq\rq{} (subset of a configuration, in the sense of \cite{Locality_riem}). For example,  all the photons released by an electron hitting a fluorescent screen. By comparing the consistency of these records, one can formulate theories about the configuration space action and the probability amplitude. This \lq\lq{}consistency of records\rq\rq{} approach, has been recently emphasized in the context of selecting the basis for branch selection in decoherence \cite{Riedel}.  Since in this paper I\rq{}m not explicitly dealing with subsystems -- a topic I have dealt with in \cite{Locality_riem} -- I will ignore this type of redundancy; although it is very important for doing science. } In other words, within $\mathcal{Q}_{(r)}$ one could have a polygamy of record relations; e.g.: $\phi_1, \phi_2\in \mathcal{Q}_{(r)}$  with $\phi_1\in \mathcal{Q}_{(2)}$. This happens for example in the case of strings of records, discussed above. Ideally, when discussing conservation of probabilities, we would be able to discard such redundancy. 

 A choice of subset of $\mathcal{Q}_{(r)}$ for which there is no such redundancy will be called \emph{a screen},  written as $\mathcal{S}_{(r)}\subset\mathcal{Q}_{(r)}$. In other words, 
\be\label{equ:screen} \mathcal{S}_{(r)}:=\{\phi_i\in \mathcal{Q}~, i\in I\, |~ \phi_k\not\in \mathcal{Q}_{(\phi_j)}, \, \forall\,  k,j \in I\}
\ee
Screens are important when we want to discuss conservation of probability. As I will now show, we can expect probabilities to be conserved between records and the total amplitude of its screen. 

According to \eqref{equ:deparametrizable}, whenever we have a good \lq{}clock\rq{} subsystem, we know that the propagator $W(\varphi_1, t_1, \varphi_2, t_2)$  will obey a Schroedinger equation with respect to clock time. It follows that suitably defined currents and probabilities will obey conservation laws. However, part of the challenge of quantum cosmology is precisely to inform us of physics when such clock systems are \textit{not} available. Therefore, here I will show how a stand in for conservation of probability can be extracted from records. 

Now, in standard quantum gravity, there is  no fixed causal structure, and thus there is difficulty in defining concepts such as conservation of probability. In our case it is clear that e.g. for actions that are of geodesic type in configuration space,  at the semi-classical (WKB) level probability conservation laws should hold,  since the form of the equations resemble that of a particle in a constant potential in a zero energy eigenstate.

For a screen as defined through \eqref{equ:screen}, by the definition of semi-classical records \ref{def:full_record},  if extremal trajectories go through a small volume $\mathcal{D}\phi_{r}$, the total semi-classical probability flux for any screen will not exceed that around $\phi_{r}$ (or of a previous record-screen). In other words:  
\be\label{record_conservation} P(\phi_r)\simeq \Delta(\phi^*,\phi_r)\geq \int_{\mathcal{S}_{(r)}}\mathcal{D}\phi\, \Delta(\phi^*,\phi)\simeq\sum_{\alpha\in I_{\mathcal{S}}}\Delta_\alpha(\phi^*, \phi_\alpha)= P(\mathcal{S}_{(r)})\ee
where the extremal paths that leave $\phi_r$ and intersect the screen are parametrized by $I_{\mathcal{S}}$. In this particular discussion, it is assumed that there is no interference at the screen; to each point on the screen corresponds a single $\alpha$. Moreover, the flux is equal to the Born volume of a (infinitesimally) thickened region.  In particular, since 
$$P(\mathcal{S}_{(r)})=\int_{\mathcal{S}_{(r)}} \mathcal{D} \phi\, \Psi(\phi)\overline{\Psi(\phi)}\simeq |\Psi(\phi_r)|^2 \int_{\mathcal{S}_{(r)}} \mathcal{D} \phi\,  |W(\phi_r, \phi)|^2\leq |\Psi(\phi_r)|^2=P(\phi_r)
$$ 
this means that 
\be\label{conserv_prob}\int_{\mathcal{S}_{(r)}} \mathcal{D} \phi\,  |W(\phi_r, \phi)|^2\leq 1 \ee
If every extremal path that crosses $\phi_r$ also intersects the screen, then, up to higher orders of $\hbar$, the inequality of \eqref{conserv_prob} is saturated.

\paragraph*{Application to timeless configuration space}
Suppose then that we were able to find a metric in configuration space for which extremal trajectories of the action are given by geodesics.\footnote{These are called Jacobi metrics, and there is wide class of dynamical systems that can be put into this form \cite{Lanczos}.} From the Jacobi procedure, this does not imply that there is no potential for the dynamical system, just that it can be reabsorbed into an effective metric in configuration space, with the use of specific conformal factors.  

Using DeWitt functional notation, let the index $a$ represent the totality of both continuous and discrete indices of the spatial field in question (thus contraction includes spatial integration). Then, for a configuration-dependent supermetric $G^{ab}(\phi)$, and configuration space curves $\gamma_a$, 
\be\label{action_geod}S[\gamma]=\int dt \left( V(\phi) G^{ab}\dot \gamma_a\dot\gamma_b\right)^{1/2}\ee
Where $V(\phi)$ is the conformal Jacobi factor.
  Reparametrization invariance implies a reparametrization constraint, since 
\be\label{momentum_geod}p^a=\frac{\partial L}{\partial \dot\gamma_a}=\frac{V(\phi)^{1/2}G^{ab}\dot \gamma_b}{\left( G^{cd}\dot \gamma_c\dot\gamma_d\right)^{1/2}}\ee
(where $\partial$ represents the functional partial derivative).
We thus get:
\be\label{Ham_geod}H\equiv p^a\,G_{ab}\,p^b-V(\phi)=0\ee%\Rightarrow p^a\,\frac{G_{ab}}{V(\phi)}\,p^b=1\ee
%where $\frac{G_{ab}}{V(\phi)}$ is just the inverse metric of the one used in \eqref{action_geod}. Everything that follows could be done using the right-most equation of \eqref{Ham_geod}, which is in fact an equation mode adapted to the Jacobi geometical origin of the system. However, to keep the reader\rq{}s familiarity with the notation, I will use the left-most version throughout. 

Note that \eqref{Ham_geod} is not a local equation like the usual scalar ADM constraint \cite{ADM}, since there is a hidden integration on all contracted indices. It is a bona fide global conservation law, and that makes all the difference. Although we are not here at a homogeneous (minisuperspace) approximation, we can still use many of its standard techniques and results. Of course, in the standard superspace approximation of homogeneous fields, equation \eqref{action_geod} would no longer contain the local (integrated over) index, and only the actual indices appearing in \eqref{Ham_geod}. %Moreover, since extremal paths in configuration space are foliation dependent, it is unlikely that anything like screens (or records) can be defined (see section \ref{sec:conclusions}) unless one is already at a minisuperspace approximation, in which case it becomes much more similar to our approach.  

For the reparametrizations generated by \eqref{Ham_geod}, the algebra is a bit more laborious, but in the end one obtains, in \eqref{boundary_transf}, for $\epsilon(t)$ a global (in space) parameter, just 
$\delta_\epsilon S=[2\epsilon(t) V(\phi(t))]|^{t^f}_{t_i}$ and $\delta_\epsilon \phi_a=2\epsilon(t) G_{ab} p^b$. Thus, from \eqref{HH_constraint}, it follows that the wave-function constructed from the path integral satisfies the constraint:
\be\label{Schro_geod}\hbar^2 \left(G^{ab} \frac{\delta^2\Psi(\phi)}{\delta \phi^a\delta \phi^b}\right)-V(\phi)\Psi(\phi)=0\ee
where I have chosen the non-self-adjoint factor ordering, with a possibly non-ultralocal supermetric $G^{ab}$ symmetric in $ab$.\footnote{Non ultralocal, means that $G_{ab}p^ap^b=\int d^3x\, d^3 y G_{ab}(x,y)p^a(x)p^b(y)$ and $ G_{ab}(x,y)$ is not necessarily proportional to a Dirac delta. Such generalization is useful in order to facilitate point-splitting regularization methods, even if one takes the ultralocal limit later.} With the polar form for a wavefunction (assuming that there is at most one extremal solution connecting to $\phi$),  
\be\label{polar_psi}\Psi(\phi)= A(\phi)\exp{(i S(\phi)/\hbar)}\ee
 with $S(\phi):=S(\phi^*,\phi)$ (the on-shell action from $\phi^*$ to $\phi$) and $A$ real functionals, 
 we obtain
 $$\left(\hbar^2G^{ab}\left(\nabla_a\nabla_b A+\frac{i}{\hbar}\left(2\nabla_aA\nabla_bS+\nabla_a\nabla_bS \right)-\frac{A}{\hbar^2}\nabla_a S\nabla_b S\right)-AV(\phi)\right)\exp{(i S(\phi)/\hbar)}=0$$
%\begin{align}\label{equ:Schrodinger}
%&-\int d^3x\, d^3 y\, \left(G^{ab}(x,y) \left(\frac{\delta S[\phi]}{\delta \phi^a(x)}\frac{\delta S[\phi]}{\delta \phi^b(y)}-\frac{\hbar^2}{A[\phi]} \frac{\delta^2A[\phi]}{\delta \phi^a(x)\delta \phi^b(y)}\right)\right)- V[\phi]\Psi[\phi]=0\\
%&\int d^3x\, d^3 y\, \left(G^{ab}(x,y) \frac{\delta}{\delta \phi^a(x)}\left( A^2[\phi]\frac{\delta S[\phi]}{\delta \phi^b(y)}\right)\right)=0\label{conserv_geod}
%\end{align}
%which are the analogous equations to:
%\begin{align}
%&\frac{1}{2m}\nabla_aS\, g^{ab}\, \nabla_b S+V-\frac{i\hbar}{2m}\nabla^2S=0\label{conserv2}\\
%&g^{ab}\nabla_a(\rho\nabla_b S)=\nabla_a j^a=0\label{conserv1}
%\end{align}
with $\frac{\delta S(\phi)}{\delta \phi^b}=\nabla_b S$. To $ \order\hbar$, it implements the Hamilton-Jacobi equation (at order 0) and  the standard current conservation law (at order $\hbar$), 
\be\label{bare_conserv}G^{ab}\nabla_a\left(A^2(\phi)\diby{S}{\phi^b}\right)=0
\ee with $A^2\rightarrow \rho\,\, ,\, \,  j_b\rightarrow A^2(\phi)\nabla_b S$.   This will hold for any such Jacobi action, and without approximations other than the semi-classical one.  %Moreover, everything would go through using the full Jacobi metric version, namely, with the substitutions: $G_{ab}\rightarrow \frac{G_{ab}}{V(\phi)}, V(\phi)\rightarrow 1$. 

   \paragraph*{Recovering Schroedinger.}
But we would like to recover an approximation of the Schroedinger equation, for a weak coupling to gravitational degrees of freedom. 
We will now roughly go over a similar derivation by Banks \cite{Banks_grav}, here adapted to our setting; namely a simple geometrodynamical model.

Suppose  that one separates two kinds of fields, $\phi=(g_{ab}, \varphi)$, i.e. $\varphi$ should be seen as some sort of source matter field for the metric. Suppose moreover that the gravitational field is mostly unaffected by the source field, since the coupling is assumed to be very weak. 
  In this approximation, for  the analogue of Hamiltonian \eqref{Ham_geod}, we would have (reinstating $\hbar$ and $G$): 
\be\label{matter_Ham}
\left(-\frac{1}{2m^2_p}\nabla^2+m_p^2V[g]+H_{\mbox{\tiny mat}}(g,\varphi)\right)\Psi[g,\varphi]=0
\ee
where  $m_p$ is the Planck mass (which will give us an ordering), basically establishing the separation of scales between the gravitational and source Hamiltonians, and where, to avoid ambiguity now that we will explicitly refer to local and functional dependence, I used square brackets to denote functional dependence. The gravitational functional Laplacian is: 
\be\label{grav_Laplacian}\nabla^2=\int d^3x\,d^3y\, \left(G_{abcd}(x,y) \frac{\delta^2}{\delta g_{ab}(x)\delta g_{cd}(y)}\right).\ee
 where the simplest example of a supermetric is $G_{abcd}(x,y)=g_{ac}(x)g_{bd}(y)\delta(x,y)$, and the potential is left completely arbitrary.

  Assuming moreover that the WKB state now takes the form
\be\label{matter_WKB}
\Psi(g,\varphi)=\exp{(im_p^2 S[g])}A[g]\psi[g,\varphi]+\order{m_p^{-2}}
\ee
Here I am assuming that the WKB part of the wave-function only holds for the gravitational field. In other words, we are at the  no interference limit \emph{for gravity}.\footnote{In the path integral context, the perturbations due to the source fields should remain \lq\lq{}small\rq\rq{}, i.e. within the extremal coarse-graining (defined in \cite{path_deco}), which define a tubular bundle around the extremal paths. The two conditions can be shown to be identical.}
Note moreover that the split between $A[g]$ and $\psi[g,\varphi]$ is largely arbitrary. This allows us to set $A[g]$ as satisfying the functional differential equation (implementing a conservation law along the direction of the momenta):\footnote{The difference between self-adjoint factor ordering and the one we chose here, amounts to a difference in the above definition; whether we put the supermetric inside or outside -- our choice -- of the functional derivatives. Had we chosen the self-adjoint one, with the supermetric inside the derivatives, this would have amounted to using a covariant divergence in superspace, instead of the simple one we used. I decided to use the non-covariant one for simplicity of the formulas, but everything here is translatable to the covariant (self-adjoint) context. Note moreover, that there are two covariances at work here. One is the one given by the principal fiber bundle structure; it ensures that the relevant structures live in the reduced configuration space. The other, which I have not given much attention to, ensures that quantities don\rq{}t depend on the \textit{coordinates in reduced configuration space}.  \label{footnote_conserv} }
\be\label{conserv_imposed} \int d^3 y\int d^3x \left(G_{abcd}(x,y) \frac{\delta }{\delta g_{ab}(x)}\left(\frac{\delta S[g]}{\delta g_{cd}(y)}A^2[g]\right) \right)=0
\ee
with the components of this (bare) current being: 
\be\label{bare_current}J^{ab}(x)=\left(\frac{\delta S[g]}{\delta g_{ab}(x)}A^2[g]\right)
\ee
and still leaving the semi-classical wave-function general. 

Expanding \eqref{matter_Ham}, at order $m_p^2$ we get the Hamilton-Jacobi equation, as expected:
$$ \int d^ 3x\,d^3y\, G_{abcd}(x,y)\frac{\delta S[g]}{\delta g_{ab}(x)}\frac{\delta S[g]}{\delta g_{cd}(y)}+V[g]=0
 $$ and  next order ($m_p^0$) we obtain:
 \begin{align}
 \int d^3x\,d^3y\, \left(G_{abcd}(x,y) \frac{\delta^2 S[g]}{\delta g_{ab}(x)\delta g_{cd}(y)}\right)A[g]\psi[g,\varphi]+2\int d^3 y\int d^3x \left(G_{abcd}(x,y) \frac{\delta S[g]}{\delta g_{ab}(x)}\frac{\delta A[g]}{\delta g_{cd}(y)} \right)\psi[g,\varphi]\nonumber\\
 +2\int d^3 y\int d^3x \left(G_{abcd}(x,y) \frac{\delta S[g]}{\delta g_{ab}(x)}\frac{\delta \psi[g,\varphi]}{\delta g_{cd}(y)} \right)A[g]-iH_{\mbox{\tiny mat}}(g,\varphi)A[g]\Psi[g,\varphi]=0\nonumber
 \end{align} 
and finally, using \eqref{conserv_imposed},  one obtains that the state functional satisfies the first order functional variational equation:
\be\label{functional_Schroedinger}
i\int d^ 3x\,d^3 y\, G_{abcd}(x,y) \diby{S[g]}{g_{ab}(x)}\diby{\psi[g,\varphi]}{g_{cd}(y)}=H_{\mbox{\tiny mat}}(g,\varphi)\psi[g,\varphi]
\ee

For  a given region in $\mathcal{Q}$, one might choose a smooth screen by finding a functional time  such that:
\be\label{equ:time}
\frac{\partial}{\partial T}=\int d^ 3x\, d^3y \, G_{abcd}(x,y)\diby{S[g]}{g_{ab}(x)}\diby{}{g_{cd}(y)}
\ee
which is a standard choice for time functions for Wheeler-DeWitt in minisuperspace (see eq 10.25, \cite{Kuchar_Time}).\footnote{This is almost the integrable arc-length parametrization (which need not exist for large regions of configuration space), but which in general should exist for given small region (around a record, for example). It is almost, but not quite, because we have not used the Jacobi metric. Had we done so, then it is trivially checked that, from the Hamilton-Jacobi equation, one would get $\frac{\partial S}{\partial T}=1$. In the Jacobi metric, the standard way of building such surfaces is the following: particular solutions to the Hamilton-Jacobi equations define a congruence of classical trajectories. By choosing an initial arbitrary surface and Lie dragging along the (arc-length parametrized) trajectories, one builds a foliation. In our case, we can Lie drag from the (small region around the)\footnote{As made explicit in \cite{path_deco}, records, and the coarse-grainings around extremal paths, are not infinitesimal, but form finite regions in configuration space, determined by the degree of distinguishability of said regions, a distinguishibility which in its turn is determined by the accuracy of the semi-classical transition amplitude.} record configuration, and thus there is little arbitrary choice of the initial surface. 
The remaining directions would be the ones orthogonal to the gradient of the time function (they would span the screen).} 
 \begin{figure}[h]
\begin{center}
\includegraphics[width=0.6\textwidth]{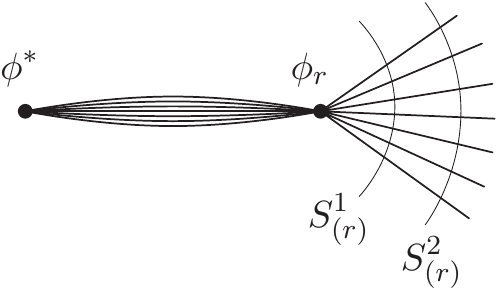}
\caption{A schematic representation of a sequence of two screens,  $S_{(r)}^1, S_{(r)}^2$ ,  for a given recorded configuration $\phi_r$ (the divergence between the rays after $\phi_r$ are exaggerated for illustration).   }
\end{center}\label{fig:screens}
\end{figure} 

Using the choice of time \eqref{equ:time} on \eqref{functional_Schroedinger}, we obtain the Schroedinger equation (on a given classical background),
\be\label{time_schroedinger}i\frac{\partial}{\partial T}\psi[g,\varphi]=H_{\mbox{\tiny mat}}(g,\varphi)\psi[g,\varphi]\ee
 It is easy to show that the probability currents and densities of gravity and sources also factorize, and the Hamiltonian is Hermitean with respect to the usual Schroedinger inner product for the source fields. I show this in appendix \ref{app:conservation}. And thus we obtain the standard Schroedinger interpretation in the given timeless background  satisfying the Hamilton-Jacobi equation.\footnote{ We could also have records obeying the approximations for the different fields. This naturally allows for tunneling of the source fields on a given background. }

For example, if we want to find the partial  infinitesimal flux of the bare probability current \eqref{bare_current}, through our smooth screen as defined by \eqref{equ:time}, we just need to take the inner product between the vectors $\frac{\partial}{\partial T}$ and the respective current (i.e. $(dT)_a\,j^a$). In components, we have: 
 \be\label{equ:normal} \left(dT\right)^{ab}(x)=\frac{1}{V[g]}\diby{S[\phi]}{g_{ab}(x)} ~~~ \mbox{and}~~~J_{cd}(x)=\int d^3 y\, G_{abcd}(x,y)\left(\frac{\delta S[g]}{\delta g_{ab}(x)}A^2[g]\right)
\ee
since then,  from the Hamilton-Jacobi equation,  $\mathbf{dT}\cdot \frac{\partial}{\partial T}=1$.

  Infinitesimally, in this approximation, the flux of current is given by the probability:
\be\label{Schroedinger_metric} \mathbf{dT}\cdot {\vec{J}}= \int d^3 x\, d^3 y\,\frac{1}{V[g]}\diby{S[\phi]}{g_{ab}(x)}G_{abcd}(y,x)\diby{S[\phi]}{g_{cd}(y)}A^2[g]=A^2[g]
\ee
where we used the Hamilton-Jacobi equation. 
This equation is approximately conserved along the flow of $\mathbf{j}$, as per \eqref{conserv_imposed}, and it represents the infinitesimal Born volume of the region corresponding to the infinitesimal thickening of the screen area element.

\paragraph*{Differences from Wheeler-DeWitt.}

This approach has profound advantages in comparison to the analogous Wheeler-DeWitt calculation -- which has many significant flaws.  Here I will list  only two of these flaws, which I believe are the most relevant (for a complete list see \cite{Kuchar_Time}, pgs 54-62), and explain how the present approach overcomes them. 
\begin{itemize}
\item \textbf{Superposition problem.}  The previous calculations, but done  in the standard -- minisuperpsace WdW -- context, are bedeviled by  a lack of uniqueness of the Hamilton-Jacobi functional. In our WKB ansatz, the only assumption we required was that we were in the no interference domain for the gravitational part (interference for the source terms was still allowed by our form of \eqref{matter_WKB}). For most gravitational systems, this just means that records are sufficiently far away (in terms of arc-length distance) from the region being considered so that gravitational decoherence has ensued (see section \ref{sec:gravity} below).\footnote{For an instantiation of this fact, and a rationale based on chaos for which such non-linear classical interactions should quickly (in terms of arc-length time along extremal paths in reduced configuration space)  should suffice for establishing decoherence (in the consistent histories sense). See also \cite{path_deco}. } With our preferred initial configuration $\phi^*$, there is no further choice that goes into determining the solutions to the Hamilton-Jacobi equation. 

This is not the case in general, and in particular it is not the case for WdW. Due to the linearity of the equations,  one could in principle have a solution in  the form: 
\be\label{bad_sum} \Psi(g,\varphi)\approx \exp{(im_p^2 S_1[g])}A_1[g]\psi_1[g,\varphi]+\exp{(im_p^2 S_2[g])}A_2[g]\psi_2[g,\varphi]
\ee 
 The Hamilton-Jacobi equation is not linear in $S$, and thus two solutions don\rq{}t necessarily yield another solution.\footnote{Not to mention they might have different boundary conditions.} 
 
 In the presence of more than one Hamilton-Jacobi solution,  one loses almost all the nice features of the semi-classical approximation above. For instance, the derivation of the Schroedinger equation \eqref{time_schroedinger} no longer holds. Even if one assumes that $S_1$ and $S_2$ separately satisfy the Hamilton-Jacobi equation, and that $A_1$ and $A_2$ separately satisfy the conservation equation, one still only obtains a sum, for $\psi_1[g,\varphi], \psi_2[g,\varphi]$: 
$$\left(i\frac{\partial}{\partial T}\psi_1-H_{\mbox{\tiny mat}}(g,\varphi)\psi_1\right)e^{(im_p^2 S_1[g])}A_1[g]+\left(i\frac{\partial}{\partial T}\psi_2-H_{\mbox{\tiny mat}}(g,\varphi)\psi_2\right)e^{(im_p^2 S_2[g])}A_2[g]=0$$
which provides many more solutions than just the individual Schroedinger equations. Moreover, such a superposition would allow for solutions with negative norm. In the WdW case, since the long-distance dynamics is silent about initial conditions (and their Hamilton-Jacobi associated solution), unless the short-distance dynamics somehow select a unique solution, the wave-function will not have a desired semi-classical interpretation.%\footnote{ In ordinary quantum mechanics, if the WKB approximation is valid in some region of configuration space, a coherent state will propagate classically in this region. However, there are many examples of multi-dimensional quantum systems whose eigenstates obey the WKB equations in some region of configuration space but are linear superpositions of WKB wave-functions corresponding to different classical trajectories. The emerging behavior is far from classical.} 
 Moreover, it seems that, no matter what the short-distance dynamics is, if $S[g]$ is a solution, so is $-S[g]$. It is easy to find solutions of the form \eqref{bad_sum} which then have zero (Klein-Gordon) norm. 

The reason we have avoided this fate is that due to our fundamental asymmetry of configuration space, we have naturally restricted the space of solutions to ones corresponding to a single Hamilton-Jacobi principal function, $S[g]=S[g^*,g]$, and by choosing a smooth screen (corresponding to a time function) transverse to the classical trajectories and with the orientation coming from the classical trajectories themselves (away from the record). The point is that these are not choices in our framework, but consequences of the given approximation and our axioms.

\item  \textbf{Many-fingered time problem.}
This is the problem of time, in the semi-classical approximation of minisuperspace. Outside of minisuperspace for standard Wheeler-deWitt, $\partial_T$ is not a global time, it is space-dependent. Fluxes, conservation laws, and the like become ill defined concepts in the full superspace, partly because of the problems outlined in section \ref{sec:symmetry} (after equation \eqref{refol}): equal time surfaces in superspace will not be gauge-invariant quantities. Moving on to minisuperspace, consequences first appear in  that the time function given in \eqref{equ:time} \lq\lq{}should correspond to a foliation in which geometry is homogeneous. It would be a consistent choice if we found that gravity had such homogeneity down to the Planck scale;  which it doesn\rq{}t\rq\rq{} \cite{Kuchar_Time}. Moreover, one should note that within gravity no space-time scalar can be built just from the spatial metric, as should be the case if \eqref{equ:time} was to have meaning in a fully covariant theory. Thus an ontological barrier between the semi-classical approximation in a homogeneous (minisuperspace) approximation and the full theory is erected, isolating the two; the approximation becomes its own consistent theory, without a strong connection to the full theory \cite{Kuchar_Time}.

As mentioned in the beginning of this section, the standard covariant interpretation of gravity has no place for (reduced) configuration space structures, such as records and screens. We do, because we have different symmetry principles and only global conservation laws associated to (global) time. In the context of timeless theories in configuration space as I have presented them here, the results of this section still  of course arise from an approximation; but only one of convenience,  supported by the hierarchy of the interactions and dynamical behavior of gravity.  
\end{itemize}

  %Although of limited validity, this approximation is more problematic for the full  WdW equation, since the kinetic term has a negative contribution coming from the trace terms of the momenta, and the potential -- i.e. the Ricci scalar $R$ -- has no definite signature (see \cite{Wald_Unruh}). This means that the one might obtain negative probabilities.\footnote{Although for Klein-Gordon this is not an issue, since one can split the frequencies into positive and negative, here the absence of a time function prohibits us to give any meaning to this.} Moreover, many natural wavefunctions (such as the Hartle-Hawking no-boundary one) have zero norm with this definition \cite{Wiltshire_intro}.  Another issue with the WdW analog of \eqref{equ:Schrodinger} is that there is no integration  in the WdW \eqref{WdW}-- since the scalar constraint is local -- which brings with it $\delta(0)$ elements that need to be regularized (see appendix \ref{app:scalar_ADM} for further technical problems). 
  
  \paragraph*{Summary of this subsection.}
  
Although the Born rule gives a volume in configuration space, in the absence of external time we need to define what \lq\lq{}conservation\rq\rq{} means. For this, I defined screens: sets of configurations with the same record, so that no element of this set is a record of another. That is, there is no redundancy of records within this set.  I then showed that for smooth screens -- defined by the flow of the extremal paths -- one can indeed find conservation of probability (for the associated foliation, where it exists). Moreover, I showed that, for weak matter couplings, one can recover the Schroedinger equation for matter fields in the classical background (of each extremal path). Unlike what is the case for minisuperspace WKB approximations of WdW, records and screens define the time functions and unique Hamilton-Jacobi functionals -- each in its domain of existence -- which allow the construction of positive currents associated to the probability densities, which can be at least formally extended to the full theory. 
 
\section{Applications to quantum gravity}\label{sec:gravity}

I started the paper discussing fault lines between the foundations of quantum mechanics and those of gravity. I went on to explain some of the problems for defining a reduced configuration space in the presence of local refoliations. Most of these problems are not visible from  minisuperspace. Thus, to highlight this aspect of the distinction between approaches,   I now briefly present a simple gravitational toy model, corresponding to a form of strong gravity\footnote{In $N=1$ gauge, but with a strictly positive kinetic term, i.e. with no negative conformal modes.} that is not a symmetry reduced model.
Indeed, usually, one must resort to minisuperspace approximations to obtain information about quantum cosmology. Here I propose a different simplification (which might have even less to do with our Universe than standard minisuperspace cosmology \cite{Kuchar_valid}). 
The proposal is to use features of the infinite-dimensional geometry --- specifically the relation between the Van-Vleck determinant and the geometric expansion scalar along a geodesic congruence \cite{Visser_Van_Vleck} ---  to directly obtain first order quantum effects for gravitational theories. Here I will give only a very brief  demonstration of this tool.  The geometric analysis used below, however, is available for a much larger set of geometric actions than the one presented here (see \cite{Michor_general}). 

\subsection{Gravitational toy model}

First, a brief word on the use of the geometry of infinite-dimensional spaces in order to study quantum cosmology in the semi-classical limit. 

\paragraph*{Geometry and dynamics.}

Dynamical systems that are quadratic in the momenta and time-independent can have their flows associated to geodesics of a Riemannian metric \eqref{equ:Jacobi}. It turns out that even more general dynamical systems have similar geometric formulations, as shown in \cite{Lanczos}, albeit not necessarily through the use of Riemannian metrics. 

This implies for example that the confluence or divergence of nearby trajectories are governed by the Jacobi equation (the geodesic deviation):
\be\label{equ:Jacobi}
\frac{D^2 J^a}{ds^2}=-R^a_{bcd}\dot \gamma^b\dot \gamma^d J^c
\ee
where $R_{abcd}$ is the Riemann tensor associated with the Jacobi metric and $D/ds = \dot\gamma^a\nabla_a$. Equation \eqref{equ:Jacobi} is easily obtained by defining a geodesic congruence $\gamma(t,s):[0,1]\times [0,1]\rightarrow \mathcal{Q}$,  
$$\frac{d}{dt} \gamma(t,s)=\dot\gamma^a(t,s)\,\, \, |\, \, \,\dot\gamma^a\nabla_a\dot\gamma^b=0\,\, \mbox{and}~~J^a(t,s):= \frac{d}{ds} \gamma(t,s)  $$ where we used abstract index notation. Definition through the map $\gamma(t,s)$ implies that $\dot\gamma^a\nabla_aJ^b=J^a\nabla_a\dot\gamma^b$, and appropriately commuting the derivative on the lhs of \eqref{equ:Jacobi} yields the rhs.

Many dynamical conclusions can be brought to bear here from geometry. For instance, the Hadamard-Cartan theorem states that if the sectional curvature is strictly negative along $\gamma(t,0)$, then there are no conjugate points to $\gamma(0,0)$; i.e. extremal paths don\rq{}t \lq\lq{}reconverge\rq\rq{} or \lq{}recohere\rq{}. The other side of this is the Bonnet-Myers theorem (similar to the standard Focussing theorem), which states that for a bounded sectional curvature along the curve $K(u,\dot\gamma)= R_{abcd} u^bu^d  \dot\gamma^a\dot\gamma^c\geq\kappa>0$ %(where $e_i^a$ are the parallel propagated orthonormal basis along $\gamma(t)$) 
then for $\beta\geq \pi/\sqrt\kappa$, $\gamma$ contains a conjugate point before reaching time $\beta$. This is important in the context of the application of the semi-classical formula and the occurrence of interference \eqref{equ:semi_field_interference}. It also suggests that interference between these extremal paths will be suppressed for many types of chaotic behavior. 

Moreover, in this geometric case, one can get many interesting relations between the Van-Vleck determinant \eqref{equ:path_Van_vleck}, the Raychaudhuri equation, the expansion scalar, Jacobi fields and the Riemann curvature itself. Reference \cite{Visser_Van_Vleck} gives a plethora of such relations, for congruences of different signatures.

\paragraph*{The toy model}
To recapitulate how my assumptions here fit the axioms of section \ref{sec:axioms}: I will take $M=S^3$, the 3-sphere,  $\mathcal{Q}$ to be given by Riem again, i.e. Riem$(M)=C^\infty_+(T^*M\otimes_S T^*M)$ and the gauge-group to be the group of 3-dimensional diffeomorphisms, Diff$(M)=\mathcal{G}$ acting through pull-back on the metric. The non-degenerate configurations with the highest isotropy subgroup are on the conformal class of $g^*:=d\Omega^3$. Although in the case of diffeomorphisms the degenerate boundaries of Riem (inside its affine vector space $C^\infty(T^*M\otimes_S T^*M)$) are accessible by the dynamics, I will for now ignore this, and later transition to the full Diff$(M)\ltimes \mathcal{C}$ group. I will thus, more or less arbitrarily for now, select $\Phi_o=\{d\Omega^3\}$ and ignore boundaries of Riem. The volume form will be dictated by $F:\mathbb{C}\rightarrow R_+$, with $F(z)=z\bar z=|z|^2$.  

For the action, given the space of 3-metrics, Riem (given in section \ref{sec:axioms}), I will choose the simplest possible dynamical system, where the action is given by the length functional, for $\gamma\rightarrow g_{ab}(t)$:
\be\label{geod_riem}
S[{g}]=\int dt \langle \dot{g}, \dot{g}\rangle^{1/2}_{{g}(t)}=\int dt\left(\int d^ 3x\, \dot {g}_{ab} \, g^{ac}\, g^{bd}\, \dot {g}_{cd}\sqrt{g}\right)^{1/2}
\ee 
This action is globally (but not locally) reparametrization invariant. Here the path integral could be defined using the arc-length gauge-fixing of the parametrization, discussed in appendix \ref{app:arc_length}.  Alternatively, I could use the Riemann-Stieltjes integration over parametrizations  (see   section {\bf IV, A} in  \cite{Chiou}). Note that we can do this here mainly because there is a single reparametrization constraint; there is no mixing of local gauge-symmetries with evolution, and the existence of a supermetric allows us to take limits of infintesimal segments to zero, i.e. there is meaning in taking the size of the \lq\lq{}mesh\rq\rq{} to zero, and no need to define the gauge-fixing before defining the path integral.\footnote{In the presence of refoliation invariance, one needs to define a physical slicing condition -- so that small steps of the skeletonization make sense -- even before defining the path integral. The challenge then is to determine how a change of gauge would affect this definition (see appendix \ref{app:issues}).}

  However, the action \eqref{geod_riem}  is still not invariant under configuration-space-dependent diffeomorphisms, i.e. under ${g}_{ab}(t)\rightarrow f(t)^*{g}_{ab}(t)$ it acquires non-covariant terms depending on $\dot f(t)$. To covariantize, I would have to add a prescription for lifting curves from superspace, $\mathcal{Q}/$Diff to $\mathcal{Q}$. This can be done by exploring the principal fiber bundle structure of  $\mathcal{Q}$, and introducing a connection form, as explained in section \ref{sec:symmetry}. In this case, one can take advantage of the metric on $\mathcal{Q}$ to introduce a connection which is roughly a projection orthogonal to the orbits of Diff(M). In other words, configuration space vectors tangent to the orbits are of the form $\nabla_{(a}\xi_{b)}$ for $\xi^b$ a vector field, and thus orthogonal vectors to the fibers  wrt to the metric \eqref{geod_riem} are  $v_{ab}\in T_g\mathcal{Q}$ such that $\nabla^a v_{ab}=0$. The connection itself is a type of Green\rq{}s function; the inverse of a second order differential operator, $\varpi^a(\dot{g})=(\delta^a_c\nabla^2-R^a_c)^{-1}(\nabla_b\dot {g}^{cb})$.\footnote{See also \cite{Aldo_HG} for a relation between a choice of connection, locality, and the representation of abstract observers. }  This implies that $\varpi^a(\mathcal{L}_\xi g)=\xi^a$ and $\varpi^a(v)=0$ iff $\nabla^a v_{ab}=0$. Since the connection-form is being defined by an orthogonality conditions for a $\mathcal{G}$-invariant metric in configuration space, it automatically satisfies both requirements,  \eqref{eq_fundamental1}-\eqref{eq_fundamental2}.

The action then becomes:
\be\label{geod_riem_connection}
S[{g}]=\int dt\left(\int d^ 3x\, (\dot {g}_{ab}-\mathcal{L}_{\varpi(\dot g)}g_{ab}) \, g^{ac}\, g^{bd}\, (\dot {g}_{cd}-\mathcal{L}_{\varpi(\dot g)}g_{cd})\sqrt{g}\right)^{1/2}
\ee 
Under a diffeomorphism, it is now easy to show that the integrand remains invariant.\footnote{Note that the terms which take into account the metric variation of $\varpi$ itself vanish by the equations defining $\varpi$.  The same sort of lift would ensue if the symmetry group included Weyl transformations, and the absence of Weyl anomalies in 3d extends, through the use of the horizontal lift, to the absence of Weyl anomaly in the lifted path integral. Note moreover that, unlike for refoliations,  in this case it is easy to show that the Fadeev-Popov determinant behaves like a true determinant, and so, taken together with the measure, is invariant under the given transformations. }

As I am mostly interested in a semi-classical approximation,  I will take a short-cut, sufficient for my purposes in this analysis. Since the orbits of Diff$(M)$ are Killing directions of the metric given in \eqref{geod_riem}, the inner product between the tangent to a geodesic and the orbit remains constant.\footnote{$\frac{D}{dt} (g^{ab}\dot \gamma_a u_b)= \dot\gamma^a\dot\gamma^b\nabla_au_b=0$ for $u^a$ Killing. This elementary fact about Pseudo-Riemannian geometry remains true also in the infinite-dimensional case \cite{Michorbook}. } Thus a geodesic with initial transverse velocity will remain transverse. I thus assume here that all initial velocities are transverse, and thus we can use for all practical purposes just \eqref{geod_riem}, not \eqref{geod_riem_connection}, since $\varpi(v^h)=0$ (see below for the appropriate Jacobian that would appear in the path integral).

The geodesic equations  are: 
 $$ \ddot {g}_{ab}=\dot {g}_{ac}\dot {g}^c_{~b}+\frac{1}{4}\dot {g}_{cd}\dot {g}^{cd} {g}_{ab}-\frac{1}{2}\dot {g}^c_c \dot {g}_{ab}
 $$
  and one can, in fact, find explicit analytic solutions to the equations of motion \cite{Gil_Medrano}. Given $g^0_{ab}\in \mathcal{Q}$ and $h_{ab}\in T_{g^0}\mathcal{Q}$, let $h^T_{ab}=h_{ab}-\frac{1}{3}h g^0_{ab}$, with $h=h_{ab}g_0^{ab}$ and $\nabla_0^a h_{ab}=0$. Then the geodesic starting out at $g^0_{ab}$ in the direction of $h_{ab}$ is given by:
\be\label{equ:geodesic_gravity}{g}_{ab}(t)=g^0_{ac} (e^{(A(t)\mbox{\tiny Id}+B(t)\mathbf{h}_T) })^c_b\ee
where $\mathbf{h}$ is the matrix given by $h^a_b=h_{bc}g_0^{ac}$,
$$A(t)= \frac{2}{3}\ln ((1+\frac{t}{4} h)^2+\frac{3}{8} h^T_{cd}h_T^{cd}t^2)~~\mbox{and}~~~B(t)= \frac{4}{\sqrt{3 h^T_{cd}h_T^{cd}}}\mbox{arctan}\left(\frac{\sqrt{3 h^T_{cd}h_T^{cd}}t}{4+th}\right)$$
Note that the geodesics are ultralocal. Since the action has no spatial derivatives, the solution ${g}_{ab}(x,t)$ depends only on ${g}_{ab}(x,0)$ and $\dot {g}_{ab}(x,0)$. In arc-length units, the velocity $h_{ab}$ has units of $1/$sec. 
 Let me briefly note that the Jacobi fields can also be explicitly solved for, but the resulting expression is too involved to be written here (see Theorem 4.7 in \cite{Gil_Medrano}). 

For standard homogeneous cosmological models, one would have $h^T_{ab}=0$, greatly simplifying the solution, as $B=0$ and $A=\frac23\ln(1+\frac{t}{4}h)^2$  where $h$ is then the instantaneous expansion.  In this case, however, the volume of the manifold, which is dictated by $A(t)$ above, will reach a singular value in finite time for $h<0$.  If $h^T_{ab}\neq 0$, however, geodesics will run forever.  %Note that since the action is not conformally invariant, directions which are initially traceless will not stay traceless, unlike for transverse directions, for which they do stay transverse.

Endowed with such a metric, $\mathcal{Q}$=Riem$(M)$ has an associated Riemann super-curvature (of the Levi-Civita connection).  Along pure trace directions (as would happen in a purely homogeneous expansion), it is  zero. Otherwise, in index-free notation it is given by: 
\be\label{curvature_infinite}
R(\mathbf{h}_1,\mathbf{h}_2)\mathbf{h}_3=-\frac14[[\mathbf{h}_1,\mathbf{h}_2],\mathbf{h}_3]+\frac{3}{16}\left(\mbox{tr}_g(\mathbf{h}_1\mathbf{h}_3)\mathbf{h}_2-\mbox{tr}_g(\mathbf{h}_2\mathbf{h}_3)\mathbf{h}_1\right)
\ee
where $\mathbf{h}$ are matrix vector fields, extending the sort of matrix vector given above for $T_{g^0}\mathcal{Q}$ to $T\mathcal{Q}$.  I have assumed that $\mathbf{h}=\mathbf{h}^T$, and the standard matrix commutator is used. The sectional super-curvature is then calculated to be:\footnote{All the formulas in this section are generalizable to $n$ dimensions. For instance, the $3/16$ above in fact comes from $\frac{n}{32}(4+n(n+1))$ \cite{Gil_Medrano}. Moreover, one should note that there are two sorts of connections: one related to the internal gauge space, and the other the Levi-Civita. This is precisely the analogous case of Kaluza-Klein, and the two satisfy relations \cite{Gil-Medrano}.} 
\be\label{Sectional}K(h_1,h_2):=\langle R(h_1,h_2)h_1, h_2\rangle_g=-\int d^3x\, \sqrt{g}\, \left(\mbox{tr}_g([\mathbf{h}_1,\mathbf{h}_2]^2)+\frac{3}{16}\left((\mbox{tr}_g(\mathbf{h}_1\mathbf{h}_2))^2-\mbox{tr}_g(\mathbf{h}_1^2)\mbox{tr}_g(\mathbf{h}_2^2)\right)\right)
\ee
For $\mathbf{h}_1, \mathbf{h}_2$  orthogonal, i.e. 
$$\int d^3x\, \sqrt{g}\,\mbox{tr}(\mathbf{h}_1\mathbf{h}_2)=\int d^3x\, \sqrt{g}\, h^1_{ab}\,g^{ac}g^{bd}h^2_{cd}=0$$ we immediately get from \eqref{Sectional} that $K(h_1,h_2)\leq 0$. By the Hadamard-Cartan theorem, this means that there will be no conjugate points for these dynamical paths. It also means that interference of alternative paths will quickly subside, and semi-classical quantum effects will become predominantly due to other fields, present on a fixed metric background (see section \ref{sec:conservation}). 
 
For quite general cases, the geodesic equation, the exponential maps, Jacobi fields, curvature, etc, have been computed \cite{Michor_general}.
Quantities of interest become geometrical; for example, the Van-Vleck matrix:
\be\label{equ:VV_gravity}\det\frac{\delta^2 S_{{g}_{\mbox{\tiny{cl}}}}(g^i, g^f)}{\delta g^i_{ab}(x)\delta g^f_{cd}(y)}=\det{ (J(g^i,g^f)_{abcd}(x,y))}^{-1}=\Delta(g_i,g_f)\ee
Where $J$ is the  Jacobi matrix associated to the geodesic variations.\footnote{Jacobi fields -  geodesic variations of ${g}_{\mbox{\tiny{cl}}}$ - are uniquely defined by its value and first derivative at $\phi_i$. Alternatively, it can be characterized by it values  at $\phi_i$ and $\phi_f$, through the two-point tensor matrix $J(\phi_i,\phi_f)$. }   The determinant here requires regularization, but in principle we can calculate everything  needed for a semi-classical path-integral within the classical  geometry of $\mathcal{Q}$, by using equation \eqref{equ:semi_classical_exp} and \eqref{equ:path_Van_vleck} together with \eqref{geod_riem} and the expression for the derivatives of the Riemann exponential map. Alternatively, we could use Barvinsky\rq{}s reduction methods for funcional determinants \cite{Barvinsky2}, which reduce the 1-loop functional determinant to a spatial one, integrated over a given time parametrization. Here I will choose the first, simpler route. 

    There are two possible procedures to obtain closed expressions for the Van-Vleck determinant: i) input equation \eqref{equ:geodesic_gravity} into the action \eqref{geod_riem}, with the end-point $g_{ab}(1)$ as a function of $h_{ab}$  (for fixed $g_0$). Or, ii) use the given closed expression for the Jacobi fields in Theorem 4.7 in \cite{Gil_Medrano} to directly write the Jacobi matrix. 

Both manners are calculable, but the resulting expression is not compact, or enlightening at face value. We pursue the first option in appendix \ref{app:Van_Vleck}. Alternatively, we can use  geometric approximations for the Van-Vleck which work, respectively, for weak field and small geodesic distance \cite{Visser_Van_Vleck} (for finite-dimensions, with $\gamma$ the geodesics):
\begin{align}
\label{equ:curv_approx}\Delta^{\mbox{\tiny weak}}_{{g}}(x,y)&=\exp{\left( \frac{1}{t}\int_0^t(t-s)(R_{ab}\dot{\gamma}^a(s)\dot{\gamma}^b(s)\, s ds +\order{[\mbox{Riemann}]}^2\right)}\\
\label{equ:distance_approx}\Delta^{\mbox{\tiny small d}}_{{g}}(x,y) &= 1+\frac{1}{6}(R_{ab}\dot{\gamma}^a\dot{\gamma}^b S[{g}]^2+\order{S[{g}]}^3)
\end{align}
The Ricci super-curvature does not properly exist, since the mapping $k\mapsto R(h,k)j$ is just the push forward of the section by a certain tensor field, a differential operator of order 0. If this is not zero, it induces a topological linear isomorphism between certain infinite dimensional subspaces of $T_g\mathcal{Q}$, and is therefore never of trace class. In this point of this approximation  we need to smuggle in some sort of regularization.\footnote{For an action with spatial derivatives, which would give Laplacians, and an expansion on eigenvalues of the tensor Laplacian would provide a cut-off for the trace inherent in the determinant. This will also be important to get some inhomogeneity in our probability for gravitational modes.}

Here, following \cite{Gil_Medrano}, I take a  shortcut  by considering the pointwise trace of the local action of the Riemann tensor:
\be\label{Ricci}
\mbox{Ric}(h_1, h_2)=-\frac32\int d^ 3x\, \sqrt{g}\, h^1_{Tab}g^{ac}g^{bd} h^2_{cd}
\ee
which is almost proportional to the supermetric, except that one of the components gains a traceless projection. Due to this traceless projection, we don\rq{}t have to worry about the extremal paths under consideration reaching the boundary of Riem (where we would have to set up boundary conditions). The problem with the super-curvature being  proportional to the supermetric (and that inherent in \eqref{curvature_infinite}) is that it means in a sense we are in a constant curvature regime, and cannot really apply \eqref{equ:curv_approx}. We are better off applying \eqref{equ:distance_approx}. For points close to $g^0$ in arc-length, i.e. for $S[{\gamma}]\sim \epsilon$, and initial velocity $\dot{g}_{ab}=h_{ab}$, we get:
\be\label{Ricci_Delta} \Delta_{g}(g^0, g(\epsilon))=1-\frac{1}{4}\left(\int d^ 3x\, \sqrt{g^0}\, h_{Tab}g_0^{ac}g_0^{bd} h_{cd}\right)\epsilon^2+\order{(\epsilon)}^3
\ee
Clearly the pure trace directions (cosmologically homogeneous) have no contribution from $\epsilon^2$ terms, and thus, at least close to $g^0$, are of greater amplitude. Note however, that here we must strike a fine balance. That is because the semi-classical approximation is valid for $S[{g}]>>\hbar$, and thus we must have an expansion for small arc-length $\epsilon$, but  $\epsilon>>\hbar$, and there is no other dimensionful constant to compare to. 

We must thus consider this Van Vleck as giving a one-loop functional determinant measure (combined with the projected Liouville measure), on the tangent space $T_{g^0}\mathcal{Q}$. I.e. on the space of initial cosmological directions at $g^0$.\footnote{ We should remind the reader that a more standard computation exists in terms of the integral of the one-loop determinant in time, which gives the Van-Vleck \cite{Barvinsky2}. Because here we are exploring the geometrical analogy, we will not pursue that strategy.}
% we should use the simple choice of measure\footnote{Note that this differs from the standard Fradkin-Vilkovisky measure \cite{Fradkin_Vilkovisky}. This was to be expected; we are not in a 4-dimensional, covariant setting, and our measure, according to \eqref{equ:measure_timeful} consists of the dynamical term and the plain term, given above (see \cite{Barvinsky}).} $$\mathcal{D}g_{ab}= \prod_{x} g_{ab}(x)$$

According to \eqref{equ:measure_timeful}, for the short distance approximation, we are in fact integrating over directions, $h_{ab}$, and such that $\nabla_{g^0}^ah_{ab}=0$. But because of the latter, \lq{}gauge-fixing\rq{} condition, we gain a Jacobian, $J_{TT}(g^0)$. This is calculated in appendix \ref{app:Jacobian}, equation \eqref{TT_Jacobian}. The determinant that emerges however, depends only on $g^0$, and thus is independent of the integration variable. 

Finally, since \eqref{Ricci_Delta} is obtained from an exponential, we obtain, for the relation between two total surface volumes (or screen-flux) for the screen regions parametrized by $H_1,H_2\in T_{g^0}\mathcal{Q}$, at a geodesic distance of $\epsilon$, and with the same nominal volume (i.e. same volume under the homogeneous measure $\int_{H_1}\mathcal{D}h=\int_{H_2}\mathcal{D}h=V_H)$: 
\begin{align}
\frac{\int_{H_1} \mathcal{D}h^1 J_{TT}(g^0)  P(g(\epsilon))}{\int_{H_2} \mathcal{D}h^2 J_{TT}(g^0) P(g(\epsilon))}&=\frac{\int_{H_1}\mathcal{D}h^1\left(1-\frac{1}{4}\left(\int d^ 3x\, \sqrt{g^0}\, h_{Tab}g_0^{ac}g_0^{bd} h_{cd}\right)\epsilon^2+\order{(\epsilon)}^3\right)}{\int_{H_2}\mathcal{D}h^2\left(1-\frac{1}{4}\left(\int d^ 3x\, \sqrt{g^0}\, h_{Tab}g_0^{ac}g_0^{bd} h_{cd}\right)\epsilon^2+\order{(\epsilon)}^3\right)}
\nonumber\\ 
&= 1+\frac{1}{4}\frac{\int_{H_2} \mathcal{D}h^2\left(\int d^ 3x\, \sqrt{g^0}\, h_{Tab}g_0^{ac}g_0^{bd} h_{cd}\right)-\int_{H_1} \mathcal{D}h^1\left(\int d^ 3x\, \sqrt{g^0}\, h_{Tab}g_0^{ac}g_0^{bd} h_{cd}\right)}{V_H}\epsilon^2+\order{\epsilon}^4\label{Relative_amplitude}
\end{align}

This is the relative probability for these two screens defined by the same arc-length distance, in an expansion  for arc-length distance from $g^0$  larger than Planck. 
Explicit numbers can be computed by using an eigenbasis of transverse $(0,2)$ transverse traceless tensors on the round 3-sphere, $g^0$. A convenient such basis is the tensor harmonic basis on $S^3$, fully described in \cite{Gerlach}. Since $S^3$ is compact without boundaries, and the standard Laplacian is a self-adjoint operator with respect to the standard inner product (see \eqref{equ:supermetric}), by the spectral theorem the eigenfunctions  with different eigenvalues are orthogonal. In terms of these specific eigenvalues, the measure becomes $\prod_x dh_{ij}(x)\rightarrow \prod_{n=1}^\infty d\lambda_{ij}^n$, if the basis is orthonormal (see \eqref{basis_S3}). If it is not, another Jacobian appears for the transformation.\footnote{These are the coefficients for the eigenfunctions described in 17-18c of \cite{Gerlach}.}  The value of \eqref{Relative_amplitude} then becomes a difference for the integrals for different ranges of the prefactors of such a harmonic basis, which can be now calculated explicitly. But this is just a homogeneous measure,\footnote{Had we chosen a basis which is not orthonormal, the Jacobian would have cancelled out with the emerging prefactors in any case.} and thus
$$\int_{H_1}\mathcal{D}h=\int_{H_2}\mathcal{D}h\Rightarrow \int_{H_2} \mathcal{D}h^2\left(\int d^ 3x\, \sqrt{g^0}\, h_{Tab}g_0^{ac}g_0^{bd} h_{cd}\right)=\int_{H_1} \mathcal{D}h^1\left(\int d^ 3x\, \sqrt{g^0}\, h_{Tab}g_0^{ac}g_0^{bd} h_{cd}\right)
$$
In other words, this measure does not care about the eigenvalues of the transverse-traceless modes. It is homogeneous, and thus would not  probabilistically favor \lq\lq{}flatter\rq\rq{} modes, as is expected of inflation. 

However, preliminary calculations show that with coupling to a metric, i.e. having instead of \eqref{geod_riem}, 
$$S[{g}]=\int dt \langle \dot{g}, \dot{g}\rangle^{1/2}_{{g}(t)}= \int dt \left(\int d^3y f(R) \sqrt{g}\right) \left(\int d^ 3x\, \dot {g}_{ab} \, g^{ac}\, g^{bd}\, \dot {g}_{cd}\sqrt{g}\right)^{1/2}$$ 
changes terms in \eqref{Relative_amplitude} in the following way:
$$h_{Tab}g_0^{ac}g_0^{bd} h_{cd}\rightarrow h_{Tab}g_0^{ac}g_0^{bd}\nabla^{2n} h_{cd}
$$
where $n$ is the degree of $f(R)$. In this case, the higher the eigenvalues of the region $H_1$ in comparison to $H_2$,  \textit{the smaller is its relative probability flux}. This would mean indeed that more homogeneous modes would be favored. Although it should be noted that the eigenvalues of the basis $\tau^{(n)}_{ij}$ in \eqref{basis_S3} are given by $\nabla^2\tau^{(n)}_{ij}=-(n^2-3)\tau^{(n)}_{ij}$, for $n\geq 3$. And thus, one wouldn\rq{}t have complete homogeneity represented in this space. 
%The last element in order to explicitly compute is the Jacobian for going from $\prod_{x} g_{ab}(x)$ to such a harmonic basis, and it is given in \cite{Fujikawa} (for vector fields, but are easily generalizable using \eqref{TT_Jacobian}, in the appendix). 

\paragraph*{Comments on the cases where $\epsilon$ is not small.}
For the full expression of \eqref{equ:VV_gravity}, we need replace the solution \eqref{equ:geodesic_gravity} into the action \eqref{geod_riem}, between $t=0$ and $t=1$. Then  the action as 
$$\frac{\delta^2 S}{\delta g_{ab}^0\delta g_{ab}(1)}=\frac{\delta}{\delta g_{ab}^0}\left(\frac{\delta S}{\delta h_{ab}}\frac{\delta  g_{ab}(1)}{\delta h_{ab}}\right)$$ 
This is a cumbersome procedure, requiring quite a bit of algebra. As the purpose of this exercise is a proof of principle for the method, we leave these calculations for the appendix \ref{app:Van_Vleck} (see equation \eqref{final_full}).

In the general case it also happens that, since there will be at most one extremal path between $g^0$ and $g$,  the semi-classical probability density is given simply by 
$$P(g)  \mathcal{D}g=F(\Delta^ {1/2}(g^0,g)e^{iS[g^0,g]/\hbar})\mathcal{D}g= \Delta (g^0, g) \mathcal{D}g$$ where $g$ lies in a classical path from $g^0$ with a horizontal initial condition, i.e. $\nabla_{(g^0)}^a\dot {g}_{ab}=0$ and $\Delta$ is defined by the determinant of the geometrical Jacobi matrix, in \eqref{equ:VV_gravity}. It is clear from \eqref{equ:geodesic_gravity} that we could not have chosen the completely degenerate metric as $g^*$.

%The measure $\mathcal{D}g$ in this case comes directly from the Liouville measure, by integrating out the quadratic momentum terms in the action (in Hamiltonian variables), e.g. by defining $\int \mathcal{D}\pi_{ab} [\exp{\int d^ 3x\,  \pi_{ab}\pi^{ab}}]=1$.

 %In this case, since the action is just given by arc-length in configuration space, and again,  there aren\rq{}t multiple extremal paths between $g^0$ and $g$, the maximum radius of the extremal coarse grainings given in definition \ref{def:ECS}  is $\rho=\hbar$.  

  There are two qualitative features for the gravitational quantum mechanics model that can be seen at this primitive stage: firstly,  there is no gravitational interference, as the sectional super-curvature is strictly negative \eqref{Sectional}.\footnote{Although, see \cite{DeWittQG1} for arguments about the naturalness of certain reflecting boundary conditions for geodesics in superspace. Here too, one can impose certain artificial boundary conditions and obtain interference patterns directly from \eqref{equ:semi_field_interference}.} Since generically interacting Hamiltonians in more than $2$ dimensions are chaotic, it seems even in the more general geometric case we can  use the Rauch comparison theorem in the other direction: generically, semi-classical interference effects will become suppressed after a short evolution time (in arc-length). Different cosmological histories will quickly decohere, with no chance of recohering. Thus, in exactly the same fashion as the Mott bubble chamber (a quantum mechanical decay process), metrics along the same geodesic (separated by more than $\hbar$ in arc-length) can serve as records for ones coming later along the geodesic. At this level in $\hbar$, with weak coupling to matter fields, the gravitational degrees of freedom  would quickly become a background for the quantum effects of these other matter fields. 
    Thus in principle we can fully determine the gravitational semi-classical theory, complete with records, interference (or lack thereof) and so on.  
  
 Secondly,  geodesics can be extended indefinitely, but it is not true that any two points can be connected by a minimizing geodesic (thus the  Hopf-Rinow theorem for finite-dimensional Riemannian geometry fails in this infinite-dimensional context). This already means that certain configurations have negligible volume from the start, without even taking into account the Van-Vleck determinant or the relevant records. It also counters some finite-dimensional intuition, that would lead one to believe that any configuration could be reached by a judicious choice of initial condition. There are, in the gravitational configuration space, deserts of very little volume. 
 
  On top of this structure, we could now proceed along the lines of section \ref{sec:conservation}. This would allow us to discuss the quantum evolution of source fields on top of each one of these classical gravitational solutions, where each classical gravitational solutions is weighed by its own probability according to \eqref{Relative_amplitude}. 
    
  %A reformulation of Maldacena\rq{}s cosmological Bell tests \cite{Maldacena_Bell} in this model would be interesting.   
       
 \paragraph*{Summary of this subsection.}
 This section was a proof of principle. Here I have described a simple gravitational toy model. With it I was able to implement the structures presented in this paper (e.g.: the field-space connection 1-form, the semi-classical approximation, etc) for building the semi-classical volume form.  Extending an equivalence between the Van-Vleck determinant and geometric objects (depending on Ricci curvature), I proposed an equation for the semi-classical amplitude in the vicinity of the round 3-sphere. With it,  I found a simple result for relative volumes of a geodesic screen close (in configuration space) to the round sphere. These represent relative transitions from a total amplitude which, according to the previous section, would be preserved in the geodesic arc-length time.\footnote{At least while these \lq\lq{}geodesic spheres\rq\rq{} in field space are integrable. In the present case, by the Gauss lemma,  their integrability could have a very large domain, since there are no focusing points.}  In this simple case, these amplitudes are just proportional to the respective \lq\lq{}functional area\rq\rq{} encompassed by two sets of TT-modes (of the round $S^3$), on the tangent space $T_{g^0}\mathcal{Q}$. Nonetheless, I indicated how, including some coupling between points (through the inclusion of spatial derivatives in the action functional), the relative transition amplitude would naturally favor more homogeneous directions in field space (departing from the round sphere). It is important to note that this was all done without a minisuperspace assumption, and the results are not strictly classical, as they require the computation of the semi-classical effects through the Van-Vleck determinant. 
     
\subsection{Regaining space-time}
    As I mentioned in the introduction,  it should be noted that even after having a good semi-classical quantum gravitational model of the sort presented here, one must still reconstruct an effective space-time  from a curve of geometries. In other words, although along a classical trajectory we can order records and obtain a global (in space) notion of time, this might not be relevant for what observers within these trajectories in Riem perceive as duration. This touches on the common question in quantum mechanics, surrounding what constitutes a relational clock. This section does not provide a definitive argument, but a sketch on different directions that could be pursued in this respect. 
    
    It should be said that all theories that refer fundamentally only to the 3-metric, --- all standard interpretations of canonical quantum gravity, such as Hartle-Hawking --- would require a  reconstruction of space-time at a hypothetical classical limit. As with the necessity of records, these issues are less pressing in the minisuperspace context, and usually glossed over. 

For shape dynamics \cite{SD_first}, the dynamics of matter fields can reproduce those of GR (at least for some finite duration). As has been shown in different ways \cite{Steve_SD, Tim_effective}, one can then reconstruct an effective slab of space-time which is indistinguishable from GR, and recover aspects of the equivalence principle \cite{Steve_SD}. But these solutions are more or less fine-tuned to match general relativity, and the way in which they acquire those features is not completely understood. 

In the more general case pursued here --- which gives a method and a paradigm, not a specific theory --- we would like to understand the nuts and bolts about how any space-time can be recovered from a curve in (conformal) superspace, and what are its properties. This, we don\rq{}t yet know how to do, but here is one attempt. I will first discuss the emergence of some limited, finite refoliation symmetry, based on scalar fields. 

\paragraph*{Refoliations}
It is not really mysterious how regaining a space-time structure from matter fields can break refoliation invariance. Indeed, the useful matter fields for this role should be expected to break refoliation invariance; after all, they can define a rest-frame, or a surface in space-time where their gradient is purely timelike.  For example, the CMB can be taken as one such reference. 

Moreover, the observability of a \lq\lq{}preferred foliation\rq\rq{}  is not obvious from the form of the action. For instance, the BSW action \cite{BSW}:
\be\label{BSW}
S_{\mbox{\tiny BSW}}=\int dt \int d^3x \sqrt{g}\sqrt{R(\dot g^H_{ab}G^{abcd}\dot g^H_{cd})}
\ee
where $\dot g^H_{cd}:=\dot g_{cd}-\mathcal{L}_\xi g_{cd}$ is the \lq\lq{}dressed\rq\rq{}  horizontal metric velocity, for $\xi^a$ a specific functional of the metric and its time metric. Equation \eqref{BSW}   is not explicitly space-time covariant --- it does not contain the lapse or the shift, components of the space-time metric in the covariant form --- and yet it is exactly the Einstein-Hilbert action, written in 3+1 form and with eliminated lapse and shift. The important aspect of GR that \textit{can be} seen in \eqref{BSW} is the fact that it contains the same amount of spatial and time derivatives.

Given these two considerations, I  thus start with an action of the form: 
\be\label{action_lapses1}S=\int dt\int d^ 3 x\sqrt{g}\left( \dot g_{ab}G^{abcd} \dot g_{cd} -R +\sum_i a_i(\dot \psi_i)^2\right)\ee
(let's ignore the shift, because that is easy to accommodate), and supposing that all scalar fields are such that $\frac{d\psi_{(i)}}{dt}>0$ in some domain, having no spatial gradients in the action. The point is that I can replace $dt$ by $d\psi_{(i)}$, and there I will have a freedom of choosing " the lapse"  in $n$ different ways, one for each field determining time. 

That is, I can rewrite the action as:
\be\label{action_lapses2}
S=\int d\psi_i\int d^ 3 x\sqrt{g}\frac{1}{N_{(i)}}\left(N_{(i)}^ 2( \frac{d g_{ab}}{d\psi_{(i)}}G^{abcd} \frac{d g_{cd}}{d\psi_{(i)}}+a_i+\sum_{j\neq i} a_j(\frac{d\psi_{(j)}}{d\psi_{(i)}})^2 )-R \right)\ee
 This is  basically the standard gravitational Einstein action with $d/dt$ replaced by the derivative along the scalar field, and with the "lapse" 
$N_{(i)}=\frac{d\psi_{(i)}}{dt}$. 
Now, one can shift from one i to the other iteratively, i.e. forgetting about their original relation to $t$,  obtaining the same results. 
I.e. by using $d\psi_k=\frac{d\psi_k}{d\psi_i}d\psi_i$ in the equations above, one gets precisely the same action, with $i\rightarrow k$, i.e. just described under evolution for a different scalar field. It is a \lq\lq{}discrete\rq\rq{}  sort of refoliation. 

  Moreover, the perturbation equations for the metric are hyperbolic, and would construct a universal gravitational light cone. I.e. 
in this extremely simple case, the equations of motion of the gravitational fields will be hyperbolic (for the unit lapse case): 
$$\ddot g_{ab}=\dot g_{ij}\dot g_{kl}\Gamma^{ijkl}_{ab}+2(R_{ab}-\frac{1}{4}g_{ab}R)
$$
where $\Gamma^{ijkl}_{ab}$ are the Christoffel symbols for the standard DeWitt supermetric. And thus, the propagation of gravitational perturbations around stationary solutions, such as $g_{ab}=\delta_{ab}$ (the flat Euclidean metric), will form null-cones. The requirement that these characteristics are indeed the null-cones of a metric, will determine the conformal class of the space-time metric (and thus a relation between the lapse and the spatial metric). 

What is most unnatural about equation \eqref{action_lapses1}, however, is the fact that there are no spatial gradients of the scalar fields. This was motivated by  the way one would use such a scalar field in space-time to define a foliation. Nonetheless, we can remove this condition by adding gradient terms --- e.g. the standard one $\nabla_a\psi\nabla^a\psi$ --- and using shift vectors adapted to each scalar field, in the following way. 
Correcting the time velocities of the scalar fields: $\dot\psi\rightarrow \dot\psi_H=\dot\psi- \chi^a\nabla_a \psi$, we obtain the following correction to the kinetic terms:
\be \dot\psi^2-g^{ab}\nabla_a\psi\nabla_b\psi\rightarrow \dot\psi^2-2\dot\psi \chi^a\nabla_a\psi+(\chi^a\nabla_a\psi)^2-g^{ab}\nabla_a\psi\nabla_b\psi
\ee
Now, choosing $\chi^a=\chi g^{ab}\nabla_b\psi$, we get a quadratic equation on $\chi$ so that 
$$\dot\psi_H^2-g^{ab}\nabla_a\psi\nabla_b\psi\rightarrow \dot\psi^2$$
namely, 
$$\chi^2(\nabla_a\psi\nabla^a\psi)-2\chi \dot\psi - 1=0$$
which can be solved for the fields, 
$$\chi= \frac{\dot\psi\pm \sqrt{\dot\psi^2-\nabla_a\psi\nabla^a\psi}}{\nabla_a\psi\nabla^a\psi}$$
It thus seems that one can choose the shift vectors to be physically given in such a way as to always \lq\lq{}select a surface of spatially constant $\psi_{(i)}$\rq\rq{}. In this way, one could in principle retain the spatial gradients of one $\psi_{(i)}$ in a foliation, but get rid of another. 

But there are many problems with this resolution however: when coupling to the metric, or with the other fields, it would imply a non-miminal coupling between $\psi$ and $g$, which would also reinstate the dependence of $\nabla_a\psi$. This perhaps could be resolved by adding the terms from the beginning and solving a possibly more complicated equation, depending on all the fields. Another problem is that the solution for $\chi_a$ doesn\rq{}t seem to define a bona-fide connection-form. I.e. it doesn\rq{}t have the right covariant transformation properties as in \eqref{eq_fundamental2} or \eqref{Lagrange_transf}. Nonetheless, this is first attempt, a proof of principle that certain things can work, while others not at this point. 

\paragraph*{Null-cone and equivalence principle.}
As a simple example, given extremal paths of \eqref{geod_riem}, we can attempt to reconstruct a space-time just with the use of scalar fields. The easiest manner to do this is to introduce a weakly back-reacting scalar field, with action in arc-length parametrization given by
$$ S_{\mbox{\tiny scalar}}=\epsilon\int dt\int d^3x\,\sqrt{\gamma} (\dot\psi^2- c^2 \gamma^{ab}\nabla_a\psi\nabla_b\psi+m^2\psi^2)
$$
for a very small $\epsilon$. An objection here is that this action is largely inspired by a relativistic one -- in that it contains the same number of spatial and temporal derivatives. Ideally, one would have more than one field, e.g.: $\psi_1, \psi_2$, and use the characteristics of one to define the propagation of the other.  We leave this for further work. Using the characteristics of the field equations (on this background $\gamma_{ab}(t)$), we could have a metric given by:
\be\label{constructed_metric} ds^2=-c^2 dt^2+ \gamma_{ab}(t)dx^a dx^b\ee where $\gamma_{ab}(t)$ is given in \eqref{equ:geodesic_gravity}. The first question one can ask is when \eqref{constructed_metric} also obeys the Einstein equations for example. 

Now, to discuss something such as the equivalence principle, one could proceed as in \cite{Steve_SD}: couple a point-like particle with trajectory $z^a(t)$ -- with the usual covariant action, parametrized by the superspace arc-length, and with an induced lapse given (in this simple case) by \eqref{constructed_metric} -- i.e. 
$$S_{\mbox{\tiny point}}= \int dt\int d^3 x\,  \delta(x^a-z^a(t))(c^2-\dot z^a \dot z_a)^{1/2}$$
The equations of motion are  the usual geodesic ones for the metric \eqref{constructed_metric}.  According to \cite{Steve_SD}, to first order, one could use time-dependent spatial diffeomorphisms to correct the frame of reference centered on such a particle to be described by geodesics in Minkowski space, apart from transformations of the lapse.\footnote{Note here that the propagation of spatial coordinates are also \lq\lq{}physical\rq\rq{}, as they are given by a choice of $\varpi$. But different choices will be related by spatial diffeomorphisms, as is the case with different charts for the manifold. } In the present simplified case, the physical lapse is already equal to one, so we regain one of the statements named \lq{}the equivalence principle\rq{} -- namely, there is a gauge symmetry that will bring an arbitrary classical trajectory to a geodesic of Minkowski space, at least infinitesimally. This issue, however, needs to be further investigated in more general cases with more general matter fields; it is not clear to me that it must always hold.

\paragraph*{Summary of this subsection.}
In this subsection I have sketched different ways one would go about reconstructing a space-time metric and regaining the equivalence principle (and other notions of refoliation invariance). This can be accomplished from the relative evolution of observables along an extremal path in configuration space. There, there can still be more than one way of defining duration, and perhaps these can replace different notions of refoliations (which is an inherently spacetime operation). Indeed, what do we physically do when we want to refoliate? We choose a different set of clocks spread out through space. 
Moreover, we still want a hyperbolic structure. Since we do observe that light has the same speed for all observers, and that can only happen if the metric has a null cone. In specific cases this translation seems to go through without much problems, but no generic statement was proven. 

\section{Conclusions}\label{sec:conclusions}

\subsection{Problems between records and  refoliation invariance.}\label{sec:WdW}

\paragraph*{Differences from Hartle-Hawking and the tunneling proposals}
As discussed in the main text,  the formalism presented here has certain features in common with the Hartle-Hawking no-boundary \cite{Hartle_Hawking}, and with Vilenkin\rq{}s tunneling \cite{Vilenkin} proposals. Vilenkin\rq{}s model in mini-superspace coincides with our model, for the reduced ADM action, as I comment below. 

There are, however, some crucial differences. The boundary conditions are based on completely different principles. In Hartle-Hawking,  the path integral integrates over all  4-metrics interpolating between a given regular Euclidean 4-metric without boundary  and a given 3-geometry with matter content. In the tunneling proposal, the boundary conditions should embody the notion that the Universe \lq\lq{}tunnels\rq\rq{} from nothing. In both cases, the resulting wave-function (in terms of its one boundary 3-geometry and  matter content) should satisfy the Wheeler-DeWitt equation.  

 Like the present work, these theories attempt to define a wave-function of the Universe from a path integral in some field configuration space. But there are a couple of important distinctions between such proposals and the one developed here: {\bf i)} Both Hartle-Hawking and tunneling proposals are based on path integrals in superspace. But supespace is not the reduced configuration space of ADM, because one still has refoliations. In our case, there must be a well-defined reduced configuration space for the volume form to have physical meaning.  Records and screens required crossing extremal curves in superspace to have objective meaning (there can\rq{}t be one if refoliations are part of the local symmetries).   {\bf ii)} In the present approach, the boundary conditions are selected uniquely, and are such that the arrow of time should point in the direction of greater inhomogeneity.

The absence  of further specification of boundary conditions is also unlike either  Hartle-Hawking or the tunneling wave-functions. In Hartle-Hawking, one usually  stipulates boundary conditions such that $\Psi(g_{ab})=0$ for the degenerate determinant $g=0$ \cite{Hartle_Hawking}, integrating only over metrics for which $g>0$ and which can be \lq{}capped-off\rq{} by part of a 4-sphere. In Vilenkin\rq{}s proposal (see also, Linde \cite{Linde}), the definition of the wave-function is given by a transition function from the completely degenerate 3-metric, $g_{ab}=0$. One also requires, however, further boundary conditions on superspace. For this, Vilenkin attempts to separate curvature singularities which would correspond to singularities of a 4-geometry -- the singular boundary of superspace -- and those which would correspond to singularities due to bad choices of slicings -- the non-singular boundary of superspace. This is a difficult distinction to make in practice, at least away from minisuperspace. Assuming that it could be done, then the proposal requires the wave-function to have only out-going modes at the singular boundary. There is also a problem here, as in general there are no time-like Killing vector fields in superspace to define positive and negative frequencies. However, the proposal can be carried through in the WKB, minisuperspace context, yielding results similar to Hartle-Hawking.

Here, the boundary (or initial) condition is motivated from an informational and timeless perspective. A reduced configuration space of observables does exist,  and in it, one looks for the most homogeneous elements -- i.e. the ones that encode the least irregularities, or \lq{}information\rq{}, in its configuration. Under certain symmetry groups, and with certain assumptions about the degrees of freedom of the theory, a unique such element can be singled out. In addition, due to the stratified nature of reduced configuration space --  with concatenated boundaries -- this point consists of a natural boundary. In a geometric sense it is the \lq{}boundary of all boundaries\rq{} \cite{Fischer}, and no further boundary conditions need to be given. Such a radical asymmetry of reduced configuration space is furthermore necessary for our notion of records (Def \ref{def:full_record}) -- the crucial element which makes the recovery of time possible. 

At least  in the  case of Diff$(M)\ltimes \mathcal{C}$ with $M=S^3$, this construction means no further boundary  conditions need to be imposed apart from our single axiom. This can be seen by parametrizing physical space with unimodular metrics, $\tilde g_{ab}:=g^{-1/3} g_{ab}$. To change the signature, the determinant must become degenerate, which disconnects the physical space of positive definite signature from those with other signatures. In other words, the  \lq{}cone\rq{}  which makes up the boundary of Riem inside the affine space $C^\infty(TM\otimes_S TM)$ becomes inaccessible from within reduced configuration space. 

Nonetheless, if one was to blindly apply the axioms of this work to Riem and  Diff$(M)$, one would recover Vilenkin\rq{}s tunneling proposal, but with an extra scalar degree of freedom. That is because the orbit of $g=0$ is connected to the rest of superspace (but would not be to conformal superspace), and $g=0$ has the highest possible isotropy subgroup of Diff(M). As is the case with the standard tunneling proposal, one would have to establish further boundary conditions on superspace, which could be taken to be the \textit{singular boundaries} of superspace \cite{Wiltshire_intro}.   In minisuperspace, the extra scalar degree of freedom does not appear, and, having provided such extra boundary conditions, the two approaches coincide.

  Needless to say,  both Hartle-Hawking  and Vilenkin\rq{}s proposals are more studied and developed theories of initial conditions than the one presented here, even if they face enduring questions about i) the complex countour integration in which the path integral should be executed \cite{Halliwell_Hartle_boundary}, ii) the meaning of the Wick rotation (see e.g. \cite{Turok_latest} and references therein), and iii) the meaning of probabilities, and their conservation.  Moreover, it should be noted that due to the algebra of constraints containing the metric, a rigorous proof of gauge-invariance requires the use of the FBV formalism \cite{Fradkin_Vilkovisky}, which for this case will include interaction terms of fourth order in ghosts, with no geometric meaning as a Fadeev-Popov determinant \cite{Hartle_Halliwell}.\footnote{ In fact, due to this complexity, the BFV covariant version of \eqref{transition_amplitude} can only be shown to be gauge-fixing independent if the gauge-parameters are field independent, a result expected from \cite{Aldo_HG}.}

In the technical front, in the present work  simple BRST symmetries with a geometric meaning as a FP determinant \cite{Jaskolski} can be derived because all constraints are linear in the momenta (see \eqref{boundary_transf} and \eqref{final_func_conserv}). The wave-function built from the path integral satisfies the given functional constraint equations (even in the case of Weyl symmetry). A reduced (physical) configuration space exists, where we can make sense of a volume-form, including the records it creates, and decoherence in the sense of consistent histories.   In the toy model, the action functional is positive-definite, with no requirements for choosing complex actions and complex contours of integration. More generally, the Wick rotation is well defined along each path. More work needs to be done in this direction to ascertain more precise properties of the semi-classical solution. 

\paragraph*{Incompatibility between records and fundamental refoliation invariance.}
It is not possible to apply the present formalism to the Wheeler-DeWitt equation in general -- i.e. outside of the minisuperspace \lq\lq{}approximation\rq\rq{} --  including its embodiment in the Hartle-Hawking no-boundary  proposal. This comes about because of refoliation invariance, which makes it impossible to use the acquired notion of time through  records, as I now comment on. 

Firstly, there is no known form of reduced configuration space for the symmetries of ADM. This creates an insurmountable obstacle for the constructions of this paper. Even if one defines an initial configuration, records, screens, and so on, with respect to a given gauge-fixing of the foliations, these structures won\rq{}t be preserved under a different choice of gauge-fixing, as I described in equations in section \ref{sec:symmetry}.

Moreover, to the extent that such foliations can be of physical character (i.e. some surface of homogeneity of a given field), they should be reproduced by the purely relational evolution of observables, which I described in  subsection \ref{sec:duration}. 

Lastly, note that if the theory possesses a single global reparametrization constraint, my notion of records remains fully functional. This makes it more akin to minisuperpsace.  A curve will go through a given point however it is parametrized. A profound asymmetry of reduced configuration space will then spontaneously form records and, consequently, an effective notion of time. This is how records untangle evolution and symmetries.

%The type of models I introduced have different \textit{local} symmetries -- local symmetries which are from the outset untangled from evolution. This allows an  (ultra)local interpretation in configuration space, making such models more intuitive and in many senses more tractable within the context of reduced configuration space. One does not need to know the entire space-time to make assertions about observables. In section \ref{sec:axioms} I have mentioned the related difficulties in the quantum treatment (e.g. BRST, BFV) of the gauge-symmetries of ADM, and the relative advantages of gauge-symmetries that respect the principles I advocate for here (namely, with a point-wise action in field space).\footnote{A subject pursued at length in \cite{Gauge_riem}. } It all hinges on relegating the status of causal relations to being of  dynamical origin, and not kinematical. 

 \paragraph*{Non-causally related observables, and its problems.}\label{sec:duration}

This paper was based   on assigning a  fundamental role to acausal relations, \emph{prior}, and not defined by,  the specific field content which one would like to superpose. % This is the reason the nomenclature \lq{}space-like separated\rq{} (which is field-dependent) will be avoided, in favor of e.g.  \emph{causally independent}. 
   Whereas in GR causality is inherited from the space-time metric --  \lq{}space-like\rq{} is a statement about causal independence which depends on the metric -- I would like to  place local causally independent observables  at the foundations of quantum gravity (QG).  Space-time physics are to be recovered only effectively. This led me to a description of quantum mechanics on timeless configuration space. 
   
   This setup is perfectly suited for relational approaches to quantum mechanics \cite{Page_Wootters, Dolby, Page_summary}, the results of which we are able to recycle.    
  Importantly, foregoing the  space-time picture off-shell (or before dynamics)  \emph{does not imply} that a preferred simultaneity surface is classically  \textit{detectable} (see  \cite{Steve_SD} for a explicit example). %This intuition assumes an underlying space-time picture, but since notions of duration are only emergent, there might be no objective meaning to preferred simultaneity surfaces. In other words, 
   In fact, models of gravity can be formulated where we get  all the advantages of a preferred notion of simultaneity -- compatibility with QM collapse and superpositions -- and  none of the drawbacks -- no detectable \lq{}ether\rq{} for classical (relativistic) mechanics. The fact that such theories need not imply a preferred choice of simultaneity goes  contrary to common wisdom.  
The misguided intuition comes from imagining such \lq{}space-like\rq{} surfaces as being embedded in space-time, in which case indeed, they do define a notion of simultaneity. However, without the prior existence of space-time, one first requires a recipe to recover duration, using the dynamics of clocks and rods and fields throughout evolution in field-space.\footnote{See \cite{Wald_Unruh} for a related view.} 

For instance, for a theory  whose field space is that of spatial metrics, Riem,  for a given action $S(\gamma)$,  a solution to the equations of motion is a curve of 3-metrics $\gamma_{ab}(t)$, such that $S\rq{}_{|\gamma}=0$. However, to find out what a  space-time corresponding to this curve is, one needs to \lq\lq{}fill in\rq\rq{} the curve with a notion of duration (or a lapse),
e.g.: 
\be\label{space-time}ds^2=-N_{\mbox{\tiny eff}}[\gamma; x)^2\,dt\otimes dt +\gamma_{ab}(x,t)\,dx^a\otimes dx^b+\varpi_a[\gamma;x) (dt\otimes dx^a+dx^a\otimes dt)\ee
where I used the mixed functional notation of DeWitt (square brackets signify functional dependence, and round brackets ultralocal dependence). 
In \eqref{space-time} both the \lq\lq{}shift\rq\rq{} $\varpi$ and the lapse $N_{\mbox{\tiny eff}}$ are \lq\lq{}effective\rq\rq{}. The important point is that they only depend on $t$ through the curve $\gamma_{ab}(t)$ and are not inherited from space-time; quite the opposite: they build space-time.\footnote{This is what Wheeler used to call the struts and rods holding up space-time from its geometrodynamical skeleton \cite{MTW}.}   One example where this  is explicitly realized, obtaining exactly the same empirical content of GR, is the BSW formulation of the Einstein-Hilbert action, \eqref{BSW}, \cite{BSW}.\footnote{Although, being essentially GR, the Hamiltonian version of the theory reverts back to the standard ADM.}

\subsection{Summary}
To summarize, in this paper I have described a theory existing in  configuration space, $\mathcal{Q}$, with no singled out time variable. In the presence of  physical subsystems that act like  clocks, the formalism is able to recover the transition amplitude of standard quantum mechanics (see section \ref{sec:config_space}). However, this clock subsystem does not need to be defined over the entire configuration space for the theory to be consistent.

Given an action on curves in $\mathcal{Q}$, and a preferred boundary condition which is part of the theory, I described how to construct a volume-form in $\mathcal{Q}$ from the path integral. The volume-form must be constructed from the path integral. Under a simple further assumption required so that my notion of records (and a local decomposition property \cite{Locality_riem}) may later emerge, such a volume-form is uniquely given by the Born density. 

What I termed \lq{}records\rq{} is basically a sort of  \lq{}information\rq{} contained in the volume-form. Such records encode what we usually ascribe to the passage of time (in terms of conditional probabilities), and statements about conservation of probability using records were also shown to hold. It would be interesting to explicitly connect this notion of records with the one emergent in \cite{Barbour_Arrow}, and there claimed to resolve the issues of an arrow of time.  

The preferred boundary condition here is an integral part of the theory, but only in the case of a topology $M\simeq S^3$ and with the group being Diff$(M)\ltimes \mathcal{C}$ have I shown  it to be unique. In this particular case, the boundary  is the unique point which constitutes the least dimensional corner to the physical state space $\mathcal{Q}/$Diff$(M)\ltimes \mathcal{C}$, and no other boundary conditions need to be given. 

This is the ultimate asymmetry of configuration space, and it is what makes most of our constructions possible. For example: the notion of conservation of probabilities, discussed in section \ref{sec:conservation}; a notion of  global time, and of a standard Schroedinger equation for source fields propagating with this time; are made possible because there is  such an asymmetry in configuration space. The unique $\phi^*$ acts  in a similar fashion to the point of emission of alpha-particles in a Mott bubble chamber experiment. Here, as there, it is such an asymmetry that allows for the formation of \lq{}directional records of time\rq{}s passage\rq{} \cite{Halliwell_Mott}.

  Lastly, I proposed a simple gravitational toy model, similar to strong gravity,  compatible with the conditions imposed by this setting, and showcasing the viability of its constructions.\footnote{There is in principle no requirement for the models to be geometrical; only that they satisfy the axioms 1-5 of section \ref{sec:axioms}. } Making use of its interpretation in terms of the geometry of superspace, there is a shot-cut way to calculate (and regularize) its emergent Van-Vleck determinant.  I went on to compute the relative semi-classical quantum gravity probabilities (or volumes) for regions in configuration space, finally obtaining the approximate \eqref{Relative_amplitude}. This equation allows us to ask precise questions such as: what is the  probability of finding a given range of perturbations of gravitational modes of the original round 3-sphere, with respect to some other range? It turns out that for this simple model any mode is just as likely as another. %However, once one introduces derivatives of the metric into the action, the expectation is that, under the same calculations, modes with higher eiganvalues (for the Laplacian) will be disfavored relative to modes with small eigenvalues. In other words, more homogeneous modes will become more likely; which is one expected result of inflation. 
  Note moreover that this is a semi-classical effect -- depending on the Van-Vleck determinant -- not a classical one.

   %Shape dynamics \cite{SD_first},  and the conformal geodesic model of \cite{Conformal_geodesic}, are such theories. For the full perturbative, gauge-fixed path integral, the symmetry transformations act as a group (not a groupoid), and the arising BRST transformations are straightforward, allowing a simple treatment of gauge-symmetries (see \cite{Conformal_geodesic}, for the emerging form of the gauge-fixed path integral).  Of course, away from the semi-classical regime, nothing like a space-time should arise. 
   
 \paragraph*{Resolutions} 
Thus, summarizing the summary, we have the following resolutions of the questions posed at the start:
\begin{itemize}
\item In this framework there is no issue arising with the \lq{}superposition of causal structures\rq{}; causality can emerge semi-classically, and we can observe interference effects between alternative emergent space-times. 
\item Non-locality is a non-issue. Configuration space is the true arena of physics, and each point in it contains the entire physical space. Locality can and does emerge dynamically however \cite{Locality_riem}. 
\item Neither is there a \lq{}problem of time\rq{} (see \cite{POT_SD}). Evolution is abstracted from records -- made possible by the asymmetry of configuration space -- and is not mixed with a local gauge symmetry. 
\item One can define positive probabilities and related probability currents in the semi-classical approximation.  Unlike the corresponding approximation for WdW (which can only be done in minisuperspace), here there are no issues with positivity because time is defined around records, and at most a single solution of the Hamilton-Jacobi functional can exist. Source fields can be shown to propagate following the Schroedinger equation with respect to such an emergent time. The asymmetry of configuration space is what dictates and allows for these constructions and their consequent properties. 
\item  There is no conflict  of the sort \lq{}unitary vs reduction\rq{} processes for the quantum wave-function. 
 Relatedly, there is no \lq\lq{}definite outcomes\rq\rq{} problem, since the set of all configurations exist, with a volume-form on top of it. 

\end{itemize}
There are however, other challenges, concerned with regaining classical space-times.

\subsection{Some future projects}
 There are standard formulations of gravity which would be amenable to this treatment. Deparametrized shape dynamics \cite{Tim_effective} is one such model. This approach would give a different take on old problems, both in cosmology and elsewhere. The use of a prescribed boundary condition will restrict quantum cosmological scenarios, in a very similar way that the Hartle-Hawking condition does. In the homogeneous, cosmological scenario, our prescription will give a  different boundary condition than the Hartle-Hawking one, but which can be uniquely specified for certain types of field spaces and symmetries acting on them.  
 
 $\bullet$ \textbf{Bianchi IX shape dynamics.} Perhaps the ideal setting to test these ideas in the cosmological scenario would be the Bianchi IX shape space, characterized in \cite{Through_BB} (see also \cite{Flavio_tutorial}). There, we have an action, we have the explicit classical trajectories once again,  and we have a uniquely specified initial point -- the round sphere. In such a reduced configuration space, we get basically a Wheeler deWitt type equation, that should be solved with our boundary conditions. This would give a first prescription of a shape quantum cosmology model.  
 %The only problem that could arise comes from the compactness of the field space; extremal paths might have larger winding numbers, and the prescribed wave-function given by  $\Psi(g)=W(g^*, g)$ might not be convergent, even semi-classically. 
 
 $\bullet$ \textbf{Reconstructing space-time.} The most straightforward project directly extending this work however, is to flesh out the outline of the last paragraph of section \ref{sec:gravity} : add a simple matter scalar field to the action \eqref{geod_riem}, perform the analogous calculations, and use the prescription of \cite{Tim_effective}  to find an effective space-time according to \eqref{space-time}, below. Then use the constructions of \cite{Steve_SD} to evaluate the standing of the equivalence principle.  In this way we can explicitly start to probe qualitative implications of quantum gravity for space-times, which this approach suggests.

 $\bullet$ \textbf{Extending the analysis to other toy models.}
Lastly,  as I mentioned, one of the disappointing features of the present toy model is that it does not contain derivatives of the metric, and thus it does not couple points. Thus, perhaps the most straightforward project, would be to apply the formalism here -- which exploits parallels between geometrical quantities and the Van-Vleck determinant --  to more complicated geometries in $\mathcal{Q}$, explicitly described in \cite{Michor_general}. A particularly interesting geodesic action to try out is 
\be\label{general_geodesic} S[{g}]=\int dt \langle \dot{g}, \dot{g}\rangle^{1/2}_{{g}(t)}=\int dt \left( \int d^3x\rq{}\, \sqrt{g}(R^2+\epsilon)^{1/2}\int d^ 3x\,\sqrt{g}\,  \dot {g}_{ab} \, g^{ac}\, g^{bd}\, \dot {g}_{cd}\right)^{1/2}
\ee 
where ${g}_{ab}$ is now unimodular, i.e. $\sqrt{{g(t)}}=\sqrt{g_0}$, and with transverse-traceless velocities. For arbitrarily small $\epsilon>0$, this is the Jacobi form of the  Einstein-Hilbert action with $N=1$ (and transvere-traceless velocities), and thus its extremal trajectories should coincide for positive Ricci scalar (but not its path integral). That is, the extremal paths of \eqref{general_geodesic} coincide with those of the spatially unimodular Einstein-Hilbert action with positive spatial scalar curvature and small cosmological constant.  In fact, much of the groundwork for this analysis has already been set. In \cite{Michor_general}, general geodesic theories, with arbitrary conformal factors depending on the Riemann tensor have been studied, and many similar results to the ones presented here obtained.

\section*{ACKNOWLEDGEMENTS}

I  would like to thank Lee Smolin, Aldo Riello, Wolfgang Wieland, J. Halliwell,  Flavio Mercati and Sean Gryb for comments.  This research was supported  by Perimeter Institute for Theoretical Physics. Research at Perimeter Institute is supported by the Government of Canada through Industry Canada and by the Province of Ontario through the Ministry of Research and Innovation.

\begin{appendix}

 \section*{APPENDIX}
 \section{ The Wheeler DeWitt equation}\label{app:scalar_ADM}

 \subsection{Difficulties with The Wheeler DeWitt equation}
 
 Classical GR admits a Hamiltonian formulation, but it is a constrained theory. The configuration variable is taken to be the spatial Riemannian metric $h_{ab}$ on the manifold ${M}$, alluded to in \eqref{transition_amplitude}. The space of all such metrics is called Riem$({M})$. The emerging system (whose specific form is not relevant to our study here) is  of a fully constrained Hamiltonian, 
 \be\label{ADM_Ham} \mathcal{H}=\int_{M} d^3 x\left(N(x)H(x)+N^i(x)H_i(x)\right)
 \ee
 The standard canonical quantization rules would suggest that the states of the system be given by wave functions of the configuration variable, $\Psi=\Psi(\tau, h_{ab})$ where $\tau$ denotes the time variable occurring in the Hamiltonian formulation. Replacing as usual $h_{ab}$ by its action through multiplication, and of its conjugate momentum with functional derivation, $\pi^{ab}\rightarrow \frac{\delta}{\delta h_{ab}}$,\footnote{More formally, we should use the Radon-Nikodym derivatives in field space, which takes the measure into account \cite{Isham_aspects}.} we could have the evolution of $\Psi$ given by the Schroedinger equation, 
 \be \label{Schroedinger}
 i\frac{d\Psi}{d\tau}= \mathcal{H}\Psi
 \ee
 Since $\mathcal{H}$ is pure constraint,  $\mathcal{H}\Psi=0$, but also $\hat H \Psi=\hat H_i \psi=0$.   
The changes generated in phase space by the flow of $\int_{M} d^3 x N^i(x)H_i(x)$ correspond simply to the action of spatial diffeomorphisms on $h_{ab}$ and its conjugate momenta. Its quantization would simply mean that the value of $\Psi(h_{ab})$ should be unchanged upon the action of an infinitesimal diffeomorphism of $M$, $f\in \mbox{Diff}(M)$, which acts pointwise on $\mbox{Riem}(M)$ (through pull-backs).  The space of metrics quotiented by this gauge symmetry, Riem/Diff, is the space of geometries, also called \textit{superspace}, and the action of  $\hat H_i $ as a quantum constraint operator implies the wavefunction lives over this superspace. This part is recovered in the present setting. 

The action of $\hat H$ on the wave-functions gives rise to the Wheeler-DeWitt equation:\footnote{ It should be noted that to obtain \eqref{WdW} from \eqref{transition_amplitude} one needs to \textit{a priori} fix refoliations at the initial and final configurations \cite{Hartle_Halliwell}, in accordance with our expectations. }
\be \label{WdW} \hat {H}\Psi=\left(G_{abcd}\frac{ \delta^2}{\delta g_{ab}\delta g_{cd}}+\sqrt{g}\left(-R(g)+2\Lambda\right)\right)\Psi[g_{ij}]=0\ee
 The analogous operator interpretation to  $H_i$ here would be invariance wrt diffeomorphisms in space-time which move points orthogonally to ${M}$. Although that interpretation is not as clear in this case, it is still compatible with the intuitive effect of the implementation of the constraint, namely $\frac{d\Psi}{d\tau}=0$.

In spite of the indefinite signature in its kinetic term (coming from the DeWitt supermetric) giving rise to quadratic terms of negative sign,  the WdW cannot be interpreted as a Klein-Gordon type of equation to be ``second-quantized", for i) it is already quantized, and ii)  there are no straightforward ways to split solutions into positive and negative frequencies to construct the annihilation and creation operators, since there is no good time function in superspace. For a review, see \cite{Kuchar_Time}. This is not to mention finer points about regularization (point-splitting seems to be the more appropriate one here, due to the occurrence of the singular $\delta(x,x)$ coming from quantization of the term quadratic in momenta \cite{Woodard}) or factor-ordering.

%In my opinion, besides the more technical problems with the Wheeler-deWitt equation,  there are also conceptual knots one needs to untangle.  We could imagine $\Psi(h,\tau)$ gives the amplitude of measuring $h$ at the time $\tau$, but the WdW equation makes this interpretation untenable. That is because $\Psi(h)$ is forced to be independent of $\tau$, even \textit{right after a measurement}.

 Moreover, the requirement that the Hamiltonian commute with all observables can be extremely restrictive. For ergodic systems, it will lead to the conclusion that the only  observables (continuous in phase space, with the Liouville-induced topology) are polynomials of the Hamiltonian itself \cite{Wald_Unruh, Tim_chaos, Tim_chaos2}; a defect which is at least camouflaged in minisuperspace approximations due to their simplicity. 

 \subsection{Issues with the path-integral definition of the wave-function}\label{app:issues}
  \be\label{transition_amplitude}W(h_1,h_2)=\int_{h_1}^{h_2}\mathcal{D}g\, \exp{\left[i\int\mathcal{L}/\hbar\right]}\ee
where $\mathcal{L}$ is the gravitational Lagrangian density, and I have omitted boundary terms, the required gauge-fixings,  Fadeev-Popov determinant and regulators (since the theory is non-renormalizable, this action should only be taken as an effective theory at a given scale). Here $h_i$ are initial and final spatial geometries, i.e. for the hypersurfaces ${M}_i$ and  the embedding \,$\imath_i:{M}_i\hookrightarrow M$ then $h=\imath^* g$, where here $g$ is the abstract space-time metric tensor field.  The issue is that  refoliations shift $h_i$ (in a $g$-dependent manner), and change \eqref{transition_amplitude}.\footnote{ As I comment on below, however, the issue is more subtle than it appears, if one implements the boundary conditions with a delta function on the boundary. Then gauge transformations don\rq{}t act directly on the boundary field. }

Such difficulties with giving a simultaneous meaning to refoliations as  a gauge-symmetry and defining a meaningful gauge-invariant transition amplitude are reflected in the canonical quantization approach, both for gravity and more basic fields.  For instance, in passing from the Hamiltonian transition amplitude to the Lagrangian path integral for field theory by integrating out the momenta, indeed one obtains, e.g. for a scalar field $\varphi$:
$$\langle\varphi\rq{}\rq{}(x),t\rq{}\rq{}|\varphi\rq{}(x),t\rq{}\rangle=\langle\varphi\rq{}\rq{}(x)|e^{-iH(t\rq{}\rq{}-t\rq{})}|\varphi\rq{}(x)\rangle=A\int\prod_x d\varphi(x^\mu)\exp{\left(i\int\mathcal{L}(\varphi)d^4x^\mu/\hbar\right)}$$
where I used a non-standard relativistic notation $x^\mu=(x,t)$, and the integration is done so that $t=x_0$ lies in $t\rq{}<x_0<t\rq{}\rq{}$, and $\varphi(x, t\rq{}\rq{})=\varphi\rq{}\rq{}(x); ~~\, \varphi(x, t\rq{})=\varphi\rq{}(x)$. The only thing in this formula which is not Lorentz invariant is the domain of integration, $t\rq{}<x_0<t\rq{}\rq{}$. To properly calculate the S-matrix (and the generating functional) this issue is remedied by taking the infinite time interval $t\rq{}\rq{}\rightarrow \infty, \, t\rq{} \rightarrow -\infty$. Of course, this is an abstraction; measurements take a finite time. 

  In  more generality, for finite intervals, local time reparametrizations -- refoliations -- don't act as a gauge group in the boundary field space, thus  obstructing the definition of a transition amplitude  between fully gauge-invariant states.  In sum, one can only show that \eqref{transition_amplitude} is independent of an arbitrary choice of \emph{bulk} space-time gauge-fixings (i.e. those that are the identity at $h_1$ and $h_2$) together with arbitrary gauge-fixings of the \textit{spatial boundary} diffeomorphisms at $h_1$ and $h_2$ \cite{Hartle_Halliwell}. The choice of the \lq{}space-like\rq{} surfaces ${M}_1, {M}_2$ is \emph{not}  a gauge-fixing in this sense; different choices \textit{will} change the amplitude. More involved treatments exist (see the review \cite{Tambornino}),  in which one defines the boundary field values relationally -- with respect to some other matter field for example (or with so-called \lq\lq{}embedding variables\rq\rq{} \cite{Kuchar_Isham}). Nonetheless, these resolutions also have their issues; especially in what refers to maintaining a good choice of such boundary-defining functions throughout field space. Without  such a relational definition working globally in field space, one cannot ensure unitarity \cite{Bojowald_effective}.

For a transition amplitude, one could define $h_1$ and $h_2$ in \eqref{transition_amplitude} as the inverse value of some scalar functional on the spatial metric (such as the value of a matter field, or the scalar curvature) $f:\mathcal{Q}\mapsto \mathbb{R}$, as  $h_i:=f^{-1}(t_i)$, with restrictions on $f$ so that $h_i$ becomes invariant with respect to refoliations. Implicitly, this is what is done to obtain the invariance relations of the path integral with respect to refoliations, and it corresponds to the \lq\lq{}embedding variables\rq\rq{} of Isham and Kuchar \cite{Kuchar_Isham}.  In this case, the embedding variables function largely as a \lq{}dust field\rq{}, which defines points on the space-time manifold. 

Here adopting $h_{ij}$ for the 3-metric and $g_{\mu\nu}$ for the 4-metric, we fix  one of the boundary conditions. From \eqref{transition_amplitude} the wavefunction of the universe on a hypersurface with intrinsic 3-metric $h_{ij}$ and matter configuration, $\varphi$, can be defined by
\be  \label{path_integral_appendix}
\Psi(h,\varphi)=\int \mathcal{D}g\mathcal{D}\varphi \exp{-I\left[g_{\mu\nu},\phi\right]}
\ee
where the sum is over a class of 4-metrics, $g_{\mu\nu}$, and matter configurations $\phi$, which when restricted to the boundary 
take values $g_{ij}$ and $\varphi$, and $I$ is the spacetime covariant action functional. 

To find the functional constraints related to the Ward-Takahashi identities such a wave-function might satisfy, one must employ the more formal Hamiltonian definitions. In \cite{Hartle_Halliwell}, Hartle and Halliwell attempt to formally clarify this. Instead of \eqref{path_integral_appendix}, one defines (dropping $\varphi$): 
\be\label{HH}
\Psi(h)=\int \mathcal{D}z^A \delta(h_{ij}(t_f)-h_{ij})\Delta_C[z^A]\delta[C(z^a)]\exp {(iS[z^A])}
\ee
where $z^A=(\pi^{ij}, h_{ij}, N^\alpha)$ for the Hamiltonian path integral, or $z^A=(h_{ij}, N^\alpha)$ if one can integrate away the momenta, $\mathcal{D}z^A$ is the measure obtained from the Liouville one (for which BRST transformations are canonical), $\Delta_C[z^A]$ is the Fadeev-Popov determinant, $C(z^A)$ is the gauge-fixing condition, and $N^\alpha$ are the Lagrange multipliers. Integration runs also over all the final variables, and there they are fixed by the Dirac delta. 

In the case of gravity, the very definition of the path integral through time slicing requires one to choose a notion of physical time already at that level, so that infinitesimal slicings are meaningful and one can take extremal approximations in between the slices. The very construction of the path integral in terms of infinitesimal slicings only makes sense in a gauge-fixing for gravity, which sets it apart from theories in which a background causal structure exists.

Clearly, from \eqref{boundary_transf}, if either the gauge transformations vanish at the boundaries, or if the Hamiltonian generators of the symmetries are linear in the momenta, then $\delta_\epsilon S=0$. Moreover, in those cases, then we can resort to the standard BRST methods to show invariance of the wavefunction \eqref{HH}. This is the main difference between the constraints advocated here and the ones which shift the boundary itself. 

By redefining the integration variables according to the $\delta_\epsilon$ transformation (and subtracting the untransformed wavefunction), one obtains 
\be\label{final_func_conserv}
0=\int \mathcal{D}z^A \left[-\int d^3x\rq{} \delta_\epsilon h_{ij}(x\rq{})\frac{\delta}{\delta h_{ij}(x\rq{}) }-\delta_\epsilon S\right]\delta(h_{ij}(t_f)-h_{ij})\Delta_C[z^A]\delta[C(z^a)]\exp {(iS[z^A])}
\ee
Note that here the refoliations act on the paths (or on the $z^A$ variables), and therefore on $h_{ij}(t)$,  but not on the boundary fields $h_{ij}$. 

Now, we must use  their assumption 5 in section II B, which translates functions on the boundary variables in the path integral to functional equations on the resulting wavefunction (ignoring Lagrange multipliers,  gauge-fixings and Fadeev-Popov determinants), i.e.: 
\be\label{HH_constraint}\int \mathcal{D}\varphi\mathcal{D}\pi_\varphi F(\varphi(t\rq{}\rq{}),\pi_\varphi(t\rq{}\rq{}))\delta(\varphi(t\rq{}\rq{})-\varphi\rq{}\rq{})  e^{iS(\varphi,\pi_\varphi)/\hbar}=F[\varphi\rq{}\rq{}, -\hbar \frac{\delta}{\delta \varphi\rq{}\rq{}}]\psi[\varphi\rq{}\rq{}]\ee
And voila, one obtains the constraint equations applied to the wave-function, by taking into account the transformation of the boundary values together with the transformation of the action functional.\footnote{
However, \textit{in the case of reparametrization invariant theories} with constraints quadratic in the momenta,  there are various unsatisfactory features in this derivation. First, as commented earlier, one would need to define the slicing already in a particular physical gauge-fixing. This makes it quite difficult to compare the resulting wavefunction given in two different choices of foliations. Secondly, in standard gauge theories, one can show that the path-integral is independent of the gauge-fixing by the use of the Fadeev-Popov determinant, which has a geometrical interpretation as a Jacobian for the gauge-fixing surface \cite{Jaskolski}, and thus embodies the correct transformation properties wrt arbitrary variable redefinitions. For theories such as GR, in which the constraint algebra is field-dependent, this is not the case, and one must resort to more complicated algorithms, such as the BFV one \cite{Fradkin_Vilkovisky}, which doubles the amount of fields. Even then, to apply the Fradkin-Vilkowiski theorem, one must have an invariant action, which is not the case here, due to $\delta_\epsilon S\neq 0$. Hartle and Halliwell get around this by at first implementing boundary conditions (see eq. 3.16 therein) which don\rq{}t allow BRST transformations at the boundary (and which are also BRST invariant conditions) and then using the FV theorem for assuming independence of the gauge-fixing conditions. However, to get the Wheeler DeWitt equation, they must relax such boundary conditions, implementing new conditions which are neither BRST invariant nor allow for the use of the FV theorem (see equation 3.19). Indeed, the  only gauge-fixing used in \cite{Hartle_Halliwell}  is the one which would not require a functional connection, as in \cite{Donnelly_2016, Aldo_HG}, namely, the field-independent boundary conditions on the gauge-parameters (see their appendix B). Disallowing these, it seems it would not be possible to move from one physical slicing condition to another, as these transformations are necessarily field-dependent.  My conclusion from these points  is that perhaps the inclusion of boundaries in the formalism is more subtle than Hartle and Halliwell allow for, as indicated in many recent papers (see \cite{Donnelly_2016, Aldo_HG} and references therein). This intuition is supported by  \cite{Barvinsky_review}, in which skeletonization issues are scrutinized. To maintain the invariance of the skeletonized Liouville measure, because of the action of refoliations, Barvinsky also restricts gauge transformations at the two end-points (see eqs 4.42 and 4.43 and comments in between). These issues are not, however, of fundamental importance for this work. }

 One of the issues here is that, unlike ordinary field theories such as Yang-Mills, the negative sign in front of the squared trace of the gravitational velocity (\lq{}the conformal mode\rq{}), implies the covariant gravitational action is not positive-definite. Thus, restricting the sum to real Euclidean-signature metrics  would not generically allow the  the path integral to converge.\footnote{While it is true the path integral for fermions is also not positive-definite, their anticommuting nature ensures that the path integral converges.} Indeed, to obligate the path integral to converge in this way,   it would be necessary to include complex metrics in the class of metrics to be integrated over \cite{Schleich}. However, as a recent renewed debate  surrounding this issue makes clear \cite{Turok_latest}, there is no unique contour to to integrate along in superspace. Moreover, the result obtained for the path integral would generically  depend on the types of countours chosen.  Such  different alternatives are intimately connected with the boundary conditions chosen for the paths to be integrated over, and  these choices should  implicitly restrict  the 4-manifolds included in the sum in \eqref{path_integral_appendix}. 
 
   However,  the boundary conditions in the path integral should have a covariant meaning, and that makes them extremely  averse to being embodied in superspace; where the individual paths of the path integral are explicitly realized.\footnote{ In practice, these problems are not seen directly, since working with the infinite dimensions of the full superspace is unfeasible, and the problems are not self-evident in minisuperspace.} One objective consequence of this problem  is that, although such contour integrations and boundary conditions for the Wheeler-DeWitt equation must be somehow related, it is nearly impossible to realize this connection in a precise manner.

\paragraph*{Physically defining boundary states} 
  This approach of physically defining the boundary states  is also related to what is known as the approach of \textit{complete observables} in canonical general relativity \cite{Rovelli_book, Brown_Kuchar}, for which a large literature exists (see \cite{Tambornino} for a review) and the use of \lq{}internal time\rq{} (see e.g. \cite{Brown_Kuchar}). In the case where only one reparametrization constraint exists, complete observables should be a one-parameter family of Dirac observables. Many technical problems arise, however, when an infinite amount of reparametrization constraints exist;  as is the case of GR, and when one would like to extend this construction to the entire phase space.\footnote{See \cite{Bianca_canonical, Bojowald_effective} for some candidate constructions.}  

Nonetheless, as shown in \cite{Bojowald_effective}, even in this case,  attempting to use an internal time for each different region in phase space, patching  the  transitions between them,  actually works well only in the purely classical regime. From the point of view of state evolution, defining time for each finite
range poses an enormous issue for unitary evolution of states.\footnote{Indeed, this happens even if classical evolution with respect to a local internal time can be made unproblematic  in the transition region between one patch and another. } The major issue is then that non-unitary quantum evolution risks  meaningless results --- even away from the boundary of the patch where it should be valid --- and  it is not clear how to define quantum observables in such a situation.

 \section{Connections on principal fiber bundles}\label{app:PFB}

The advantage of working directly with principal fiber bundles (PFB's) is that structures are simpler. 
For example, it is easier to work out particular features of associated vector bundles, or features of gauge-fixed structures directly from the general formalism of PFB\rq{}s than vice-versa.
The standard example of a PFB is the bundle of linear frames over a spacetime manifold $M$, with structure group $\mathrm{GL}(n)$.

%---------------------------------------------

A principal fiber bundle $P$, is a smooth fiber bundle, for which a Lie group $G$ has an action from the right $G\times P \rightarrow P$,\footnote{This means that in the coordinate construction of $P$ the transition functions act from the left.} which we denote by $R_g p=: p\triangleleft g$, for $g\in G$, $p\in P$. 
The action is assumed to be free, so that $ p\triangleleft g=p $ iff $g=\mbox{id}_G$, the identity of the group.  
The quotient of {$P$} by the equivalence relation given by the group, that is $p\sim p\rq{}$ iff {$p'=p\triangleleft g$} for some $g\in G$, is  usually identified with the spacetime manifold, i.e. $M=P/G$.
We denote the projection map by $\mathsf{pr}:P\rightarrow M$. where $R_g$ denotes the right action of $g$ on $P$ (which should be distinguished from the action of $g$ on the group $G$ itself). The set $O_p=\{p\triangleleft g\, |\, g\in G\}$,  is called the orbit through $p$, or the fiber of $[p]=\mathsf{pr}(p)$. The action of $\mathsf{pr}$ projects $O_p$ to $[p]\in M$.

%Let $U\subset M$ be an open subset of $M$, then a smooth embedding $\sigma:U\rightarrow P$ is a section (or a ``gauge-fixing'') if $\mathsf{pr}\circ \sigma = \mbox{id}_M$. 
The section then gives rise to a trivialization $G\times U\simeq \mathsf{pr}^{-1}(U)$, given by the diffeomorphism $F_\sigma:U\times G\rightarrow \mathsf{pr}^{-1}(U)$, $([p], g)\mapsto\sigma([p])\triangleleft g\,$.
 Generically, there are no such $U=M$ and the bundle is said to be non-trivial. {In the functional, i.e. field space, case, the base manifold is a modular space and  the obstruction to triviality is known as the Gribov ambiguity}.

%-------------------------------------------------
\paragraph*{PFB connections}

Before going on to define a connection, we  require  the concept of a vertical vector in $P$. 
Let $\exp:{ \text{Lie}(G)=\mathfrak{g}}\rightarrow G$ be the group exponential map. 
Then by dragging the point with the group action we can define a vertical vector at $p$, related to $X\in \mathfrak{g}$, as 
\be 
\label{fundamental_vf} 
X_p^\#:= \frac{\d}{\d t}_{|t=0}\Big(p\triangleleft \exp{(tX)}\Big)\in \mbox{T}_pP
\ee
The vector field $X^\#\in \Gamma(\mbox{T}P)$ is  called the \emph{fundamental vector field} associated to $X$. 
The vertical space $V_p$ is defined to be the span of the fundamental vectors at $p$. 

From \eqref{fundamental_vf} one has that 
fundamental vector fields are ad--equivariant, in the following sense: 
\be
\label{ad_transf}
 ({R_g})_*X^\#_p=( \mbox{Ad}_{g^{-1}} X)^\#_{ p \triangleleft g}
\ee
where ${R_g}_*:\mbox{T}P\rightarrow \mbox{T}P$ denotes the pushforward tangential map associated to $R_g$ and $ \mbox{Ad}_g:\mathfrak{g}\rightarrow \mathfrak{g},~X\mapsto g X g^{-1}$ is the adjoint action of the group on the algebra. 
Now, using the fact that the Lie derivative of a vector field $Y^\#$ along $X^\#$ is defined as the infinitesimal pushforward by the inverse of $\exp(tX)$ evaluated at $p$, by setting $g\to \exp(-tX)$ and $p \to p\triangleleft g^{-1}$ in \eqref{ad_transf} and deriving, we obtain
\be
\pounds_{X^\#} Y^\# = [X^\#,Y^\#]_{\mathrm{T}P}= \frac{\d}{\d t}_{|t=0}\Big({R_{\exp{(-tX)}}}_* Y^\# \Big)= \frac{\d}{\d t}_{|t=0} \Big({\mbox{Ad}_{ \exp{(tX)}}} Y \Big)^\#=(\mbox{ad}_X Y)^\# = ([X,Y]_{\frak g})^\#.
\label{eq17}
\ee
This shows that  the vertical bundle, $V\subset \mbox{T}P$, is an integrable tangential distribution. 

The definition of a connection amounts to the determination of an equivariant algebraic complement to $V$ in $\mbox{T}P$, i.e. an $H\subset \mbox{T}P$ such that
\begin{subequations}
\begin{align}
& H_p\oplus V_p=\mbox{T}_pP \\
&  {R_g}_*H_p= H_{  p \triangleleft g}\,,\, ~ \forall p
\end{align}
\end{subequations}
 One can equally well define $H$  to be the kernel of a $\mathfrak{g}$-valued one form $\omega$ on $P$, with the following properties:
\begin{subequations}
\begin{align}
 &  \omega(X^\#)=X   \label{omega-a}  \\
 & R_g{}^*\,\omega=\mbox{Ad}_{g^{-1}}\omega  \label{omega-b}
\end{align}
\label{omega}  
\end{subequations}
where $ {R_g}^*\omega\equiv \omega\circ {R_g}_*$. 
 In other words, equation \eqref{omega-b} intertwines the action of the group $G$ on $P$ (lhs of the equation) and its action on $\mathfrak{g}$ (rhs). Notice that these equations hold only for a global action of $G$ on $P$. 
For any $v\in\mbox{T}P$, its vertical projection is given by $\hat V(v)=\omega(v)^\#$. %
  
The horizontal derivative is the generalization of a covariant derivative (in the associated vector bundle) to the PFB context.  The role  of a covariant derivative is to exactly cancel out the non-equivariant terms  obtained for derivatives (either in in field space or not).    
In the presence of a principal fiber bundle structure in field space,  what plays the role of the Lie group before is now a diffeomorphism $f: M \to M$, $x\mapsto f(x)$. It acts on the metric $g\in\mathcal{Q}$ via pullback,%
\be
(A_{f^*} g)_x := (f^*g)_x.
\ee
where I have taken pains to distinguish the action ($A$) from the particular group element ($f$). We can transplant the previous structures to the infinite-dimensional case: 
\be
\label{hor_field}
\frac{d}{dt}\leadsto \frac{D}{dt}= \hat{H}\frac{d}{dt}=
\frac{d}{dt} - \fI_{\hat V}\frac{d}{dt}\;,
\ee 
where  $\fI_{\cdot}$ is the inclusion operator in the field-theory context, $\hat V$ is the vertical projection in $\mathcal Q$ and   $\varpi$ is a Lie-algebra-valued one-form, i.e. $\varpi\in\Lambda^1(\mathcal{Q}, \Gamma(\mbox{T}M))$, which obeys {the analogue of }\eqref{omega}, that is
\begin{subequations}
\label{connection1and2}
\begin{align}
&\varpi(X^\#) =X\label{connection1}\\
&A_{\varphi^*}^*\,\varpi=(\varphi_*\varpi\circ \varphi^{-1}) \label{connection2}  
\end{align} 
\end{subequations}
 As before, the horizontal projection $\hat H$, is defined via the kernel of $\varpi$. With the substitution \eqref{hor_field} in the action, one automatically obtains general gauge-invariance for time-dependent gauge transformations.  See \cite{Aldo_HG} for a complete account on the relevance of this connection in the field space context, and how a choice of connection can embody a choice of abstract observer.  
Both for the functional case and in the finite-dimensional one, we have the infinitesimal form  of the equations, generalized for possibly field-dependent gauge transformations:
\begin{align}
 \fI_{\xi^\#}\varpi & = \xi  \label{eq_fundamental1} \\
\qquad\fL_{\xi^\#}\varpi & = [\varpi, \xi ] + \delta \xi \qquad \label{eq_fundamental2}
\end{align} 
where the commutator is the Lie algebra one. From these equations it is easy to see that a connection fulfills the purpose of canceliing out terms that don\rq{}t transform homogeneously under the group. 

\section{The timeless  transition amplitude in quantum mechanics}\label{app:timeless_transition}

%As one does not have access to coordinate time, in the presence of a Hamiltonian one should average over this inaccessible variable to get a density matrix for the physical variables. One needs to use the projector on the density matrix and on the other operators, where $\hat H$ is the canonically quantized  Hamiltonian, with a given choice of operator ordering. For instance, $\hat A\mapsto \hat A_{\mbox{\tiny phys}}:=\hat P \hat A\hat P$.  Note that since one integrates over all $\tau$, the projector is parametrization independent. For a careful analysis of the emerging quantum mechanical description of \lq\lq{}timeless\rq\rq{} physical systems, see \cite{Dolby}.% There, it is shown explicitly how to make sense of the probability mentioned in the introduction, of $P(A_2~~ \mbox{when}~B_2~| ~A_1 ~~\mbox{when}~B_1)$, in a timeless fashion (including a rebuke to Kuchar\rq{}s objection \cite{Page_summary}) and how to recover the standard quantum mechanics transition amplitude in the presence of a good clock subsystem.  

%Here, I am interested in a timeless transition amplitude for path integrals  in configuration space, which is more directly applicable to the axioms and motivations of this work. Let me briefly review the transposition of the work of Page  \cite{Page_summary} and others to that setting, in the  particle mechanics case. 

\paragraph*{ Quantum mechanical timeless transition amplitude in configuration space}
We start with a finite-dimensional system, whose configuration space, $\mathcal{Q}$, is coordinitized by $q^a$, for $a=1,\cdots, n$.  Let us start by making it clear that no coordinate, or function of coordinates, singles itself out as a reference parameter of curves in $\mathcal{Q}$. The systems we are considering are not necessarily `deparametrizable' -- they do not necessarily possess a suitable notion of time variable. 

Now let $\Omega=T^*\mathcal{Q}$ be the cotangent bundle to configuration space, with coordinates $q^a$ and their momenta $p_a$. For a reparametrization invariant system the classical dynamics is fully determined once one fixes the Hamiltonian constraint surface in $\Omega$, given by $H=0$.
A curve $\gamma\in \mathcal{Q}$ is a classical history connecting  $q^a_1$ and $q^a_2$ if there exists an \emph{unparametrized} curve $\bar\gamma$ in $T^*\mathcal{Q}$ such that the following action is extremized:
\be\label{equ:phase_action} S[\bar\gamma]=\int_{\bar\gamma}p_adq^a 
\ee
for curves lying on the constraint surface $H(q^a, p_a)=0$, and such that the projection of $\bar\gamma\in T^*\mathcal{Q}$  to $\mathcal{Q}$ is $\gamma$, connecting $q^a_1$ and $q^a_2$. 
 By parametrizing the curve with an innocent parameter $t$, we can get to the familiar form:
\be\label{equ:parametrized_action}
 S[\bar\gamma]=\int dt \left(p_a\dot{q}^a-N(t)H(q^a, p_a)\right)
\ee 
where the $N$ is a Lagrange multiplier (and we are considering only one constraint).

One can now define the fundamental transition amplitudes between configuration eigenstates.%\footnote{As much as possible, I want to avoid technicalities which won't be required here. Having said this, formally one would have had to define the so-called kinematical Hilbert space $\mathcal{K}$ for the quantum states over $\mathcal{Q}$ by using a Gelfand triple over $\mathcal{Q}$  with measure $d^dq^a=dq^1\cdots dq^d$, i.e.  $\mathcal{S}\subset\mathcal{K}\subset\mathcal{S}'$. This is not necessary in my case, because we did not require a Hilbert space.}
\be \label{equ:transition} W(q,q\rq{}):=\langle q|\hat P|q\rq{}\rangle\ee 
where the projector can be defined by:
\be\label{equ:projector}
\hat P:=\lim_{t\rightarrow\infty}\frac{1}{2t}\int_{-t}^t t\rq{}\, e^ {-it\rq{} \hat H}
\ee
with the standard particle Hilbert space inner product, $\langle q|q\rq{}\rangle =\delta(q,q\rq{})$. 

 In \cite{Chiou}, following previous studies of relational quantum mechanics (see \cite{Rovelli_book} for a review),  Chiou  studied the relational timeless transition amplitude between configurations, 
$W(q_1,q_2)$ in the path integral formalism.\footnote{Chiou calls  it a 'curve' integral, as opposed to a path integral, because the integral only requires unparametrized paths in phase space. I will assume that there exists a metric in configuration space, for which I can then parametrize paths by arc-length.}
  The standard proof of equivalence between non-relativistic Schroedinger quantum mechanics and the path integral formulation relies on refining time slicings -- partitioning paths into arbitrarily small segments. Without absolute time,  a parametrized curve $\bar\gamma:[0,1]\rightarrow \Omega$ need not be injective on its image (it may go back and forth).  This issue is related to the one of integrating over only positive lapses, or over both positive and negative. One yields a propagator, and not a projection onto the constraint surface.

 Chiou chooses the latter, and uses a Riemann-Stieltjes integral to make sense of the limiting procedure to infinite sub-divisions of the parametrization. Furthermore, one must then sum over all parametrizations, at which point one obtains $\delta[ H]$, and the entire transition amplitude \eqref{equ:transition}.
 
  Crucially, Chiou uses a sub-division of the parametrization (available only when there is a single time direction).  The \textit{mesh} of the parameter sequence is then defined by  the largest interval encompassed by the parameter sequence. The parameter sequence is fine enough if its mesh is smaller than a given small number $\epsilon$.\footnote{ Still, here  the integration still needs to be regularized:
$
W(q_1^a,q_2^a)\sim \lim_{\lambda\rightarrow\infty}\frac{\int_{-\lambda}^\lambda d t \langle q_1^a|e^{-i t \hat{H}}|q_2^a\rangle}
{\int_{-\lambda}^\lambda dt}$
with a cut-off $\lambda$, satisfying some hierarchical condition with respect to the mesh.} 

The transition amplitude can then be written as:
\be
\label{equ:path_integral_Chiou}
W(q_1,q_2)=A\int \mathcal{D}q^a\int \mathcal{D}p_a\, \delta[H] \exp{\left[\frac{i}{\hbar}\int_{\bar\gamma}p_adq^ a\right]}
\ee
where the path integral sums over those unparametrized curves $\bar\gamma$ in phase space, whose projection to configuration space begins at $q_1$ and ends at $q_2$.\footnote{ In the presence of gauge symmetries, if it is the case that these symmetries form a closed Lie algebra, one can in principle use the group averaging procedure, provided one uses a  translation invariant measure of integration. }

 I will not require this condition, but if configuration space can be decomposed wrt a \lq{}time\rq{} variable, $q=(t,\bar q)$, such that  the total Hamiltonian can be written as $\hat{H}=\hat{p}_t+\hat{H}_o(\bar q, {p}_{\bar q})$, with  $H_o$  not explicitly dependent on $t$,  then the timeless path integral amplitude kernel is equal to the standard non-relativistic one,  up to an overall constant:\footnote{In \cite{Briggs_Rost},  the conditions under which Briggs and Rost derive the time dependent Schroedinger equation from the time-independent one, corresponds to the system being deparametrizable. I.e. one can isolate degrees of freedom (the environment, in Briggs and Rosen) which are heavy enough to not suffer back reaction. The proposal here is not dependent on this condition.}
\be\label{equ:deparametrizable}
W(q_1,q_2)=G((t_1,\bar q_1),(t_2,\bar q_2))=\langle\bar q_1|e^{-i\hat{H}_o(t_2-t_1)}|\bar q_2\rangle
\ee
 Note however, an important difference: in the lhs of \eqref{equ:deparametrizable}, time is just part of the configuration, it has no role as a \lq\lq{}driver of change\rq\rq{} \cite{CFP_essay}. In the remaining parcels  of the equation, it is interpreted as an absolute time.  

Here I take \eqref{equ:path_integral_Chiou} for granted as an initial starting point. The difference is that I assume the system is written in the Lagrangian framework, and thus can be completely encapsulated by curves in configuration space. This would be equivalent to the starting point of Chiou if the momenta could be integrated out in the respective Hamiltonian. It would also yield the configuration space measure from the Liouville one, as in \eqref{equ:measure_timeful}.

\section{Timeless semi-classical composition law.}\label{app:semi_classical_proof}
Here I prove the theorem \ref{theo:record} in more generality. 
Assuming the action is additive, for a given path
\be\label{equ:additive}
S_{\gamma_{\mbox{\tiny{cl}}}}({\phi^*},\phi)=S_{\gamma^1_{\mbox{\tiny{cl}}}}({\phi^*},\phi_m)+S_{\gamma^2_{\mbox{\tiny{cl}}}}(\phi_m,\phi)
\ee
We need to show that 
\be\label{equ:VV_comp}\det{\left(-\frac{\delta^2S(\phi^*,\phi_m)}{\delta\phi^*\delta\phi_m}\right)}\det{\left(-{\frac{\delta^2S(\phi_m,\phi)}{\delta\phi_m\delta\phi}}\right)}\det{\left(\frac{\delta^2S(\phi^*,\phi_m)}{\delta\phi^2_m}+\frac{\delta^2S(\phi_m,\phi)}{\delta^2\phi_m}\right)}^{-1}
=\det{\left(-\frac{\delta^2S(\phi^*,\phi)}{\delta\phi^*\delta\phi}\right)}\ee
which is the required composition law of determinants to obtain \eqref{equ:semi_classical_record} (see Kleinert's textbook \cite{Kleinert} for a proof). It is easiest to see this relation by transforming to the on-shell momenta and using the interpretation of focusing of the Van-Vleck determinant. Then, 
$$\det{\left(\frac{\delta^2S(\phi^*,\phi_m)}{\delta\phi^2_m}+\frac{\delta^2S(\phi_m,\phi)}{\delta^2\phi_m}\right)}=\det{\left(-\frac{\delta\pi_f(\phi^*,\phi_m)}{\delta\phi_m}+\frac{\delta\pi_i(\phi_m,\phi)}{\delta\phi_m}\right)}$$ becomes the relative amount of focusing that occurs at the intermediary point, which we are integrating out.

Using \eqref{equ:additive}, from the  stationarity condition at an intermediary field configuration, we have (dropping the subscripts for clarity):
\be\label{equ:intermediary_condition}\frac{\delta S(\phi^*,\phi_m)}{\delta \phi_m}+\frac{\delta S(\phi_m,\phi_f)}{\delta \phi_m}=-\pi_f(\phi^*,\phi_m)+-\pi_i(\phi_m,\phi)=0
\ee this requires the momenta to be continuous at $\phi_m$, setting $\phi_m$ to be along the extremal path $\phi_m=\phi^\gamma_m$. Thus, given a classical path $\gamma$,  $\phi_m$ also depends on $\phi$ and $\phi^*$. Thus, deriving \eqref{equ:intermediary_condition} by $\phi$:
$$-\frac{\delta\pi_f(\phi^*,\phi_m)}{\delta\phi_m}\frac{\delta\phi_m}{\delta\phi}+\frac{\delta\pi_i(\phi_m,\phi)}{\delta\phi_m}\frac{\delta\phi_m}{\delta\phi}+\frac{\delta\pi_i(\phi_m,\phi)}{\delta\phi}=0
$$ 
\be\label{1}\Rightarrow~~-\frac{\delta\pi_f(\phi^*,\phi_m)}{\delta\phi_m}+\frac{\delta\pi_i(\phi_m,\phi)}{\delta\phi_m}=-\frac{\delta\pi_i(\phi_m,\phi)}{\delta\phi}\left(\frac{\delta\phi_m}{\delta\phi}\right)^{-1}
\ee
Writing equation  \eqref{equ:VV_comp} in terms of momenta we have: 
\be\label{equ:intermediary} 
\det{\left(-\frac{\delta \pi_i(\phi^*,\phi_m)}{\delta\phi_m}\right)}\det{\left(-{\frac{\delta\pi_f(\phi_m,\phi)}{\delta\phi_m}}\right)}\det{\left(-\frac{\delta\pi_f(\phi^*,\phi_m)}{\delta\phi_m}+\frac{\delta\pi_i(\phi_m,\phi)}{\delta\phi_m}\right)}^{-1}=\det{\left(-\frac{\delta\pi_i(\phi^*,\phi)}{\delta\phi}\right)}
\ee
Using the product rule for determinants we have that:
$$\det{\left(\frac{\delta \pi_i(\phi^*,\phi_m)}{\delta\phi_m}^{-1}\frac{\delta\pi_i(\phi^*,\phi)}{\delta\phi}\right)}=\det{\frac{\delta\phi_m}{\delta\phi}}$$
Finally, substituting this and  \eqref{1} into \eqref{equ:intermediary}:
\begin{align} 
& \det{\left(-\frac{\delta \pi_i(\phi^*,\phi_m)}{\delta\phi_m}\right)}\det{\left(-{\frac{\delta\pi_f(\phi_m,\phi)}{\delta\phi_m}}\right)}\det{\left(-\frac{\delta\pi_i(\phi_m,\phi)}{\delta\phi}\left(\frac{\delta\phi_m}{\delta\phi}\right)^{-1}\right)^{-1}}\nonumber\\
=&
\det{\left(-\frac{\delta \pi_i(\phi^*,\phi_m)}{\delta\phi_m}\right)}\det{\left(-{\frac{\delta\pi_f(\phi_m,\phi)}{\delta\phi_m}}\right)}\det{\left(-\frac{\delta\pi_i(\phi_m,\phi)}{\delta\phi}\left(\frac{\delta \pi_i(\phi^*,\phi_m)}{\delta\phi_m}^{-1}\frac{\delta\pi_i(\phi^*,\phi)}{\delta\phi}\right)\right)^{-1}}\nonumber\\
=&\det{\left(- \frac{\delta \pi_i(\phi^*,\phi_m)}{\delta\phi_m}\right)} \end{align}
where we note that an extra cancellation occurs since 
$\frac{\delta \pi_i(\phi_m,\phi)}{\delta\phi}=\frac{\delta \pi_f(\phi_m,\phi)}{\delta\phi_m}$ because  they are equal  as second derivatives of the action. 

Now, the continuity equation \eqref{equ:intermediary_condition}, turns the double sum of \eqref{rhs} into a single sum over the continuous extremal paths. Each overall extremal path decomposes into a unique composition  of extremal paths,  $\gamma_{\alpha'}=\gamma_{\alpha'_1}\circ\gamma_{\alpha'_2}$ with final (resp. initial) endpoints on $\phi_r$.   Putting it all together we obtain:
\be  \sum_{\alpha_1}\Delta_{\alpha_1}^{1/2}\exp{\left(i S_{\alpha_1}({\phi^*},\phi_r)/\hbar\right)}\sum_{\alpha_2}\Delta_{\alpha_2}^{1/2}\exp{\left(iS_{\alpha_2}(\phi_r,\phi_f)/\hbar\right)} = \sum_{\alpha}\Delta_{\alpha}^{1/2}\exp{\left(iS_{\alpha}(\phi^*,\phi)/\hbar\right)}
\ee
up to orders $ \order{\hbar^2}$. 

 ~~~~~~~ ~~~~~~~~~~~~~~~~~~~~~~~~~~~~~~~~~$\square$
\section{Probability currents for weakly coupled systems.}\label{app:conservation}
 Following section \ref{sec:conservation}, here I will briefly go over the derivation of the decomposition of currents and probabilities, under an inner product which is orthogonal in terms of gravity and its source fields.  
 \be\label{matter_WKB_app}
\Psi(g,\varphi)=\exp{(im_p^2 S[g])}A[g]\psi[g,\varphi]+\order{m_p^{-2}}
\ee 
I will assume the conservation law for $A$,  \eqref{conserv_imposed}, and  a specific form for the matter (or source) Hamiltonian given in \eqref{matter_Ham}, 
\be\label{matter_Ham_app}
\hat{H}_{\mbox{\tiny mat}}=-\frac{1}{2m}\partial_\varphi^2+\frac12 m v
\ee
where $v(\varphi)$ is the potential for the sources,  the quadratic momentum term in the Hamiltonian turns into
\be\label{source_laplacian}
\partial_\varphi^2=\int d^3x \left(g^{ab}(x) \frac{\delta^2}{\delta \varphi^a(x)\delta \varphi^b(x)}\right).
\ee
following \eqref{grav_Laplacian}  and, again, I have taken a simple factor ordering.\footnote{This is the ordering ignoring Christoffel symbols, i.e. ignoring covariance issues, both in configuration space and the spatial manifold. But everything goes through with $\frac{1}{|\det h|^{1/2}}\partial_a|\det h|^{1/2} h^{ab}\partial_b$, which makes everything covariant. The same is true for the gravitational modes. See footnote \ref{footnote_conserv}.}

For the (block) diagonal metric, we decompose the current in components as:
\be\label{currents_app}
J_{\mbox{\tiny grav}}^{ab}(x)=\frac{1}{2i\, m_p}\left(\bar{\Psi}\left(\frac{\delta \Psi}{\delta g_{ab}(x)}\right)-{\Psi}\left(\frac{\delta \bar\Psi}{\delta g_{ab}(x)}\right)\right); \qquad j_{\mbox{\tiny source}}^{a}(x)=\frac{1}{2i\, m}\left(\bar{\Psi}\left(\frac{\delta \Psi}{\delta \varphi^{a}(x)}\right)-{\Psi}\left(\frac{\delta \bar\Psi}{\delta \varphi^{a}(x)}\right)\right)
\ee
 for $\Psi$ given in \eqref{matter_WKB_app}. For a particle, we have 
 $$j_{\mbox{\tiny source}}^{a}(x)\rightarrow \delta(x-x^a)\left(\bar\Psi\partial_a\Psi- \Psi\partial_a\bar\Psi\right)
 $$ and $\partial_\varphi^2$ turns into the standard spatial Laplacian. Let\rq{}s look at this case for simplicity. In this case, the two currents of \eqref{currents_app}, reduce, with the approximation \eqref{matter_WKB_app}, to lowest order:
 \be\label{currents_app}
J_{\mbox{\tiny grav}}^{ab}(x)\rightarrow |A[g]|^2\frac{\delta S[g]}{\delta g_{ab}(x)}|\psi[g,\varphi]|^2; \qquad j^{\mbox{\tiny source}}_{a}\rightarrow \frac{1}{2i\, m}\left(\bar{\psi}\partial_a \psi -{\psi}\partial_a\bar\psi\right)
\ee
and, using \eqref{conserv_imposed}, the full continuity equation reduces to: 
 \be\label{conserv_approx_app}
 |A[g]|^2\left( \frac{\partial}{\partial T}|\psi|^2+\partial_a j^a\right)=0
 \ee
 where we have used \eqref{equ:time} as the arc-length parametrization (surfaces of equal on-shell action).  
 
 Now if we use the definition of the Klein-Gordon inner product, the normal to equal time surfaces is given by \eqref{equ:time}, i.e. 
 $$n_{ab}(x)=\frac{1}{V[g]}\diby{S}{g_{ab}(x)}$$ and from the block diagonal form of the supermetric for gravity and sources, we get: 
 \be\label{KG}
 \langle\Psi, \Psi\rangle_{\mbox{\tiny KG}}= \int_{\mathcal{S}_{(r)}} \mathcal{D}g\int d^3x\,  n_{ab}(x) J_{\mbox{\tiny grav}}^{ab}(x)= \int_{\mathcal{S}_{(r)}} \mathcal{D}g  |A[g]|^2 \int d^3x |\psi|^2
 \ee
where $\mathcal{S}_{(r)}$ is the screen,  relating the Klein-Gordon inner product and the Schroedinger one. 
\section{Jacobian}\label{app:Jacobian}
 \subsection{The Jacobian for the TT decomposition}
The spin decomposition wrt the background $  g_{ab}$:
\begin{align}
h_{ab} = h^\st{TT}_{ab} + \nabla_a X_b + \nabla_b X_a + 2 \, \nabla_a \nabla_b \theta - {\sfrac 2 3} \,   g_{ab} \, \Delta \theta   + {\sfrac 1 3} \, h \,   g_{ab} \,, 
\end{align}
where
\begin{align}
  g^{ab} \, h^\st{TT}_{ab} = 0 \,,  & &    \nabla^b h^\st{TT}_{ab} = 0 \,, & &   \nabla^a X_a = 0 \,,
\end{align}
%and we pick up the following Jacobian
%\begin{align}
%\mathfrak D[h_{ab}]
%=& ~ \mathfrak D[h^\st{TT}_{ab} , Z_a , h]
%\left| \text{det}  \, {M^a}_b\right|^{\frac 1 2} \\
%=& ~\mathfrak D[h^\st{TT}_{ab} , Z_a, h]
%\left| \text{det}  \, {M^a}_b\right| \, \left| \text{det}  \, {M^a}_b\right|^{- \frac 1 2} \nonumber 
%\\
%=& ~ \mathfrak D[h^\st{TT}_{ab} , Z_a , h] \int
%\mathfrak D[k_a , b_a] \exp�\left[ - {\sfrac 1 2} \int d^3 x \left( k_a \, {M^a}_b \, k^b + \, \bar b_a \, {M^a}_b \, b^b \right) \right] \nonumber 
%\end{align}
%where $k_a$ is a bosonic vector field, and $b_a$ is a fermionic vector field, and
%\begin{equation}
%{M^a}_b =  {\delta^a}_b \, \Delta - {\sfrac 1 3} \, \nabla^a \nabla_b - {R^a}_b \,,
%\end{equation}
%is the vector Laplacian, and we used the following Gaussian measure:
%\begin{equation}
%1 = \int \mathfrak D [h] \exp�\left[ - {\sfrac 1 2} \int d^3 x \sqrt g  \, h^2  \right]
%\end{equation}
%\begin{equation}
%1 = \int \mathfrak D [Z_a] \exp�\left[ - {\sfrac 1 2} \int d^3 x \sqrt g  \,  g^{ab} \, Z_a \, Z_b  \right] 
%\end{equation}
%\begin{equation}
%1 = \int \mathfrak D [h^\st{TT}_{ab}] \exp�\left[ - {\sfrac 1 2} \int d^3 x \sqrt g  \,  g^{ac} g^{bd} \,h^\st{TT}_{ab} h^\st{TT}_{bd}  \right]
%\end{equation}
we pick up a Jacobian
\begin{align}
1 =& \int \mathcal D [h_{ab}] \exp \left[ - {\sfrac 1 2}\int d^3 x \sqrt g\, h^{ab} h_{ab} \right] \\
\nonumber
= & J_\st{TT} 
\int \mathcal D [h^\st{TT}_{ab},X_a,\theta,h]  \exp \bigg{[}  - {\sfrac 1 2} 
\int d^3 x \sqrt g  \left( h^{ab}_\st{TT} h_{ab}^\st{TT} + {\sfrac 1 3} \, h ^2 - 2 \, X_a ( R^{ab}  + g^{ab} \, \Delta ) X_b  \right. \\
\nonumber
& \qquad \qquad  \qquad \qquad \qquad \qquad \qquad \qquad
\left. 
+ \sigma \,  ( {\sfrac 8 3} \Delta + 4 \, R_{ab}{}^{;b} \, \nabla^a + 4 \, R_{ab} \, \nabla^b \, \nabla^a ) \sigma
\right. \\
\nonumber
& \qquad \qquad \qquad \qquad \qquad \qquad \qquad \qquad
\left. + 4\,  \sigma \, (  R^{ab}{}_{;b} +  R^{ab} \, \nabla_b ) X_a - 4 \, X_a \, R^{ab} \, \nabla_b \sigma 
\right) \bigg{]}
\end{align}
defining
\begin{align}
1 &= \int \mathcal D [h] \exp\left[ - {\sfrac 1 2} \int d^3 x \sqrt g  \, h^2  \right] = \int \mathcal D [\theta] \exp\left[ - {\sfrac 1 2} \int d^3 x \sqrt g  \, \theta^2  \right]
\\
1 &= \int \mathcal D [X_a] \exp\left[ - {\sfrac 1 2} \int d^3 x \sqrt g  \,  g^{ab} \, X_a \, X_b  \right] 
= \int \mathcal D [h^\st{TT}_{ab}] \exp\left[ - {\sfrac 1 2} \int d^3 x \sqrt g  \,  g^{ac} g^{bd} \,h^\st{TT}_{ab} h^\st{TT}_{bd}  \right]
\end{align}
we get
\begin{equation}\label{TT_Jacobian}
J_\st{TT} = \left| \det \left(  
\begin{array}{cc}
- 2 ( R^{ab}  + g^{ab} \, \Delta) &  - 4 \, R^{ab} \, \nabla_b 
 \\
 4 (  R^{ab}{}_{;b} +  R^{ab} \, \nabla_b )  & {\sfrac 8 3} \Delta + 4 \, R_{ab}{}^{;b} \, \nabla^a + 4 \, R_{ab} \, \nabla^b \, \nabla^a
\end{array}
\right) \right|^{-1} \,.
\end{equation}

\subsection{Arc-length parametrization}\label{app:arc_length}

If we  gauge-fix the time-reparametrizations,  we choose to be an arc-length one: 
 \be\label{FP_arc}
G_t:=\sqrt{\int_M
\sqrt{g}d^3x \,\,(\dot {g}^{cd}(t)\dot {g}_{cd}(t))}-1=0
 \ee
 The infinitesimal version, i.e. for $\delta_\epsilon t= c+\epsilon t+\order{\epsilon^2}$, 
 $$G_{t^\epsilon}= |\epsilon|\sqrt{\int_M
\sqrt{g}d^3x \, \,(\dot {g}^{cd}\dot {g}_{cd})}-1$$
which gives the variation: 
\be \label{equ:FP_time}
\frac{\delta G_{t^\epsilon}}{\delta\epsilon}_{|G_t=0} = \pm 1
\ee
This gauge choice is responsible for the absence of the global square root.   
In other words, with this gauge choice one can express the action as an energy functional (in the Riemannian geometry sense) as opposed to a length functional:
$$\int ds \|\dot{g}\|^2\, ~\mbox{as opposed to}~~\int ds \|\dot{g}\|$$ The main feature however of the energy functional is that extremals wrt it automatically implement arc-length parametrization as one of the extremum conditions. In other words, since the length and energy functional coincide for arc-length parametrization, and this is implied by the extrema of the energy functional, in the semi-classical approximation one can directly use the energy functional as opposed to the length.  In a way, this gauge fixing `trades' global time reparametrization invariance for locality of the action functional.

\subsection{Eigenfunction basis}
In terms of  eigenfunctions of the Laplacian, using the notation of \cite{Gerlach}, we write 
$$\nabla^2 G^{(n)}_{ij}=\lambda_n G^{(n)}_{ij}$$
For the 3-sphere, one finds that $\lambda_n=-(n^2-3)$, for $n\geq 3$. The solution to these equations, in standard spherical coordinates is given by a combination of $n^2-4$ terms: 
\be\label{eigenbasis} G^{(n)}_{ij}(\chi,\theta,\phi)=\sum_{l=2}^{n-1}\sum_{m=-l}^l C^n_{lm}(G_{ij})^n_{lm} 
\ee
where $C^n_{lm}$ are the Fourier coefficients, and for each $ij$, $(G_{ij})^n_{lm} $ is a TT tensor harmonic, explicitly computed in \cite{Gerlach} (eqs 17-18c). Since $S^3$ is compact without boundary, and $\nabla^2$ is self-adjoint with respect to the standard supermetric 
\be\label{equ:supermetric}\langle u|\nabla^2 v\rangle_{g}= \int d^ 3x\, u_{ab} \, g^{ac}\, g^{bd}\,\nabla^2 v_{cd}\sqrt{g}=\langle \nabla^2 u| v\rangle_g
\ee by the spectral theorem for bounded operators, eigenfunctions with distinct eigenvalues are orthogonal. 

Now, for simplification, instead of \eqref{eigenbasis}, for a normalized basis (under the standard integral  and sphere metric), 
$$\int d^3x \sqrt{g_0} \kappa^n_{ij}g_{0}^{ik} g^{jl}_{0}\kappa^m_{kl}=\delta^{n,m}$$
  we simply write,  
\be\label{basis_S3} h_{ij}(x)=\sum_n\lambda_{ij}^n \kappa_{ij}^n(x)\, \qquad \prod dh_{ij}(x)\rightarrow \prod_{n=1}^\infty d\lambda_{ij}^n
\ee 
If the basis is not normalized, a Jacobian appears for the transformation. 

\section{The full Van-Vleck determinant}\label{app:Van_Vleck}
I start by rewriting, for the reader\rq{}s convenience, \eqref{geod_riem}:
$$S[\gamma]=\int dt\left(\int d^ 3x\, \dot \gamma_{ab} \, g^{ac}\, g^{bd}\, \dot \gamma_{cd}\sqrt{g}\right)^{1/2}
$$ 
and \eqref{equ:geodesic_gravity}
$$\gamma_{ab}(t)=g^0_{ac} (e^{(A(t)\mbox{\tiny Id}+B(t)\mathbf{h}_T) })^c_b$$
where $\mathbf{h}$ is the matrix given by $h^a_b=h_{bc}g_0^{ac}$, such that $\nabla_0^ah_{ab}=0=h_{ab}g_0^{ab}$ (the traceless condition is taken for simplicity), and 
$$A(t)= \frac{2}{3}\ln (1+\frac{3}{8}bt^2)~~\mbox{and}~~~B(t)= \frac{4}{\sqrt{3 b}}\tan^{-1}\left(\frac{\sqrt{3b}\,t}{4}\right)$$
where $b= h^T_{cd}h_T^{cd}$. The strategy will be to compute the Van-Vleck matrix with the following chain rule:
\be\label{chain}\frac{\delta^2 S}{\delta g_{cd}^0(x)\delta g_{ab}(1)(y)}=\frac{\delta}{\delta g_{cd}^0(x)}\left(\int d^ 3 x\rq{} \frac{\delta S}{\delta h_{ij}(x\rq{})}\cdot \frac{\delta  \gamma_{ij}(1)(x\rq{})}{\delta h_{ab}(y)}\right)\ee 
I start by noticing that the usual trick to deal with the square root in \eqref{geod_riem} is to use instead \textit{the energy functional}, i.e. the version without the square root. It is easy to show that the equations of motion then give two types of equations: one says that the paths should be arc-length parametrized, and the other that the curve will be geodesic. Moreover, I also note that the conservation arguments section \ref{sec:conservation} go through, slightly modified -- the global Hamiltonian is conserved in time, thus giving a non-local conserved energy for each curve.\footnote{Note however that in this case the Hamiltonian is conserved along each trajectory, but it is not a primary constraint, as it is in the case of the geodesic action. }

To simplify the notation, I define the matrix $\mathbf{M}:=(A(t)\mbox{Id}+B(t)\mathbf{h}) $.  Its derivative is:
\be\label{equ:Mdot}
\dot M^a_b=\frac{24 bt}{48+18bt^2} \delta^a_b +\frac{16}{16+3bt^2} h^a_b
\ee
whose dimensions in arc-length units can be checked to be of $1/$sec.  From the geodesic equation \eqref{equ:geodesic_gravity}, the inverse metric along it is
\be\label{inverse}
g^{ab}=g_0^{ad}((e^{\mathbf{M}})^{-1})^b_d
\ee
and the metric velocity is given by
\be\label{geod_velocity} \dot\gamma_{ab}=g^0_{ac}\dot M^c_d (e^{\mathbf{M}})^d_b
\ee
thus:
$$
 \dot \gamma_{ab} \, g^{ca}\, g^{bd}\, \dot \gamma_{cd}=g^0_{bf}\dot M^f_m (e^{\mathbf{M}})^m_a g_0^{ci}((e^{\mathbf{M}})^{-1})^a_i
 g_0^{bj}((e^{\mathbf{M}})^{-1})^d_j
 g^0_{ck}\dot M^k_l (e^{\mathbf{M}})^l_d=\dot M^a_b\dot M^b_a
$$ and, since we are assuming the trace of $h_{ab}$ vanishes:
\be\label{kinetic}
\dot M^a_b\dot M^b_a=3\left(\frac{24 bt}{48+18bt^2}\right)^2+\left(\frac{16}{16+3bt^2}\right)^2b
\ee
where the traces are taken with respect to $g^0$. 

The change in volume form (even for an initial traceless velocity)  is 
\be\label{volume_form}\sqrt{g}d^3x(t)= \ln (1+\frac{3}{8}bt^2) \sqrt{g^0}d^3x
\ee
and 
$$ \frac{\delta S}{\delta h_{ij}(x)}=2\int d^3 x\rq{}\frac{\delta  S}{\delta b(x\rq{})} \frac{\delta b(x\rq{})}{ \delta h^{ij}(x)}=2\frac{\delta  S}{\delta b(x)}  h^{ij}(x)
$$ since in these simple cases all quantities are ultra-local. Thus from \eqref{kinetic} and \eqref{volume_form} we calculate
\begin{align}
\frac{\delta S}{\delta h_{ij}(x)}&=2\int_0^1dt\,\, h^{ij}(x)\frac{\delta}{\delta b(x)}\int d^3 x\rq{} \sqrt{g^0}\left(3\left(\frac{24 bt}{48+18bt^2}\right)^2+\left(\frac{16}{16+3bt^2}\right)^2b\right) \ln (1+\frac{3}{8}bt^2)\nonumber\\
&=96\int_0^1dt\,  h^{ij}(x) t^2 \left(\frac{b^3 t^2+\frac{48 \left(3 b t^2+8\right)^2}{\left(3 b t^2+16\right)^2}}{\left(3 b t^2+8\right)^3}+\left(\frac{b^2 \left(b t^2+8\right)}{\left(3 b t^2+8\right)^3}-\frac{96}{\left(3 b t^2+16\right)^3}\right) \ln \left(\frac{3 b t^2}{8}+1\right)\right)\label{delta S}
\end{align}

For the other component of \eqref{chain}, we have
\be  \frac{\delta  \gamma_{ij}(1)(x)}{\delta h_{ab}(y)}=g^0_{ic}\frac{\delta M^c_d(x)}{\delta h_{ab}(y)} (e^{\mathbf{M}})^d_j (x)
\ee
where 
\be\label{delta_M1}\frac{\delta M^c_d(x)}{\delta h_{ab}(y)}=2\frac{\delta M^c_d(x)}{\delta b(y)}  h^{ab}(y)+\frac{16}{16+3bt^2}g_0^{c(a}\delta ^{b)}_d\delta(x,y)
\ee
and
$$ \frac{\delta M^c_d(x)}{\delta b(y)}=\left(\frac{2 t^2}{3 b t^2+8}\delta^c_d+\left(\frac{8 t}{3 b^2 t^2+16 b}-\frac{2 \tan ^{-1}\left(\frac{1}{4} \sqrt{3} \sqrt{b} t\right)}{\sqrt{3} b^{3/2}}\right) h^c_d\right)\delta(x,y)
$$
Putting these two together we get:
\be\label{delta_M2}
 \frac{\delta  \gamma_{ij}(1)(x)}{\delta h_{ab}(y)}
 =\left(\left(\frac{4 t^2}{3 b t^2+8}g^0_{id}+2\left(\frac{8 t}{3 b^2 t^2+16 b}-\frac{2 \tan ^{-1}\left(\frac{1}{4} \sqrt{3} \sqrt{b} t\right)}{\sqrt{3} b^{3/2}}\right) h_{id}\right)  h^{ab}+\frac{32}{16+3bt^2}\delta_i^{(a}\delta ^{b)}_d\right) (e^{\mathbf{M}})^d_j\delta(x,y)
\ee
Since there were no derivatives, the contraction between \eqref{delta_M2} and \eqref{delta S} required by \eqref{chain} is a trivial multiplication. Finally, 
\begin{align}
&\frac{\delta}{\delta g_{cd}^0(x)}\left(\int d^ 3 x\rq{} \frac{\delta S}{\delta h_{ij}(x\rq{})}\cdot \frac{\delta  \gamma_{ij}(1)(x\rq{})}{\delta h_{ab}(y)}\right)=\nonumber\\
&\frac{\delta}{\delta g_{cd}^0(x)}\Big[ 96\int_0^1dt\,  t^2 \left(\frac{b^3 t^2+\frac{48 \left(3 b t^2+8\right)^2}{\left(3 b t^2+16\right)^2}}{\left(3 b t^2+8\right)^3}+\left(\frac{b^2 \left(b t^2+8\right)}{\left(3 b t^2+8\right)^3}-\frac{96}{\left(3 b t^2+16\right)^3}\right) \ln \left(\frac{3 b t^2}{8}+1\right)\right)\times\nonumber\\
&\left(\left(\frac{4 t^2}{3 b t^2+8}h^j_d+2\left(\frac{8 t}{3 b^2 t^2+16 b}-\frac{2 \tan ^{-1}\left(\frac{1}{4} \sqrt{3} \sqrt{b} t\right)}{\sqrt{3} b^{3/2}}\right) h_{id}h^{ij}\right)  h^{ab}+\frac{32}{16+3bt^2}h^{j(a}\delta ^{b)}_d\right) (e^{\mathbf{M}})^d_j(y)\Big]\label{final_full}
\end{align}
For the last step in evaluating \eqref{chain}, one applies the chain rule for the derivative in \eqref{final_full}, using that $h^i_j=g^{ia}_0h_{aj}$, and 
$$\frac{\delta (e^{\mathbf{M}})^c_b}{\delta g^0_{ij}}= \frac{\delta{M^c_d}}{\delta g^0_{ij}} (e^{\mathbf{M}})^d_b$$ and implementing 
$$\frac{\delta}{\delta g_{cd}^0(x)}b(y)=2 h_{ac}h^{a}_d\delta(x,y)
$$
One finds that even for this simple case, the result is an immensely complicated function of the the initial velocity $h_{ab}$. Taking the determinant would once again require us to perform some regularization, as was implicitly done by taking the pointwise trace in the approximation in \eqref{Ricci}.

\section{Locality}\label{app:locality}
There is a type of  reciprocity between physical space and configuration space which can be described as follows. 
Whereas fixing the entire field configuration (non-locally on $M$) defines a point in the configuration space $\mathcal{Q}$, fixing only a partial field configuration on a subset of $M$ determines an entire submanifold of $\mathcal{Q}$.\footnote{To be more careful, I should only ascribe to them the title of `subsets', not submanifolds. However, under reasonable assumptions, as shown in the accompanying paper \cite{Locality_riem} they indeed form submanifolds.} Such submanifolds are formed by all of the configurations which have that same fixed field, let's say $\phi_{\mbox{\tiny{O}}}$ defined on $O\subset M$,  i.e. those fields $\phi$ which coincide with $\phi_{\mbox{\tiny{O}}}$ on  $O$ but are arbitrary elsewhere: 
$$\mathcal{Q}_{\phi_{\mbox{\tiny{O}}}}:= \{\phi\in \mathcal{Q}~|~\phi_{|\mbox{\tiny{O}}}= \phi_{\mbox{\tiny{O}}}\}
$$

Extremal curves in the configuration space of the complete Universe will be in general non-local on $M$, although of course local in $\mathcal{Q}$. In the accompanying paper \cite{Locality_riem}, we provide criteria for the sort of locality we are accustomed to, in the physical space $M$. 

Namely, suppose that the action admits a Jacobi-length representation, and  let $O$ be a region  in $M$,\,  $O\subset M$, defining an embedded submanifold $\mathcal{Q}_O$ (with appropriate boundary conditions, in a specific way we spell out in \cite{Locality_riem}). We show that for  semi-classical clustering -- i.e.  decoupling of $O$ from the rest of $M$   --  to occur on a given region of configuration space,  $\mathcal{U}\subset\mathcal{Q}$, we must have that  $\mathcal{Q}_O\cap \mathcal{U}$ is  \emph{a totally geodesic submanifold of $\mathcal{U}$}. We show that with this condition the semi-classical path integral kernel (in the arc-length gauge) decouples (here written for regions $O_k$), i.e. the kernel over the union of the regions turns into a product over each individual region:
\be\label{equ:cluster}W^{\bigcup_k O_k}(\phi^i_{|\dot\bigcup_k O_k}, \phi^f_{|\dot\bigcup_k O_k})\approx \prod_k W^{O^k}_{\mbox{\tiny{cl}}}(\phi^i_{O_k}, \phi^f_{O_k})\ee
giving us our version of clustering decomposition.

I should also note that there is an inherent non-locality with the Jacobi form of reparametrization invariant actions (they possess a global square root). However,  there is a preferred gauge in which they can become local: the arc-length gauge. In this gauge, the action becomes an energy functional (in Riemannian geometry terminology), and it is possible to show that the Fadeev Popov determinant is field independent. In other words, this particular gauge-fixing can trade time reparametrization invariance for locality.

     \end{appendix}

%\bibliographystyle{ieeetr}

%\bibliography{decoherence}

\end{document}